\documentclass[10pt,twocolumn,letterpaper]{article}

\usepackage[pagenumbers]{iccv} % To force page numbers, e.g. for an arXiv version

\newcommand{\TODO}[1]{\textbf{\color{red}[TODO: #1]}}
\newlength{\imagewidth}

\usepackage{graphicx}
\usepackage{afterpage}
\usepackage{tabularx}
\usepackage{colortbl}
\usepackage{multirow}
\usepackage{subcaption}
\usepackage{adjustbox}
\usepackage{array}
\usepackage{makecell}
\usepackage{booktabs}
\usepackage{wrapfig}
\usepackage{diagbox}
\definecolor{iccvblue}{rgb}{0.21,0.49,0.74}
\usepackage[pagebackref,breaklinks,colorlinks,allcolors=iccvblue]{hyperref}
\usepackage{stfloats}

\newcommand{\eat}[1]{}
\def\paperID{9147} % *** Enter the Paper ID here
\def\confName{ICCV}
\def\confYear{2025}

\title{Synthetic Video Enhances Physical Fidelity in Video Synthesis}

\author{Qi Zhao$^{1}$
\qquad
Xingyu Ni$^{2,1}$
\qquad
Ziyu Wang$^{3,1}$
\qquad
Feng Cheng$^{1}$
\qquad
Ziyan Yang$^{1}$\vspace{0.2em}
\\
Lu Jiang$^{1}$\textsuperscript{*}
\qquad
Bohan Wang$^{4}$\textsuperscript{*}\vspace{0.4em}
\\
{\normalsize
$^{1}$ByteDance Seed\qquad
$^{2}$Peking University\qquad
$^{3}$ShanghaiTech University\qquad
$^{4}$National University of Singapore
}\vspace{-0.5em}
}

\begin{document}
\twocolumn[{%
\renewcommand\twocolumn[1][]{#1}%
\maketitle

% \eat{
\includegraphics[width=\linewidth]{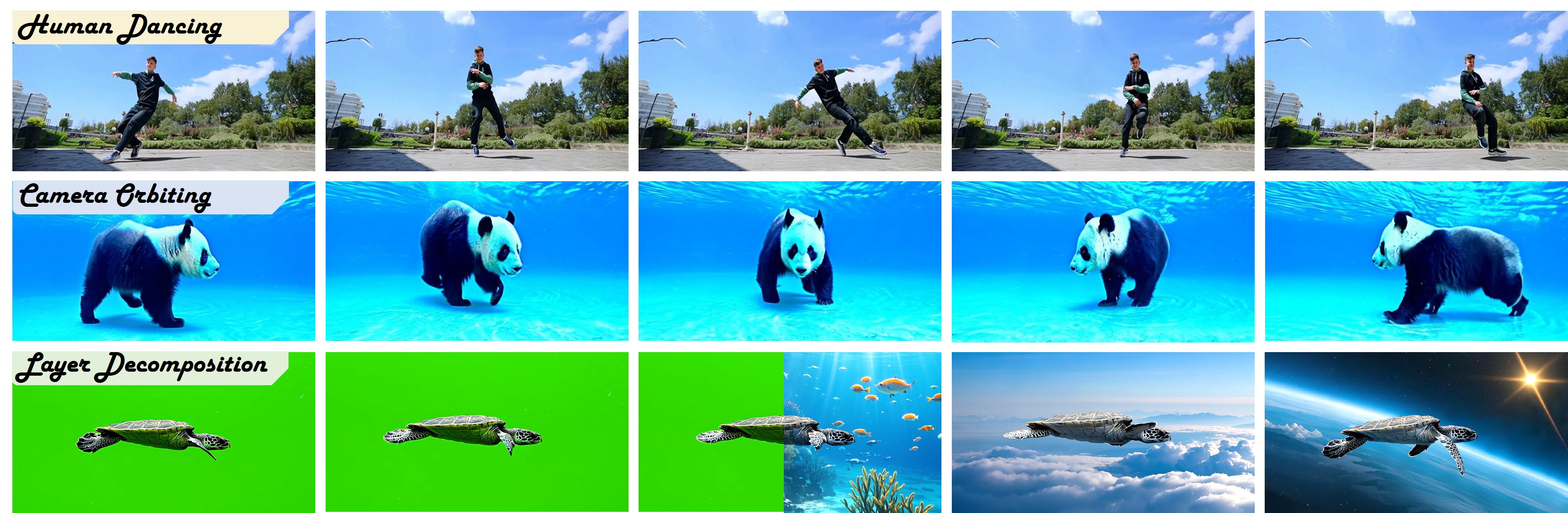}
% \vspace{-2em}
\captionof{figure}{Our synthetic-data-enhanced video generation model is capable of producing videos depicting human dancing (rows 1), scenes featuring large camera orbiting around the object (row 2), and animals against solid-color backgrounds for matting (row 3).}
\label{fig:teaser}
\vspace{0.2in}
% }
}]
\renewcommand{\thefootnote}{*}\footnotetext{Corresponding author}
% \maketitle
% \begin{@twocolumnfalse}
% \begin{figure*}[t]
% \includegraphics[width=\textwidth]{visualizations/bannerv2.png}
% \label{fig:banner}
% \end{figure*}
% \end{@twocolumnfalse}
% \eat{
\begin{abstract}
% \TODO{redo}
% We present a study on augmenting real-world video data with synthetic data to train high-quality video generative models. While synthetic data augmentation is widely employed in various machine learning domains, relatively little research has focused on effective strategies for generating and integrating synthetic video data for this purpose. By leveraging 3D rendering techniques from computer graphics, we not only produce large-scale synthetic videos but also identify key insights for generating and incorporating them into model training pipelines. 
% % \rule{0pt}{35em}
% Through both qualitative and quantitative evaluations, we demonstrate that strategically adding synthetic videos significantly expands a model's capabilities, enabling 3D-consistent, physically realistic video generation -- including humans with large motions, significant camera movements, and scene decomposition -- tasks that have been challenging with real-world data alone. 
We investigate how to enhance the physical fidelity of video generation models by leveraging synthetic videos derived from computer graphics pipelines. These rendered videos respect real-world physics, such as maintaining 3D consistency, and serve as a valuable resource that can potentially improve video generation models.
To harness this potential, we propose a solution that curates and integrates synthetic data while introducing a method to transfer its physical realism to the model, significantly reducing unwanted artifacts. Through experiments on three representative tasks emphasizing physical consistency, we demonstrate its efficacy in enhancing physical fidelity. While our model still lacks a deep understanding of physics, our work offers one of the first empirical demonstrations that synthetic video enhances physical fidelity in video synthesis. Website: \href{https://kevinz8866.github.io/simulation/}{https://kevinz8866.github.io/simulation/}

% Through extensive experiments on three prepresentative generation that emphsize phydical consistency in the video generation, 
% we identify key insights for generating and incorporating them into model training pipelines,
% and demonstrate that strategically adding synthetic videos significantly improve the physical fidelity of the generated video on tasks including humans with large motions, 
% significant camera movements, 
% and vide layer decomposition --- 
% tasks that have been challenging with real-world data alone.
\end{abstract}

\section{Introduction}
% \TODO{mention not aesthetics, improve physics, don't overclaim} 
Video generation models~\cite{videoworldsimulators2024, veo2,kling,ehtesham2024movie,kong2024hunyuanvideo} have demonstrated strong capabilities in producing high-quality and visually compelling videos of real-world scenarios.
% , particularly after pretraining on massive real-world video data. 
Despite their remarkable progress, 
these generation videos often struggle to respect the underlying physical laws of the real world, 
indicating a significant gap in applications where physical fidelity is essential~\cite{kang2024far, yang2024video, yang2024direct}.
For instance, while a video generation model can generate realistic-looking objects or humans within a scene, it may fail to maintain 3D consistency when the camera moves or when the subjects undergo deformation.
%Such shortcomings highlight the necessity for improving physical fidelity in video generation models.

In this paper, we explore whether synthetically generated videos can enhance the physical fidelity of video generation models. Specifically, we utilize synthetic videos rendered through modern computer-generated imagery (CGI) production pipelines used in gaming and film, such as Blender~\cite{Blender} and Unreal Engine~\cite{UE5}. By utilizing standard computer graphics techniques, we can generate high-quality, physically consistent video content at scale. CGI production pipelines generate videos via precise 3D asset modeling, animation, and rendering based on predetermined physical rules~\cite{de2022next}. This approach allows for highly accurate scene configuration and ensures that the rendered videos intrinsically respect real-world physics, provided the setups and parameters are properly specified. As such, synthetic video is highly configurable, allowing precise control over scene setup, objects, and motion. Additionally, ground-truth descriptions can be easily obtained based on the specifications of the 3D environment.

However, training video generation models using synthetic video data presents several challenges. Synthetic videos inherit an appearance gap, making them easily distinguishable from real videos. Further, the limited availability of 3D assets, together with the complexity of their composition, restricts the diversity of synthetic video content. As a result, leveraging synthetic video to enhance model understanding remains an active area of research~\cite{liu2025generative,Kim2022Video,liang2020simaug}. Regarding video generation, to the best of our knowledge, no prior work has specifically explored the use of synthetic videos to enhance video generation models.

Therefore, we present an investigation into how synthetic video enhances the physical fidelity of video generation models. As a pilot study, we examine three representative tasks known to be challenging even for state-of-the-art video generation models. \Cref{fig:teaser} illustrates their generated videos which include: 1) Large human motion generation, where significant movements cause noticeable shape deformations in body parts, such as breakdance or backflip. 2) Wide-angle camera rotation, where the camera spins around a specific axis, capturing a broader field of view of the object or actions. 3) Video layer decomposition, where the model must generate a subject or motion against a green screen background. This task evaluates whether the model can effectively disentangle the subject from the background during generation. These tasks are not exhaustive but serve as a reasonable starting point for studying physical fidelity in video generation.

We propose a solution that uses synthetic videos to enhance video generation models. At the data level, based on computer graphics techniques, we begin by constructing a synthetic video generation pipeline that offers diverse scene configurations, assets, and animations. 
Next, we explore the curation and integration of synthetic videos to transfer their physical fidelity to the video generation model.
% facilitating the generation of real-world videos that better adhere to predefined physical properties.
% We also study how to effectively train video generation models with synthetic videos, 
% focusing on seamlessly integrating synthetic data with the real-world videos. 
Through extensive analysis and ablations, we identify key factors that govern how well synthetic videos transfer physical fidelity to real-world video generation, including visual distribution, asset quality, rendering quality, the role of synthetic captions and the best blending strategy of synthetic videos with their real counterparts. %Determining the  best to blend synthetic videos with real-world examples during model training remains an open research question.

At the model level, we propose a novel approach \emph{SimDrop} to reduce the introduction of undesirable rendering artifacts into the final generation model by training a synthetic reference model that solely captures the visual patterns of synthetic video data. 
%We found that special tags in synthetic captions have a strong effect in distinguishing the correct properties of synthetic video data to transfer. 
%Therefore, we train synthetic reference models that aims to capture only the visual properties of synthetic data through special tags. 
We show that with classifier-free guidance~\cite{ho2021classifier}, the reference model can work in auxiliary with the generation model to remove the visual artifacts from synthetic data but keeps the physical fidelity.

To verify the effectiveness of our solution, we employ two measurements inspired by related works~\cite{khirodkar2024sapiens, agarwal2025cosmos}, assessing fidelity in terms of 3D consistency and human pose integrity. While these measurements are not perfect, they offer meaningful indicators of the physical fidelity of video generation. Additionally, human evaluations are incorporated to ensure alignment with human perception.
% we evaluate our models on three representative tasks known to be difficult for current video generation models: large human motion, large camera motion, and video layer decomposition.
% These tasks underscore the limitations of existing state-of-the-art models with respect to physical fidelity. 
Our experiments demonstrates that by carefully crafting and integrating synthetic video data, 
video generation models can significantly reduce collapse and distortion in human motion and improve 3D consistency~\cite{agarwal2025cosmos} of objects under large camera movements. 
Moreover, our approach enables models to generate backgrounds of uniform color while maintaining clearly separated, 
dynamically moving objects in the foreground. It is worth noting that while our model improves physical fidelity, it still lacks an understanding of the underlying principles of physics, leaving significant room for further improvement

In summary, we make the following contributions:
\begin{itemize}
    \item We present a computer graphics-based synthesis pipeline to generate videos for training video generation models.
    \item We identify key factors in curating synthetic video data and propose strategies for effectively training video generation models on these datasets.
    \item To the best of our knowledge, our work provides one of the first empirical demonstrations that incorporating synthetic video data can improve the physical fidelity of video generation models.
\end{itemize}

\section{Synthetic Video Generation using Computer Graphics Techniques}
\label{sec:engine}
Augmenting datasets with synthetic data has been widely adopted in the field of machine learning.
Specifically, standard CGI production pipelines, 
such as those implemented in Blender~\cite{Blender} or Unreal Engine~\cite{UE5},
have long been employed to synthesize highly controlled and visually realistic image and video data.
By explicitly modeling objects, cameras, environments, and illumination,
they offer fine-grained control over every aspect of a scene,
enabling the generation of large-scale, diverse, and visually realistic video datasets.

Our data synthesis pipeline is built on such a CGI production pipeline.
We build a procedural 3D scene generator driven by a carefully chosen set of parameters,
enabling diverse 3D scene generation.
Then, we couple it with the open-source rendering engines Unreal Engine and Blender to generate high-quality video outputs.
Based on the three aforementioned challenging tasks,
we focus on generating videos containing a single object per scene and 
aim to maximize diversity in both appearance and motion.
Following standard practice, we consider a 3D scene to include four key components:
(1) the 3D object, (2) the camera, (3) the lighting conditions, and (4) the environment.
Each component is fully customizable through a set of predefined parameters, as detailed in~\cref{sec:engine_details}.
Then, our pipeline automatically converts the parameters into a 3D scene and renders them into a video.
Next, we will explain how we effectively sample the parameter space to achieve our goal.

\begin{figure*}
    \centering
    \includegraphics[width=1\linewidth]{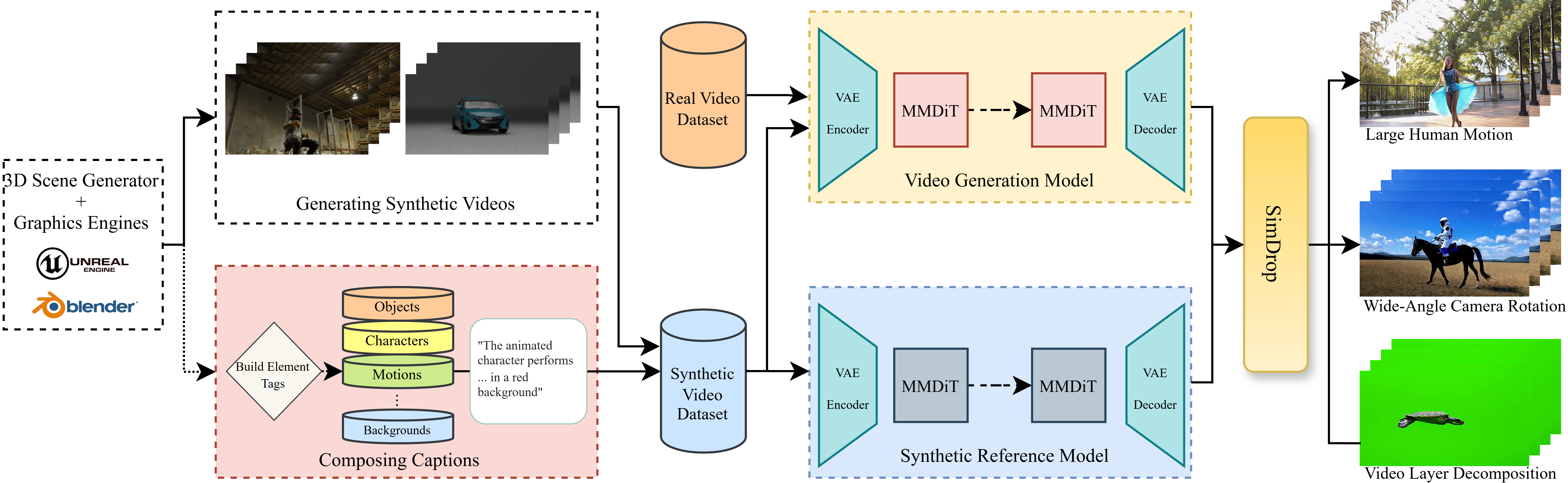}
    \caption{Visualization of the pipeline to augment video generation model with synthetic video data. We first plan the synthetic videos and generation descriptive tags for each elements (e.g. object, character, motion, etc). Then we combine the element descriptions to form the caption for synthetic videos. During training, we mix the synthetic videos with real-world video data to improve physics fidelity in challenging video generation tasks.}
    \label{fig:main_vis}
\end{figure*}

\section{Method}
\label{sec:syn-design}

Our goal is to investigate how data augmentation with synthetic videos can enhance a video generation model to produce physically consistent videos. As a pilot study, this paper focuses on three specific generation tasks, each representing a challenging generation task even for state-of-the-art video generation models: large human motion, wide-angle camera rotation, and video layer decomposition. We assess quality primarily based on physical fidelity (see \cref{sec:physics_metrics} for metric definition), rather than the commonly used visual fidelity or aesthetics.

Training video generation models with synthetic video data presents challenges due to the distributional gap between synthetic and real videos. Our method addresses the gap between synthetic and real videos through three key techniques: data curation, a captioning strategy, and a novel training approach. \Cref{fig:main_vis} provides an overview of our method, illustrating how these components work together to enhance video generation. In the following, \Cref{sec:curate} presents the curation of the synthetic video pixels. \Cref{sec:3_2} explains how we caption the synthetic videos. Lastly, \Cref{sec:3_3}  details our strategy and method to incorporate the synthetic data.

% \lu{reduced the following to a couple of sentences}
% \subsection{Training Video Generation Models with Synthetic Data}

% In the past, these foundation models have relied almost exclusively on large collections of real-world videos~\cite{kong2024hunyuanvideo,ehtesham2024movie}, 
% which tend to exhibit rich visual diversity.

% Nonetheless, harnessing synthetic datasets carries the promise of enhancing the physical realism of generated outputs, 
% as these carefully designed scenes can contribute to strengthening the physical fidelity of video generation models.

% A primary advantage of using synthetic data is the ability to generate videos under precisely controlled setups,  allowing for fine-grained manipulation of the synthesis.
% From our extensive experiments with synthetic videos, 
% we explore and analyze several critical aspects of augmenting real video data with synthetic videos that contribute to the improvement of physical fidelity of video generation models.
% \Cref{fig:main_vis} shows an overview of our method.

\subsection{Curating Synthetic Pixels}
\label{sec:curate}
% The design of the data distribution is pivotal for developing high-quality video generation models. 
This section explores strategies for narrowing the gap between real and synthetic videos by refining synthesis configurations -- including camera, background, object, lighting, and other visual factors -- as well as a study examining the impact of visual appearance brought by asset quality and rendering quality.
% Generating synthetic videos that pose significant challenges for video generation models, 
% then leveraging these videos for training, 
% can substantially enhance the physical fidelity of the resulting outputs.
% In this work, 
% we address two difficult scenarios: 

\begin{table}
\centering
\small
\begin{tabular}{l|c}
\toprule
Training Data & Human Motion Collapse Rate \\ 
\midrule
(a) Random  & 87\%  \\
(b) Forward shot only  & 42\%  \\
(c) Forward + following shot & 23\%  \\
\bottomrule
\end{tabular}
\caption{Randomly chosen camera configurations (a-b) lead to high collapse rate for generated videos. 
Using configuration (c) aligning with the real world greatly reduce the rate.}
\label{tab:camera-selection}
\vspace{-0.1in}
\end{table}

% using noident textbf for tight formatting
\noindent\textbf{Synthesis configurations\ }
Our generation tasks require producing videos that maintain 3D consistency for objects and ensure body coherence in human motion. To achieve this, we synthesize videos that emphasize these aspects by incorporating large object deformations (\eg, human dance) and significant camera rotations (\eg, orbiting around objects).
% We also enhance the diversity of synthetic videos by varying backgrounds, camera setups, and objects (see \cref{sec:engine_details}).
Additionally, it is beneficial to incorporate characteristics of real videos such as common camera setups. For instance, professional videographers often capture a subject’s upper body from frontal angles when filming humans. To align with this practice, we ensure that a significant portion of our synthetic data follows similar configurations.

To demonstrate this, we examine the effectiveness of synthetic videos with different camera configurations: random, forward-shot only, frontal, and following shots. As shown in \Cref{tab:camera-selection}, we find that synthetic videos incorporating both forward and following shots, which closely align with real-world camera setups, significantly enhance the video generation model. This approach notably reduces the collapse rate -- defined as the proportion of generated videos that exhibit body collapse -- leading to more physically realistic outputs.

We find that synthesizing objects against a clean background allows the model to focus on the subject without diverting capacity to modeling the noisy backgrounds that are inevitable in most real videos. However, using a monotonous background with little variation can lead to overfitting or undesirable associations between the background and the foreground objects. To address this, we adopt a simple yet effective approach by incorporating diverse backgrounds with variations in color, texture, transparency, lighting conditions, and environments (e.g., indoor and outdoor settings). A similar strategy is applied to the camera and object (see \cref{sec:engine_details}). Empirically, we find that this increased diversity leads to stronger model performance, particularly in previously unseen scenarios.

\noindent\textbf{Appearance Gap\ }
Ideally, we would like the appearance of a rendered video to match that of real videos. 
However, achieving this is challenging as the appearance gap arises from multiple factors.
First, real videos are captured by physical cameras,
which introduces imperfections such as lens distortions. 
Second, inaccuracies in the rendered materials and object shapes in virtual environments create additional discrepancies.
Finally, rendering algorithms themselves approximate real-world lighting physics, 
further contributing to the mismatch.
In principle, one could hire a large team of skilled artists to overcome these discrepancies,
but such a process would be highly resource-intensive.

\begin{table}
\centering
\footnotesize
\begin{tabular}{l|c|c|c}
\toprule
Training Data & Gym  & Layer & Spin shot \\ 
\midrule
Default  & 83.3\%  & 95\%  & 85\% \\
Low-quality asset & - & 92.5\%  & 22.5\%\\
Low-cost rendering & 41.7\%  & 17.5\%  & - \\
\bottomrule
\end{tabular}
\caption{Success rates illustrating how asset and rendering quality in synthetic videos affect physical fidelity. When asset or rendering quality is low, the physical fidelity in these synthetic videos is less likely to transfer effectively to video generation models.}
\label{tab:appearance-gap}
\vspace{-0.2cm}
\end{table}

\begin{figure}
    \centering
    \includegraphics[width=1\linewidth]{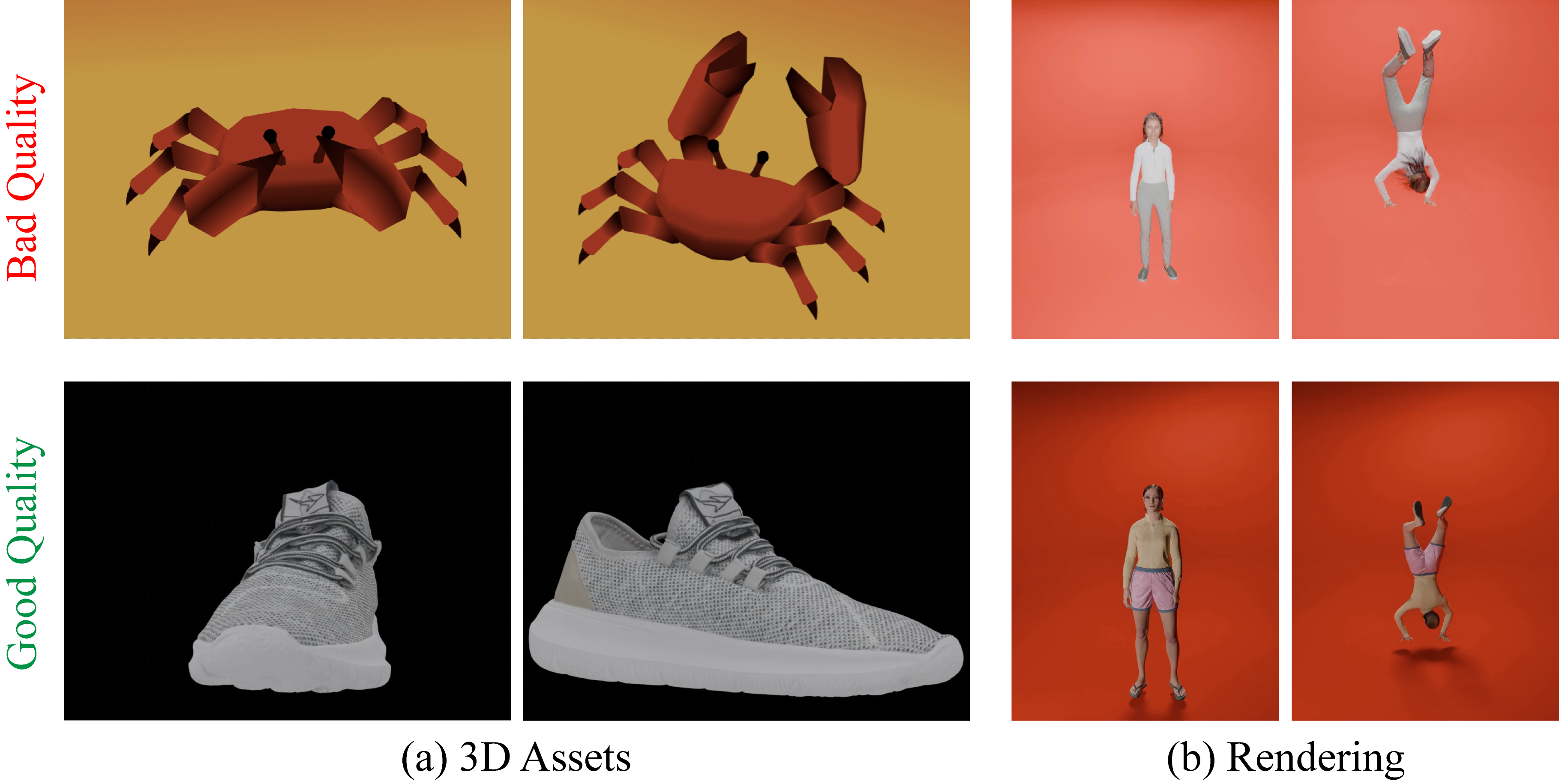}
    \vspace{-0.5cm}
    \caption{Visualizations of synthetic videos highlighting both good- and poor-quality 3D assets (a) and rendering (b).}
    \vspace{-0.2cm}
    \label{fig:appearance-gap}
\end{figure}

% Therefore, our goal is to transfer physical fidelity (e.g., 3D consistency) to the foundation model without 
% causing it to inherit the synthetic appearance traits or incurring excessive time on creating synthetic videos.

To this end, we explore several rendering settings to balance this trade-off. 
Our experiments 
indicate that both low-quality 3D assets and low-cost rendering quality (\Cref{fig:appearance-gap} top) 
significantly decrease the success rate of the generated videos, as shown in~\Cref{tab:appearance-gap}.
When generating videos with spin shot,
the success rate is greatly decreased.
For the layer decomposition task, 
even though the success rate of generating a pure color background remains high, 
the objects that appear in the output videos often look cartoonish (See \cref{sec:ablation_details}).
\Cref{tab:appearance-gap} also illustrates that ensuring sufficient quality in both the 3D assets and the rendering settings (\Cref{fig:appearance-gap} bottom) is essential to achieve a high success rate.

\subsection{Crafting Captions for Synthetic Videos}
\label{sec:3_2}
% We find the video caption method significantly affects the performance of the video generation model 

% By vitural of the syntheic rendering, we can create accurate captions 
% the captions accompanying each video significantly influence the performance of the video generation model.
% We propose an efficient and effective captioning strategy that enhances model training outcomes.

%\noindent\textbf{Efficient and accurate captioning}
Conventional pipelines for building large-scale video-caption datasets is to collect videos first and then generate captions using Vision-Language Models (VLMs). 
In contrast, as synthetic videos are created from a cross combination of 3D objects, scenes, and camera movements during video synthesis,
we caption each element separately and then merge the descriptions into a final caption for the rendered video.
This method is efficient and accurate: 
if we have $N$ objects, $M$ scenes, and $C$ camera setups, it requires only $(N+M+C)$ captions, 
whereas an existing approach would need to caption $N\times M\times C$ distinct videos.
% For human videos, we can further separate human appearance from motion to allow combinations, thus further boosting captioning efficiency. 
Such decomposition also improves accuracy and the granularity of the generated caption, 
as VLMs may produce inconsistent or vague descriptions when confronted with challenging lighting conditions or camera viewpoints in real videos.
In contrast, our method ensures consistency by keeping descriptions for the same element regardless of final scene.

% \begin{figure}
%     \centering
%     \includegraphics[width=1\linewidth]{visualizations/artifacts.png}
%     \caption{Visualization of generated outputs from video generation models trained with synthetic videos 
%     that use low-quality assets on layer decompostion task (a), over-trained with synthetic videos (b),
%     and trained with synthetic videos that use low-quality assets on large camera motion task (c).}
%     \vspace{-0.5cm}
%     \label{fig:artifacts}
% \end{figure}

%\noindent\textbf{Incorporating distinguishable tags}
As synthetic and real videos exhibit distinct visual characteristics, 
we hypothesize that embedding \emph{special tags} (\eg, ``animated'' or ``rendered,'' as shown in Figure~\ref{fig:caption-showcase-c}) within synthetic video captions helps the model distinguish the two domains and transfer only the desired physical fidelity into the generation.
Through our ablations, 
we find that explicitly tagging synthetic data promotes more effective cross-domain knowledge transfer (See~\cref{sec:4_3}). 

\subsection{Training with Synthetic Videos}
\label{sec:3_3}
We employ a diffusion transformer model based on the MMDiT architecture~\cite{esser2024scaling}, trained on real videos at native resolutions~\cite{dehghani2023patch} within the latent space of a variational autoencoder (VAE)~\cite{kingma2013auto}. The model is pretrained using the flow-matching objective.

To enhance the physical fidelity for video generation, we explore incorporating synthetic video data. While training on a mix of synthetic and real videos can improve fidelity, its effectiveness depends on the synthetic-to-real ratio and training steps. Too much synthetic data risks introducing artifacts, while too little yields minimal improvement. Similarly, excessive training can cause overfitting, whereas insufficient training fails to leverage synthetic data effectively.

Even with a well-tuned mixing ratio, synthetic video can still introduce distinctive patterns and artifacts in the generated outputs. To mitigate this, we draw inspiration from \cite{friedrich2023fair, ruiz2023dreambooth, sohn2023styledrop} and propose \emph{SimDrop}. Based on~\cite{ho2021classifier}, we can guide the diffusion generation process toward the overlapping distribution of synthetic and real videos while reducing the influence of synthetic artifacts.

SimDrop begins by training a reference model, $V_{\sigma}$, which aims to capture the unique patterns (\eg, blinkering, animated facials) of synthetic data that pair with rendering engines rather than the clearly defined visual concepts like objects or scenes. Therefore, in training the reference model, we build different captions that only ignores the desired aspect of the synthetic videos (\eg, human motion). This reference model then work in auxiliary with the generation model $V_{\theta}$ trained on mixture of synthetic and real data. Then the reference model can output only the visual patterns but not interfering the objects or human body formation in the video during inference.
Formally, let $l_{k}$ denote the denoised latent at step $k$, and we have:
\begin{equation}
\begin{aligned}
\small
l_k = V_{\theta}(l_{k-1}, t) - \alpha f\bigl(V_{\sigma}, l_{k-1}, \hat{t}, \hat{n} \bigr)
         + \beta f\bigl(V_{\theta}, l_{k-1}, t, n \bigr),
         \nonumber
\end{aligned}   
\end{equation}
where $t$, $\hat{t}$ (respectively $n$, $\hat{n}$) are positive (negative) prompts for the synthetic-mixed and reference models, and $f_\theta(V_{\theta}, l, t, n) = V_{\theta}(l, t) - V_{\theta}(l, n)$.
The terms $\alpha$ and $\beta$ control the influence of each guidance. Using the special tags discussed in \Cref{sec:3_2}, we can incorporate them into the negative prompts. Adding such tags to negative prompts further offers additional benefits, albeit limited.

\subsection{Evaluating Physical Fidelity}
\label{sec:physics_metrics}

% lu: this statement is too strong
%Instead of a general evaluation~\cite{huang2024vbench}, 
Since there is no common standard on evaluating physical fidelity in videos, we adopt the following metrics to assess the physical fidelity, inspired by related work~\cite{khirodkar2024sapiens,agarwal2025cosmos,schoenberger2016sfm}. Although these quantitative metrics are not perfect, combining them with human evaluation can provide useful signals.

\noindent\textbf{Human pose estimation confidence\ }
% \xy{
We employ a state-of-the-art human vision model, 
Sapiens \cite{khirodkar2024sapiens}, 
to evaluate the physical fidelity of the generated human motion. 
We use a 2B-parameter, 17-keypoint checkpoint to estimate the pose of single-human motion outputs from each model on a per-frame basis.
The average confidence score $\epsilon_\mathrm{conf}$ per keypoint per frame ranges from $0$ to $1$.
Based on the assumption of human vision models, 
a motion sequence with more realistic body structures and clearer poses gives a higher confidence score.
%The following two metrics are computed:
%(1) mean square error (MSE) of keypoint distances between adjacent frames, with coordinates normalized by the human bounding box, and
%(2) average confidence score per keypoint per frame, ranging from 0 to 1.
%Following the original method, keypoint data with a confidence score below 0.3 are discarded in advance.
% }
%\TODO{dataset, citation, and explanation}. 

\noindent\textbf{3D reconstruction error\ } 
%Intuitively, any 3D object should not be distorted during camera movement. The closer the generated video remains to preserving 3D consistency of objects, the higher its physical fidelity. 
% \xy{
Based on a widely used 3D sparse reconstruction tool, 
COLMAP~\cite{schoenberger2016sfm,schoenberger2016mvs}, 
we evaluate the physical fidelity of the static objects in the videos with large camera motion.
Using a single pinhole camera model and a sequential feature matching mode, 
COLMAP reconstructs the scene from the video frames.
Similar to the work of \cite{agarwal2025cosmos},
we use the following metrics as indicators of physical fidelity:
(1) the number of matched feature points ($\mathcal{N}$),
(2) the average track length ($\mathcal{T}$), and
(3) the average re-projection error ($\epsilon$).
In general, if the video frames generated by a model provide a greater diversity of camera viewpoints yet still maintain the 3D consistency, the number of matched feature points tends to increase. 
However, the mean track length of each feature point is expected to decrease due to the faster camera motion. 
Furthermore, a model that is more physically consistent will yield a lower re-projection error in the resulting 3D reconstruction.
% }
%\TODO{dataset, citation, and explanation}. 

\noindent\textbf{Human evaluation\ }
For each prompt, we generate two videos of different random seeds.
%resulting in 20 videos for each model per given task. 
For human evaluation, we instruct our annotators to examine the outputs from different models side-by-side for each prompt strictly following the guideline that focus on physical fidelity of the video.
In general, a successfully generated video refers to one that follow the text prompt without visible artifacts.
For large human motion, we let the human annotators focus on the integrity of human body, such as limbs, hands, and neck, during the large motion.
% while making sure the visual of the video does include patterns from the synthetic videos.
For large camera motion, the annotators will determine whether the camera motion is performed according to the prompt and
examine the object quality. 
For the layer decomposition task, annotators judge the videos based on two criteria: 
object quality and background quality. 
The details of the guideline is in~\cref{sec:human_details}.
Afterwards, we compute the successful rate and average the results across all human evaluators.
\vspace{-0.1in}
\section{Experiments}
\begin{figure*}
  \centering
  \setlength{\imagewidth}{.123\linewidth}
  \includegraphics[width=\imagewidth]{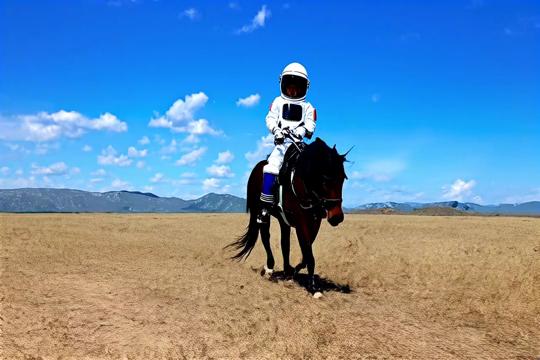}\hfill%
  \includegraphics[width=\imagewidth]{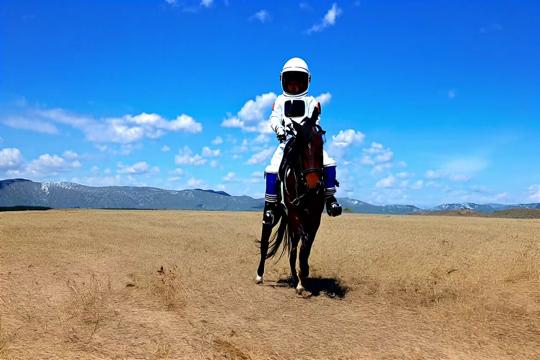}\hfill%
  \includegraphics[width=\imagewidth]{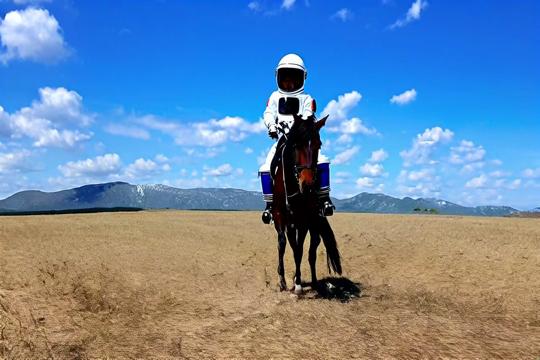}\hfill%
  \includegraphics[width=\imagewidth]{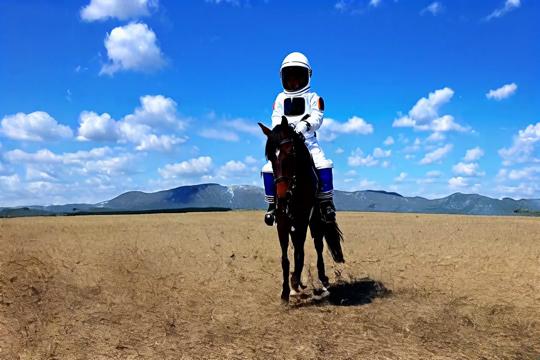}\hfill%
  \includegraphics[width=\imagewidth]{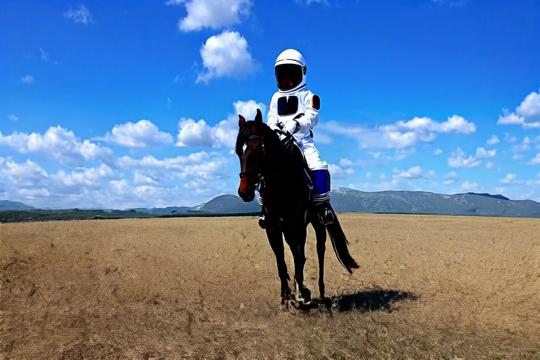}\hfill%
  \includegraphics[width=\imagewidth]{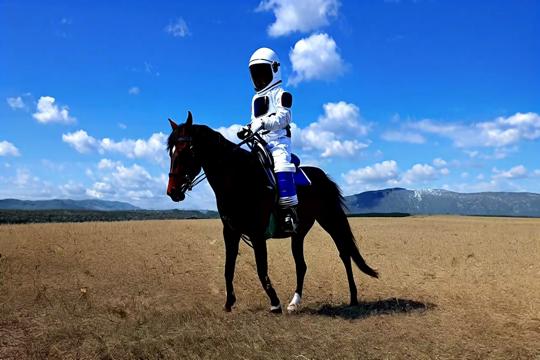}\hfill%
  \includegraphics[width=\imagewidth]{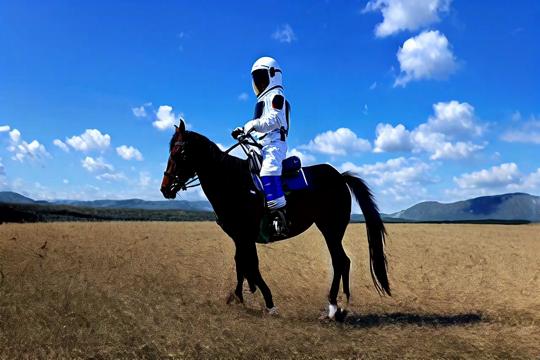}\hfill%
  \includegraphics[width=\imagewidth]{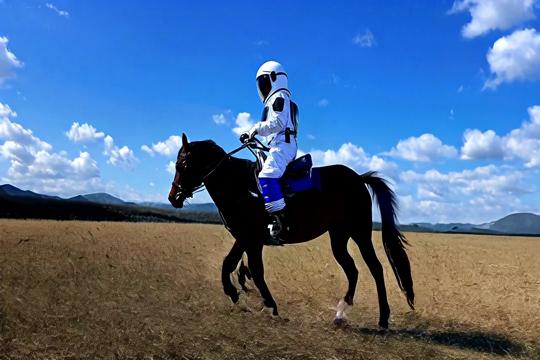}\\%
  \includegraphics[width=\imagewidth]{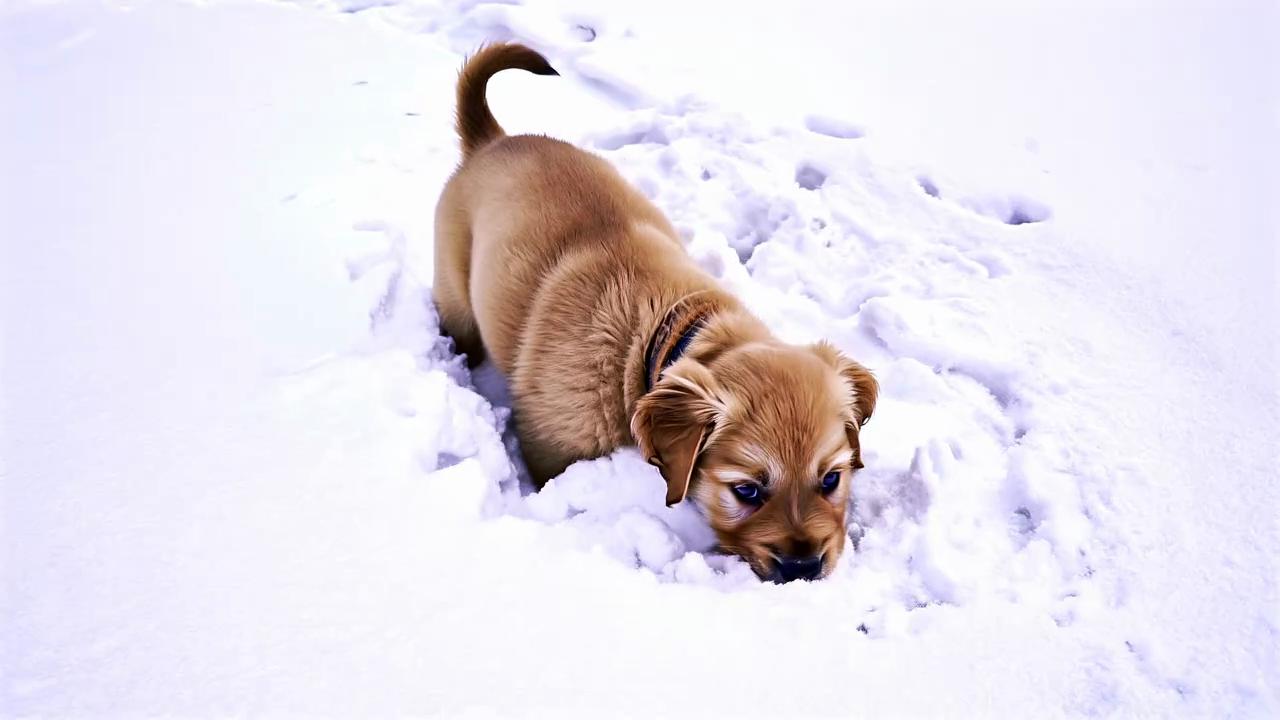}\hfill%
  \includegraphics[width=\imagewidth]{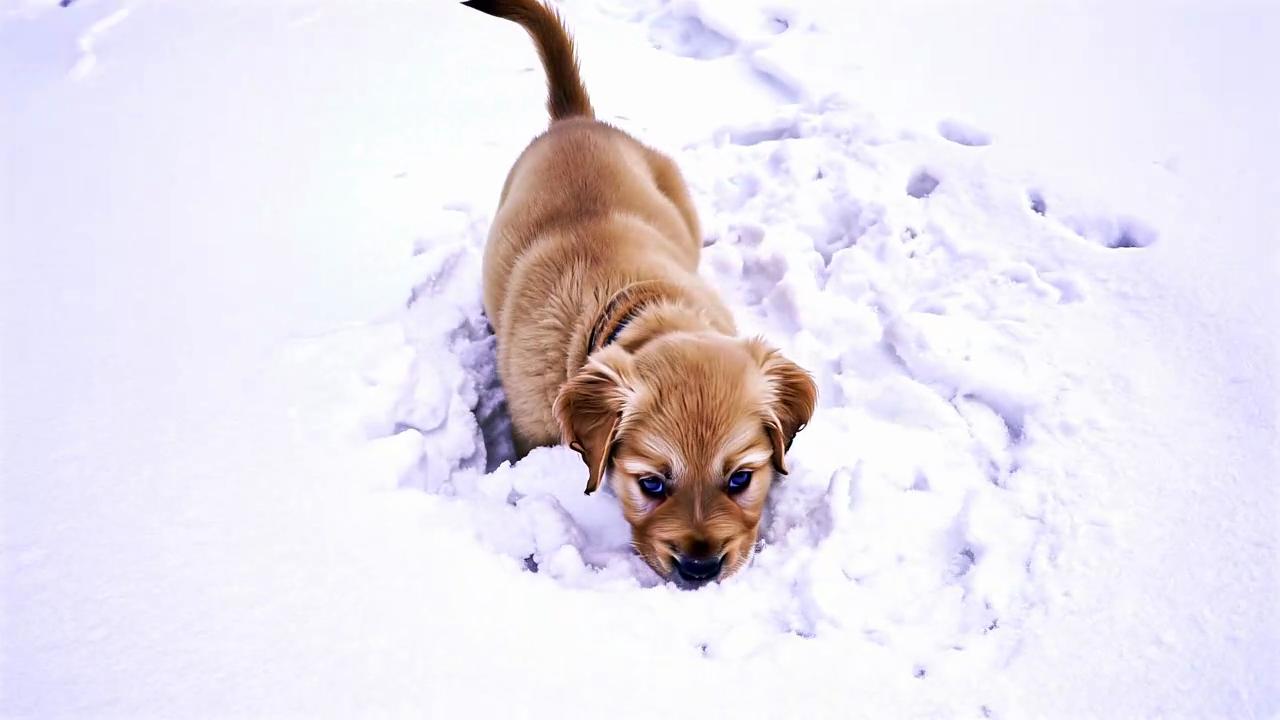}\hfill%
  \includegraphics[width=\imagewidth]{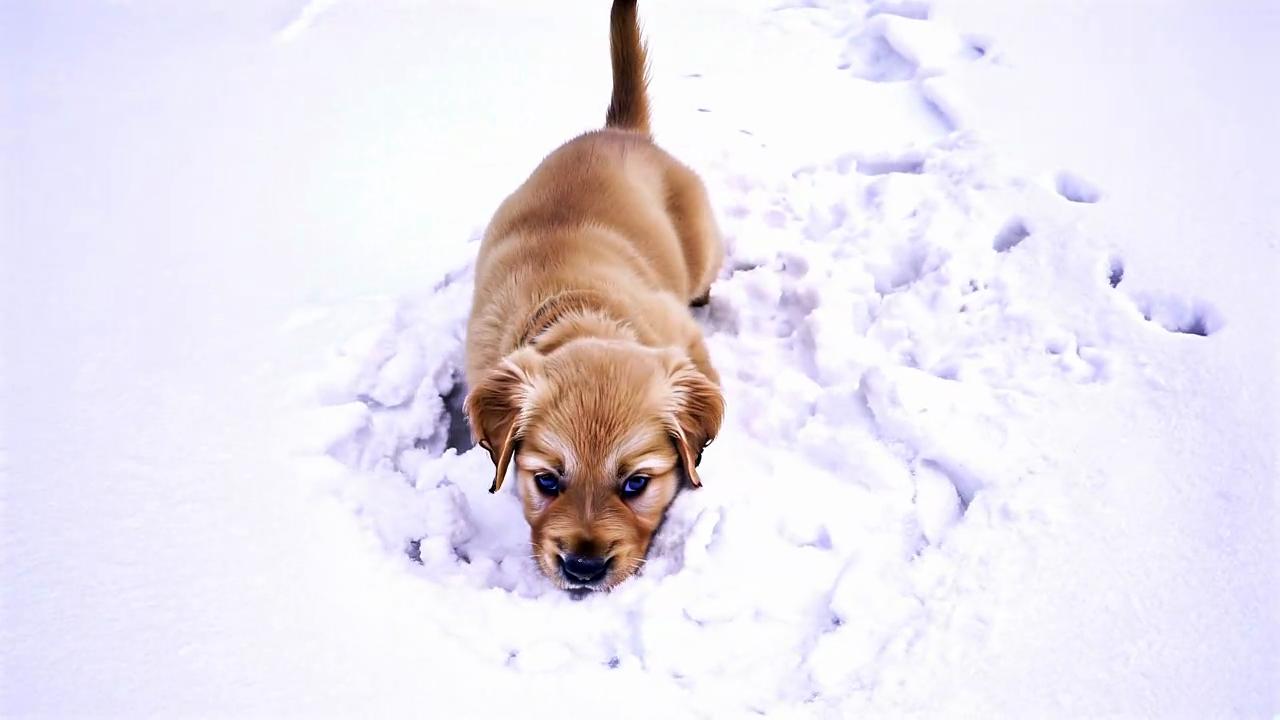}\hfill%
  \includegraphics[width=\imagewidth]{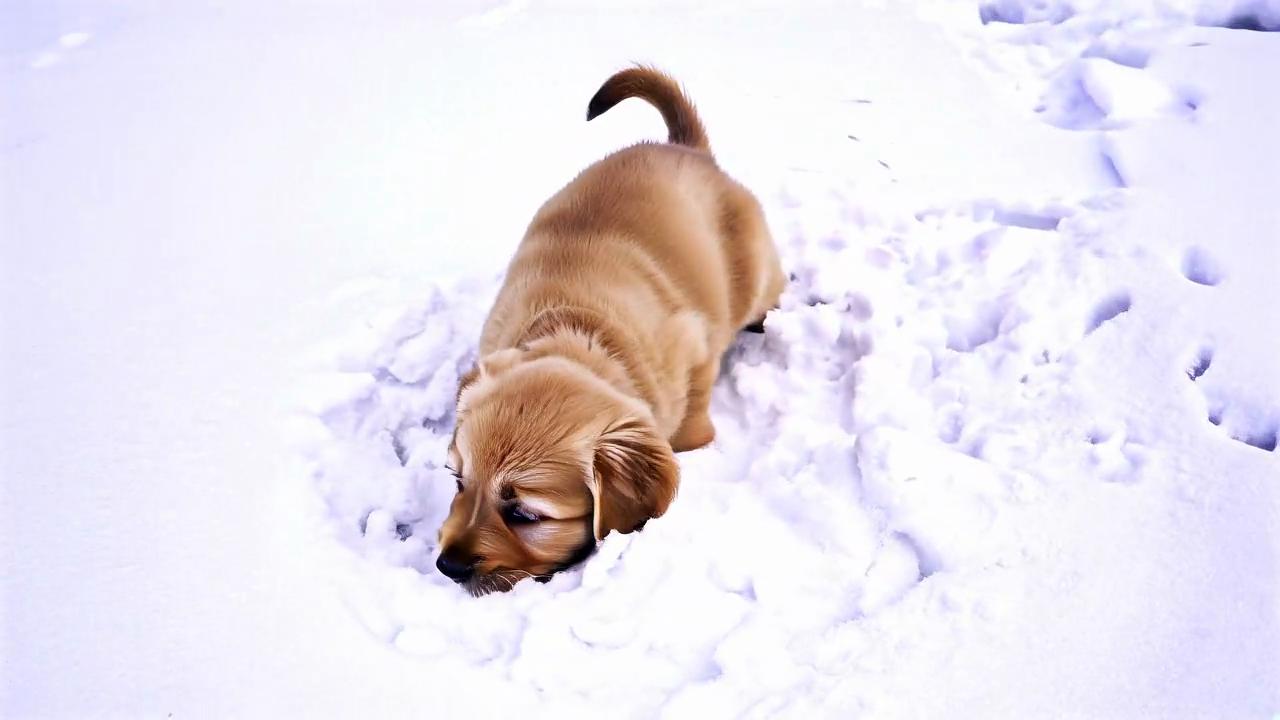}\hfill%
  \includegraphics[width=\imagewidth]{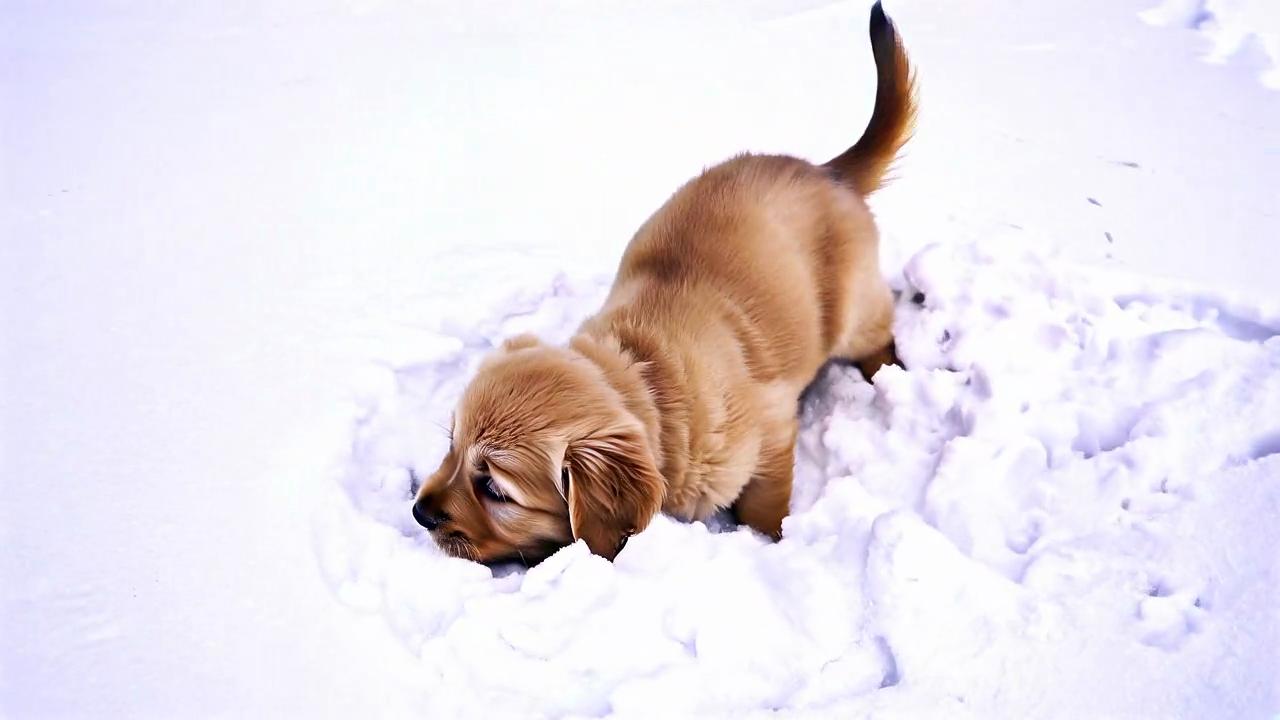}\hfill%
  \includegraphics[width=\imagewidth]{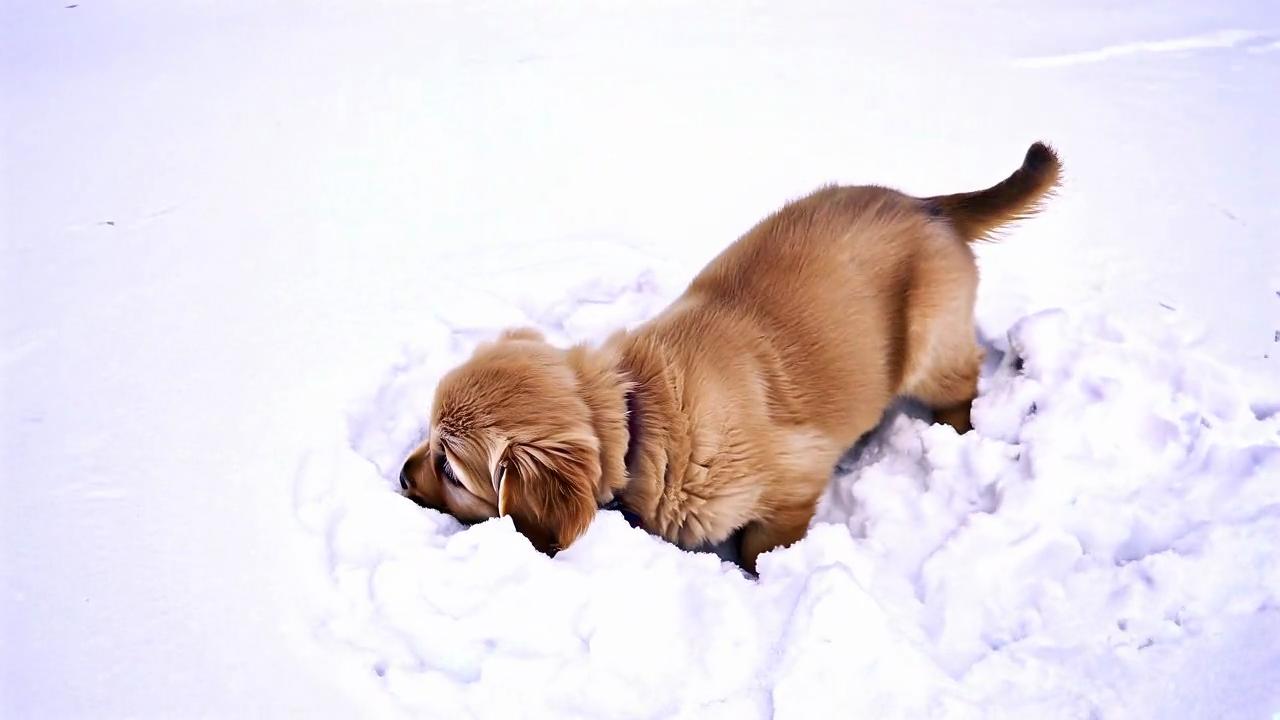}\hfill%
  \includegraphics[width=\imagewidth]{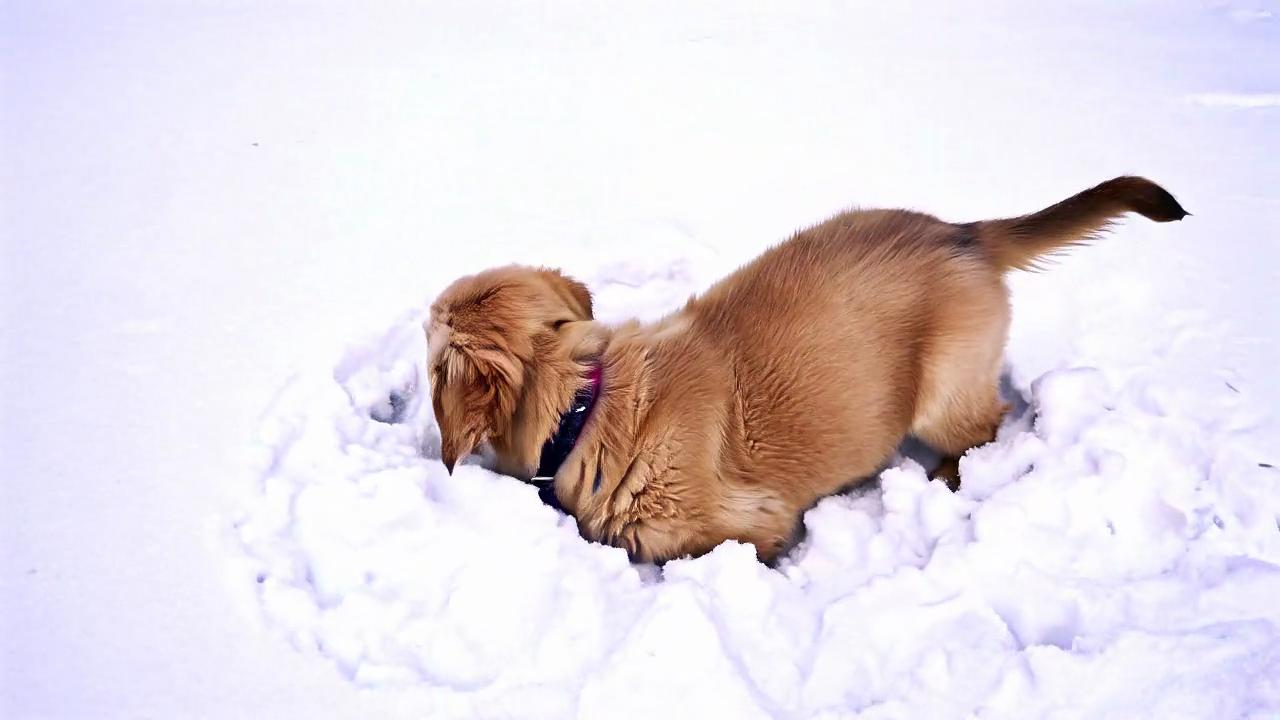}\hfill%
  \includegraphics[width=\imagewidth]{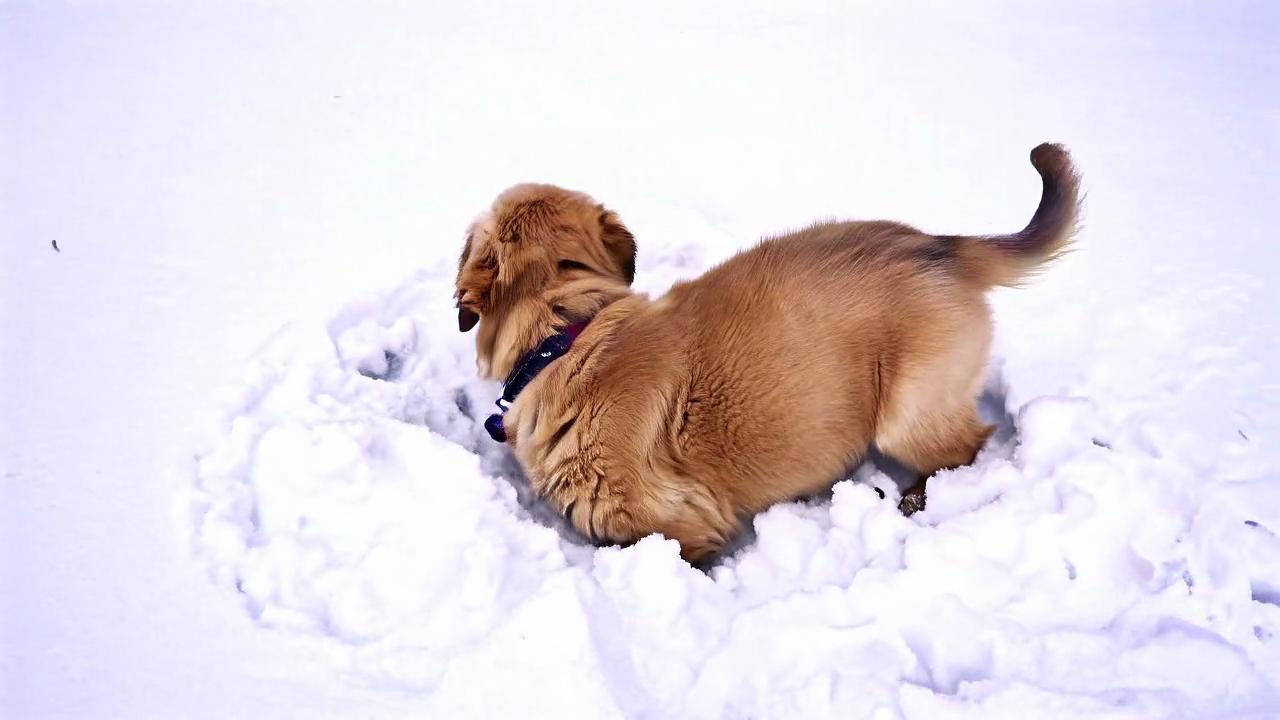}\\%
  \includegraphics[width=\imagewidth]{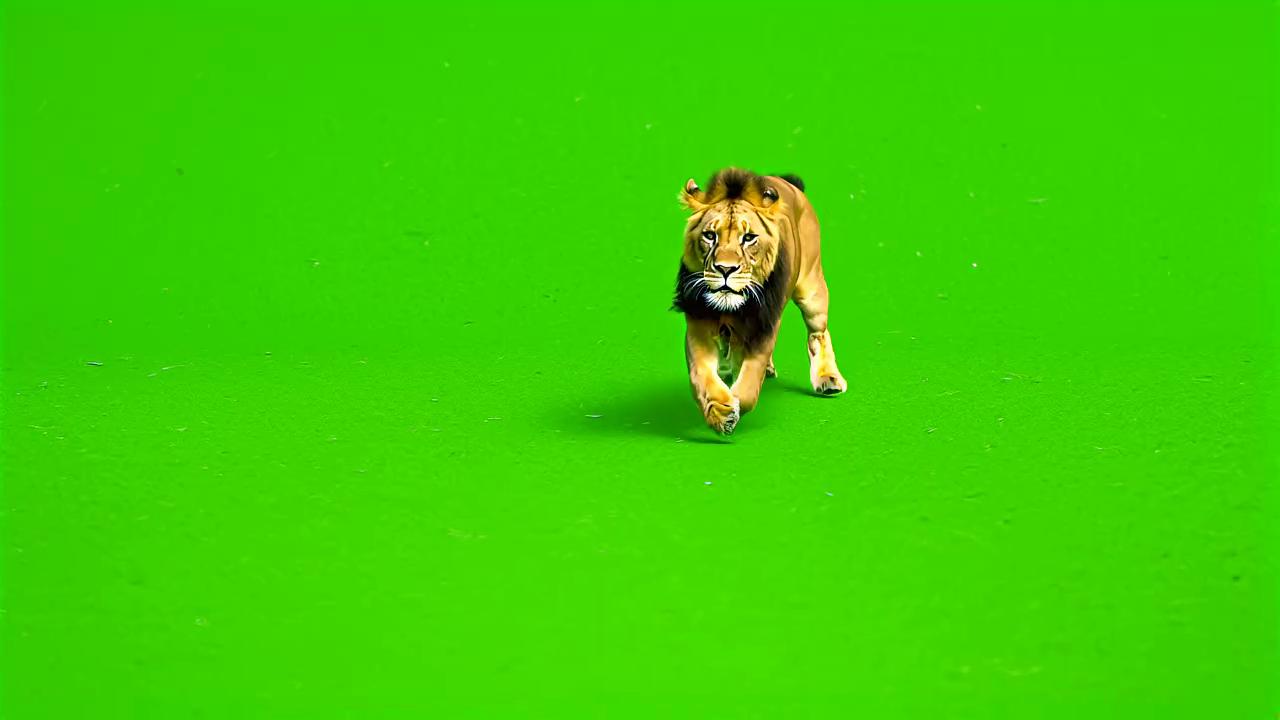}\hfill%
  \includegraphics[width=\imagewidth]{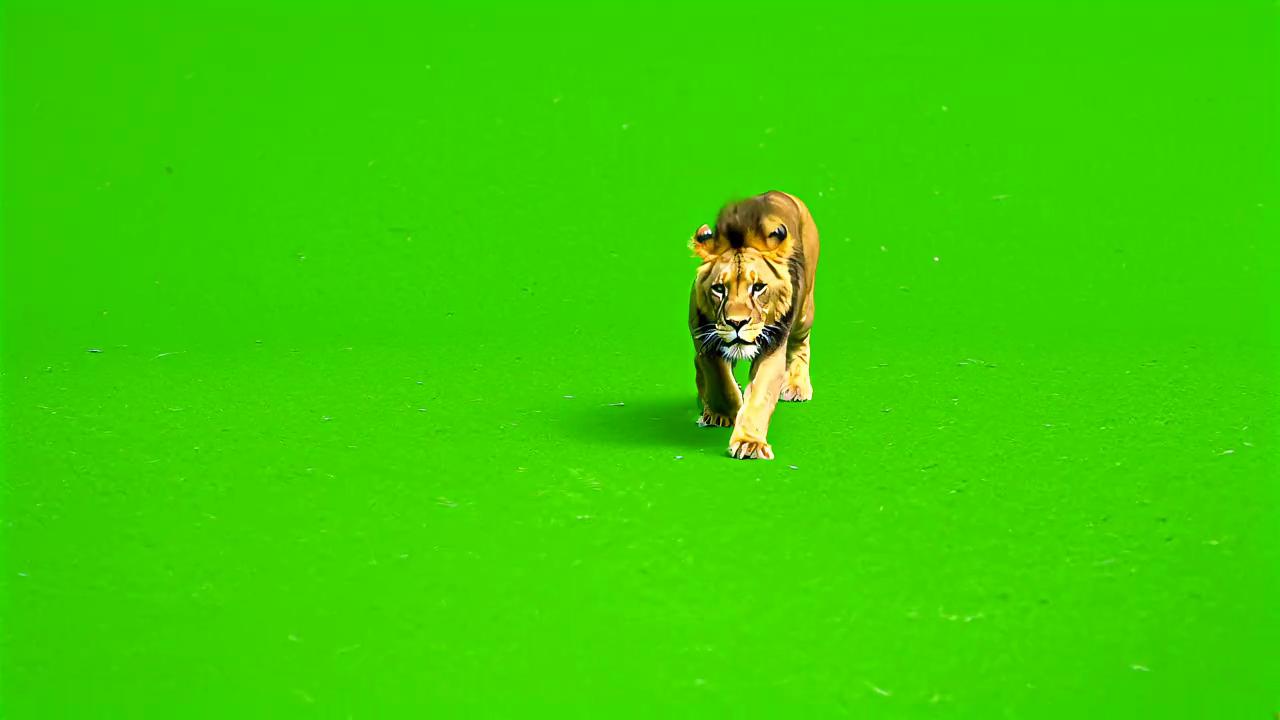}\hfill%
  \includegraphics[width=\imagewidth]{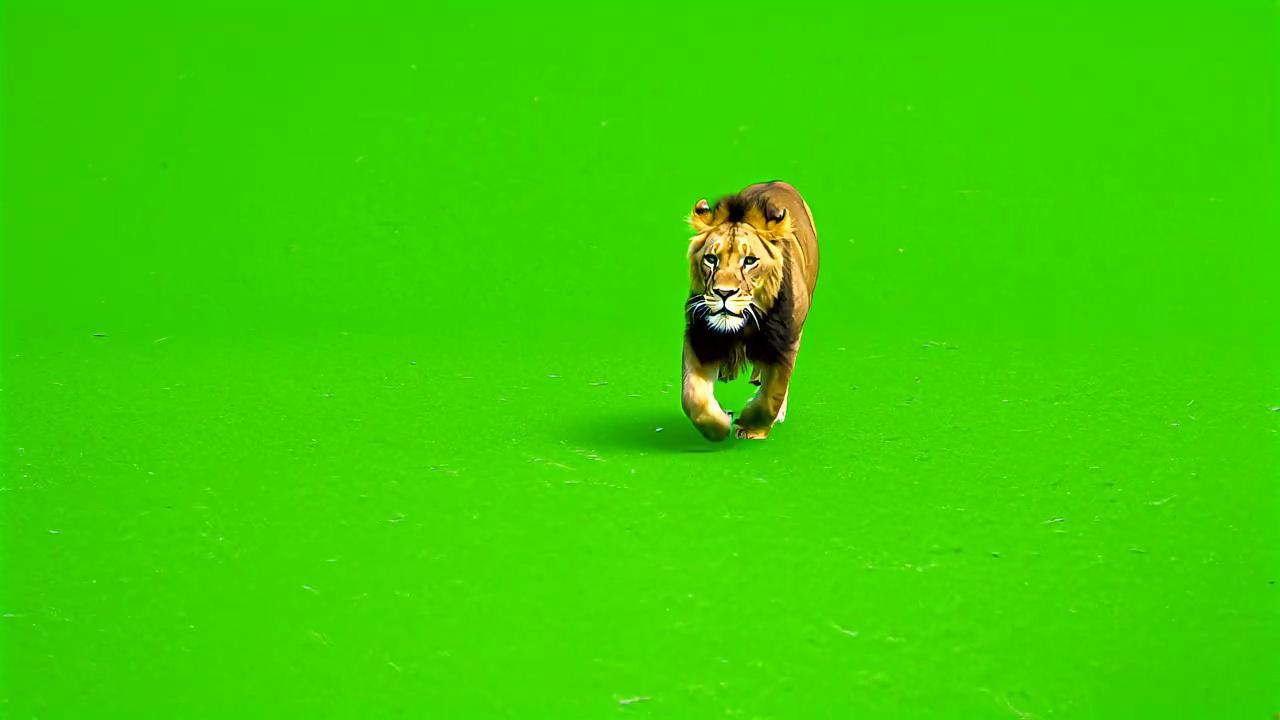}\hfill%
  \includegraphics[width=\imagewidth]{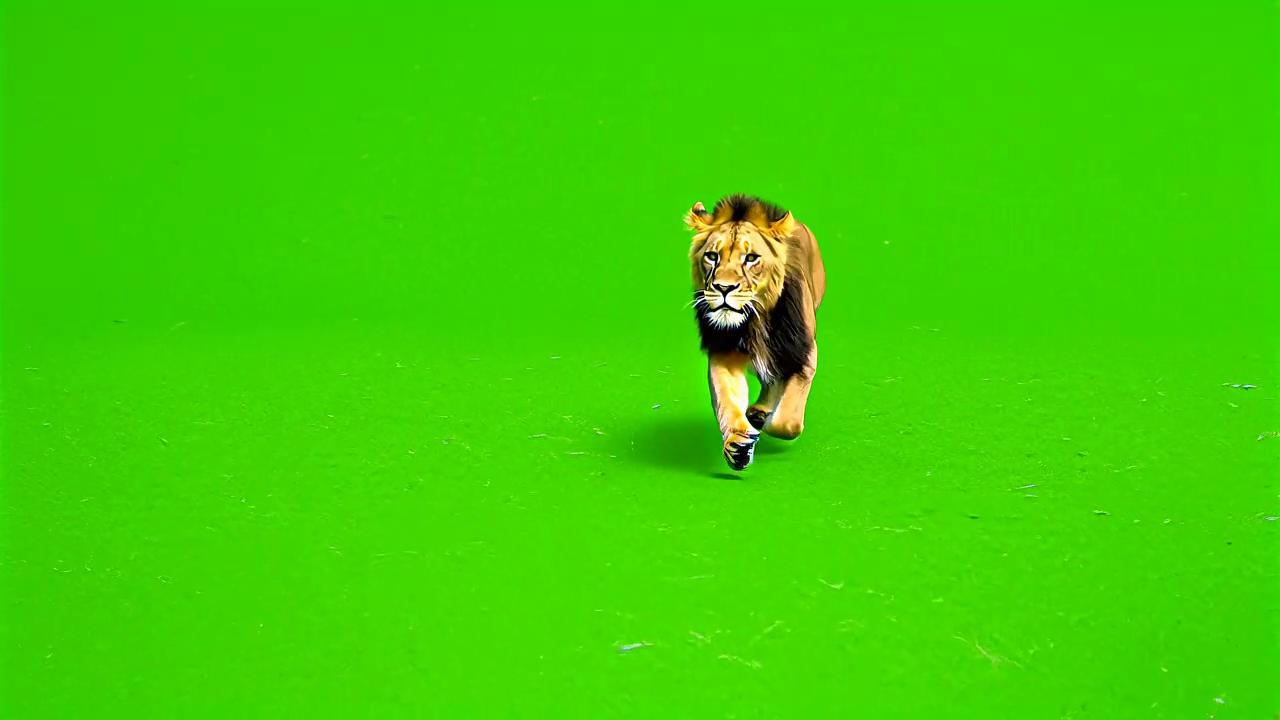}\hfill%
  \includegraphics[width=\imagewidth]{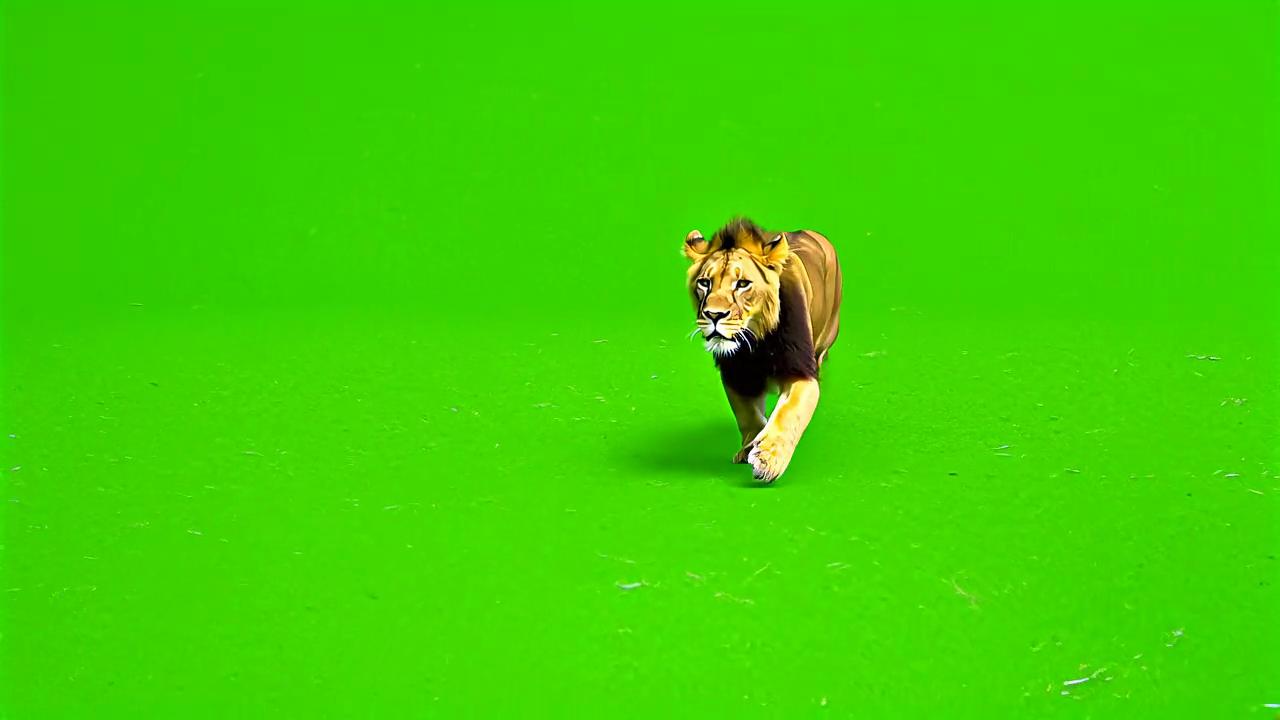}\hfill%
  \includegraphics[width=\imagewidth]{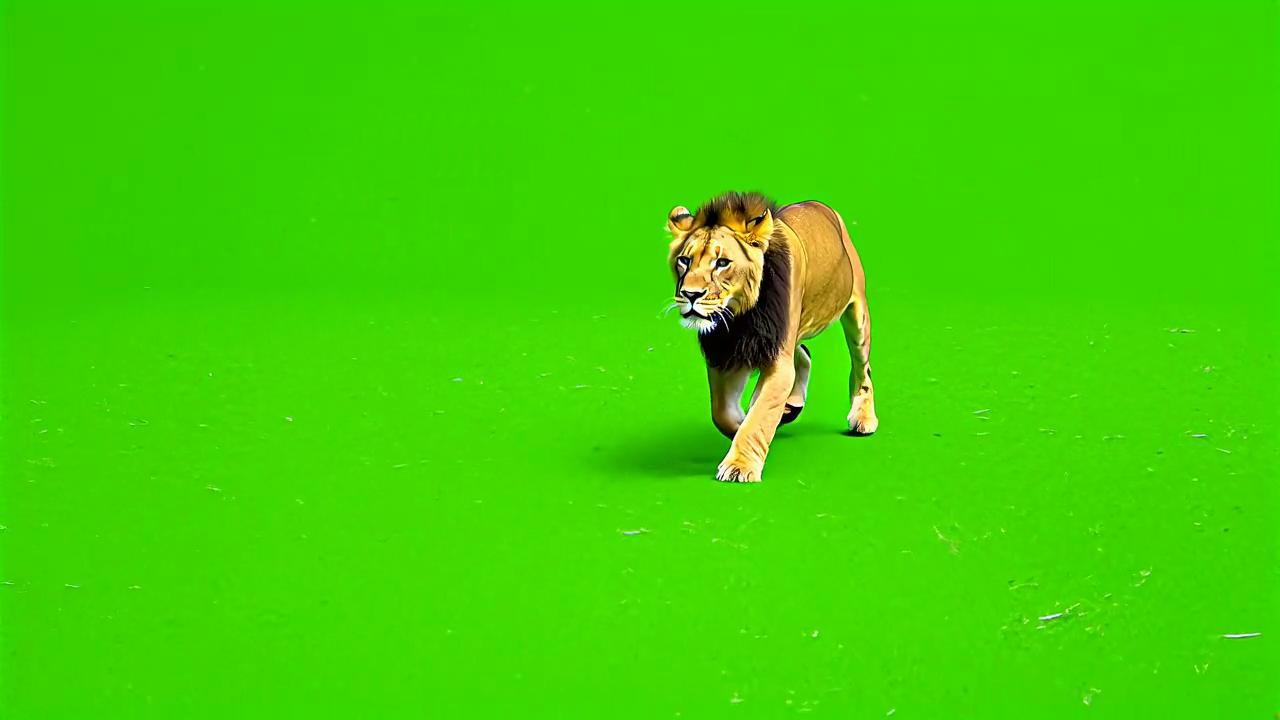}\hfill%
  \includegraphics[width=\imagewidth]{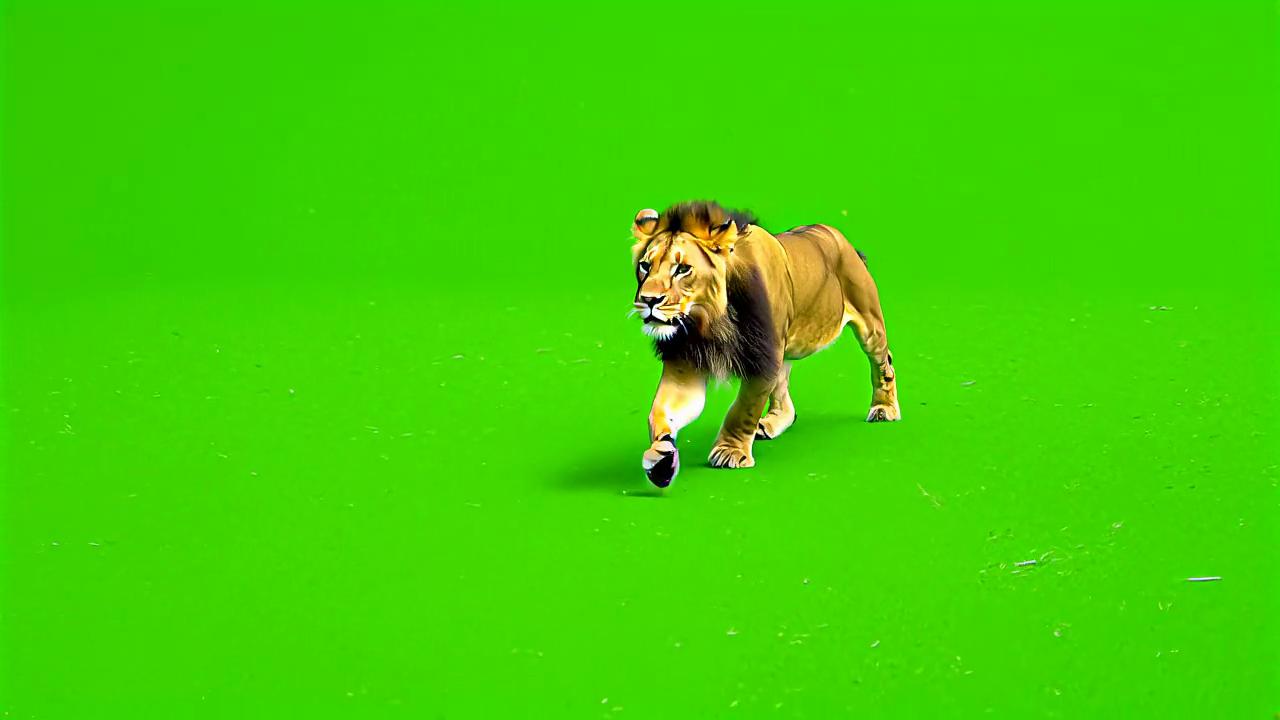}\hfill%
  \includegraphics[width=\imagewidth]{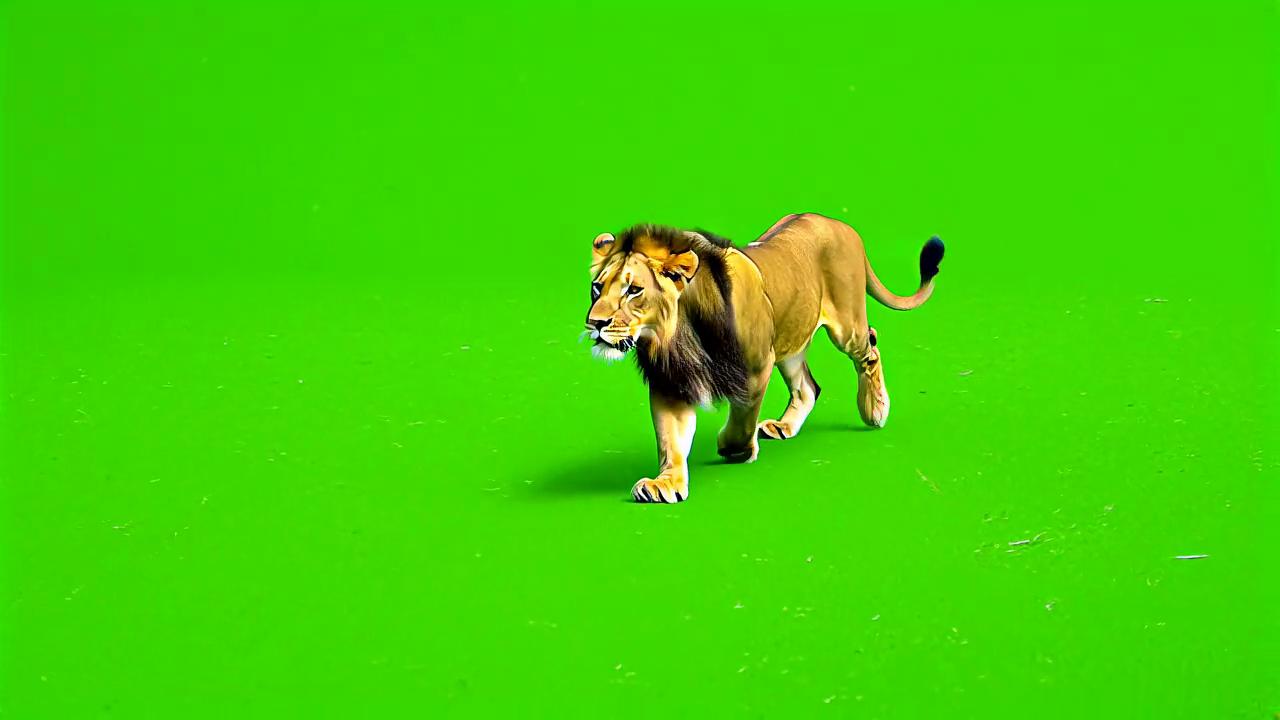}\\%
  \includegraphics[width=\imagewidth]{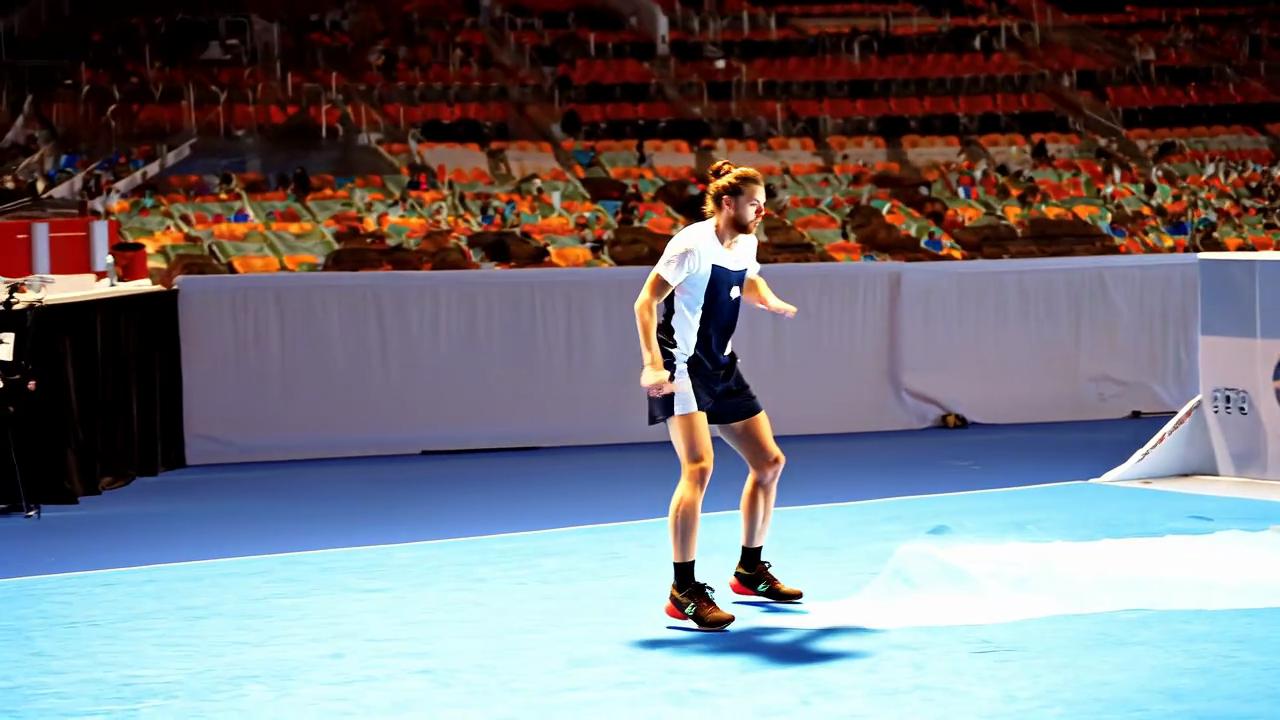}\hfill%
  \includegraphics[width=\imagewidth]{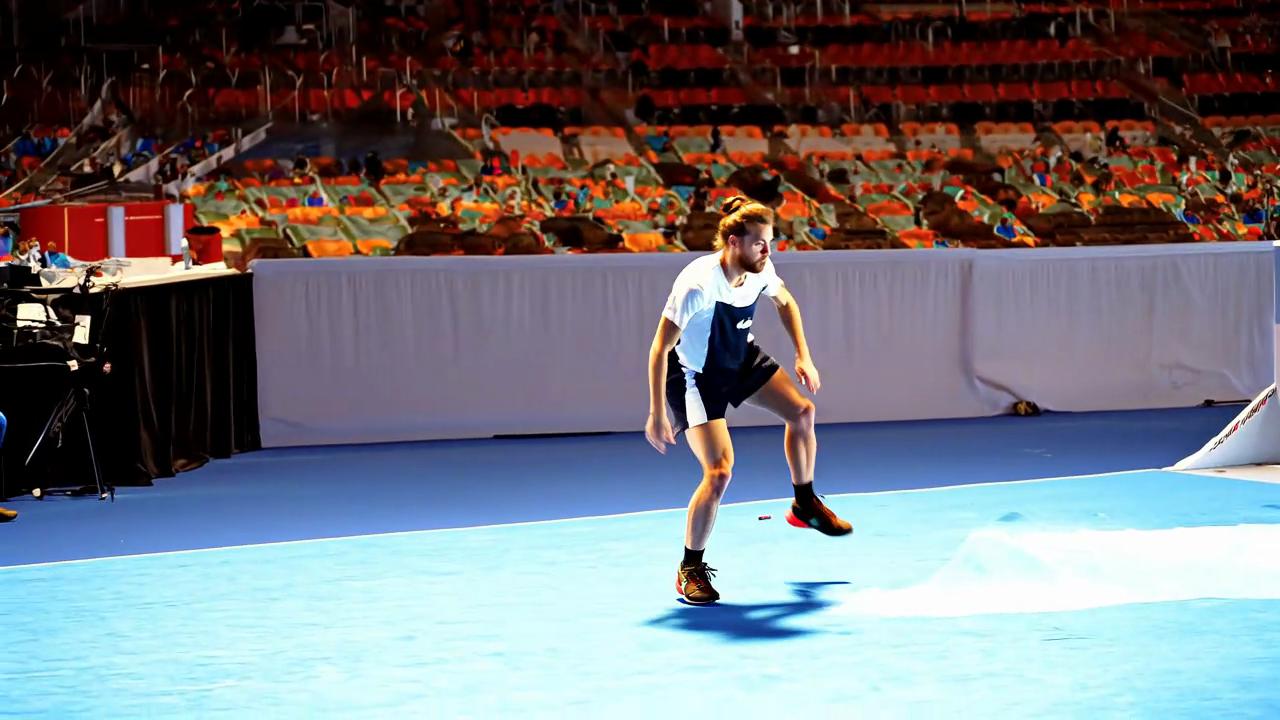}\hfill%
  \includegraphics[width=\imagewidth]{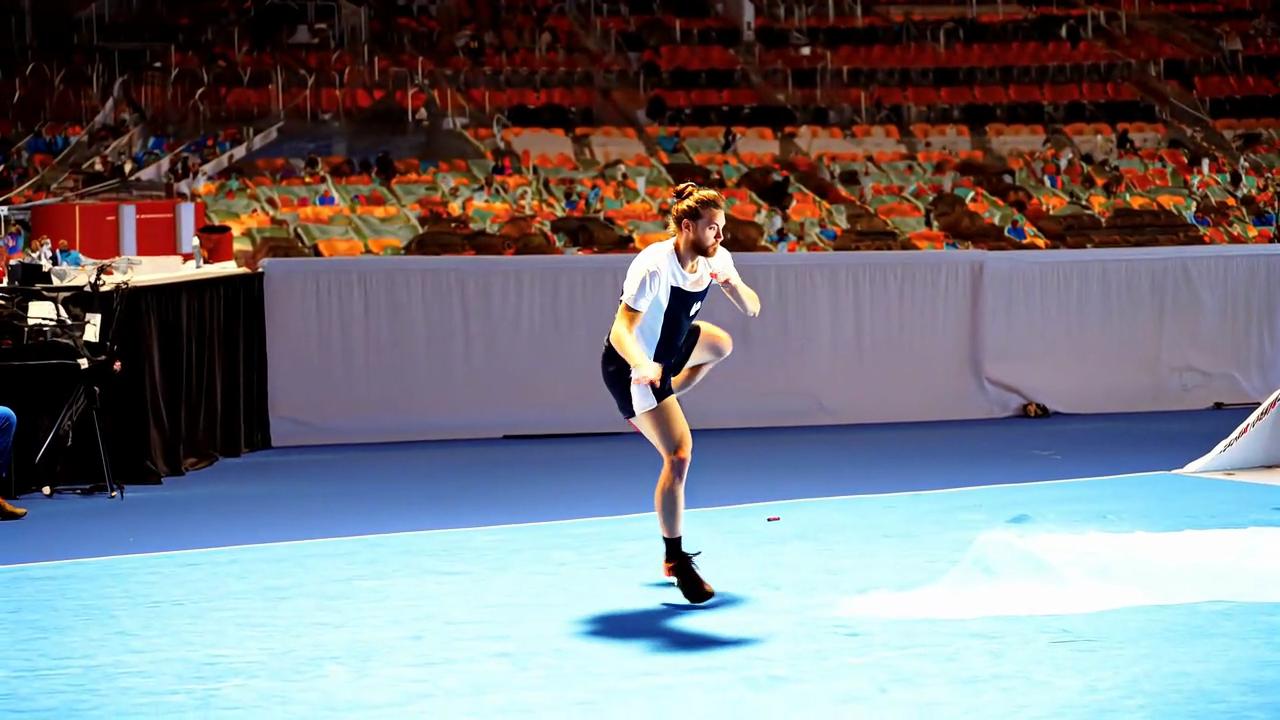}\hfill%
  \includegraphics[width=\imagewidth]{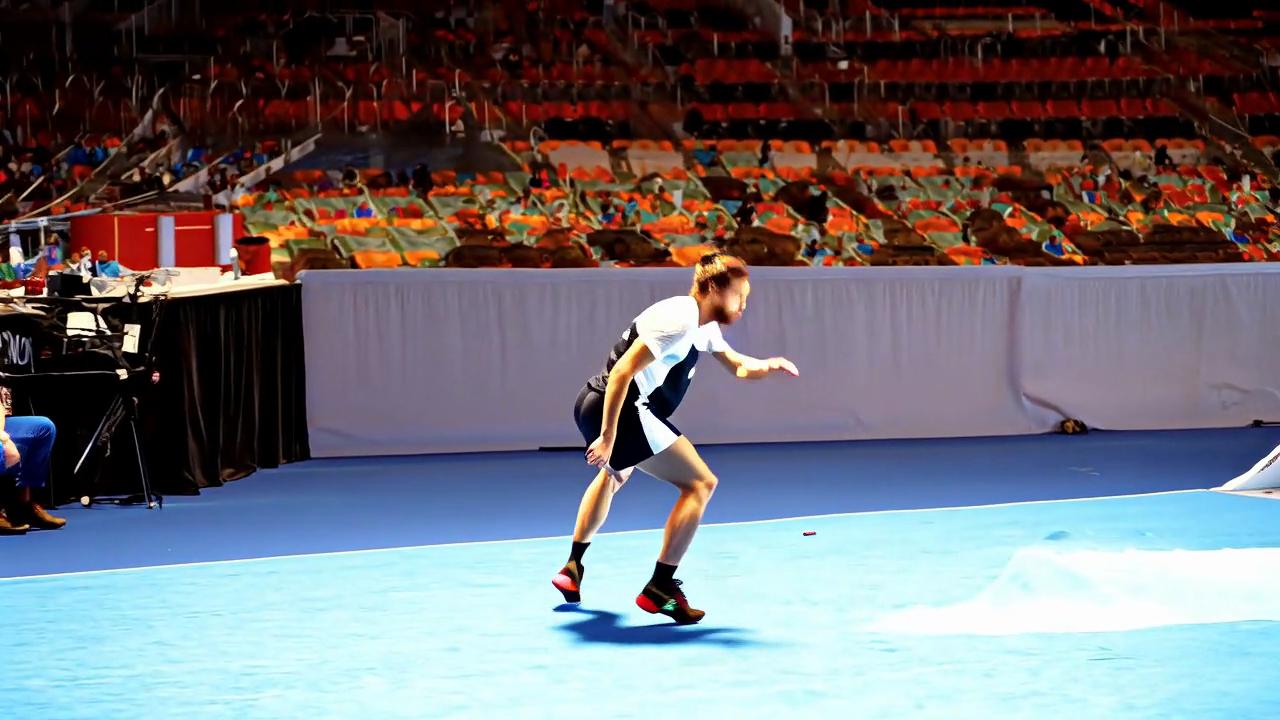}\hfill%
  \includegraphics[width=\imagewidth]{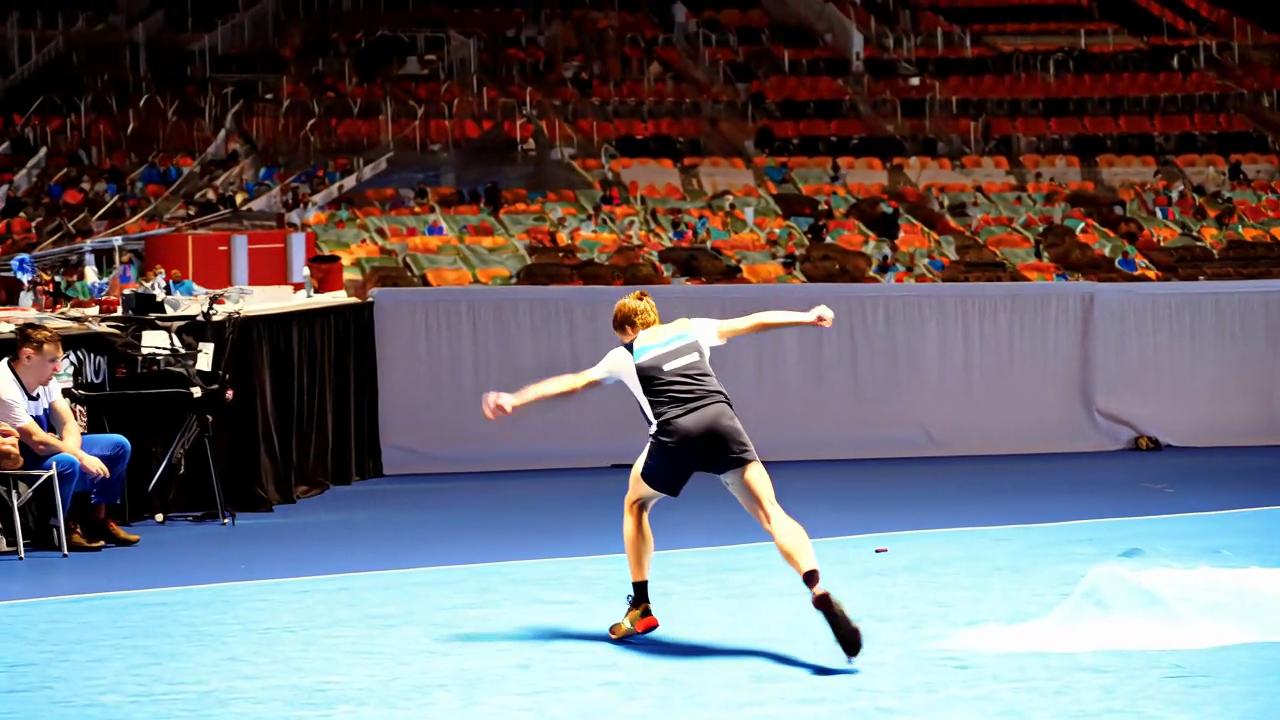}\hfill%
  \includegraphics[width=\imagewidth]{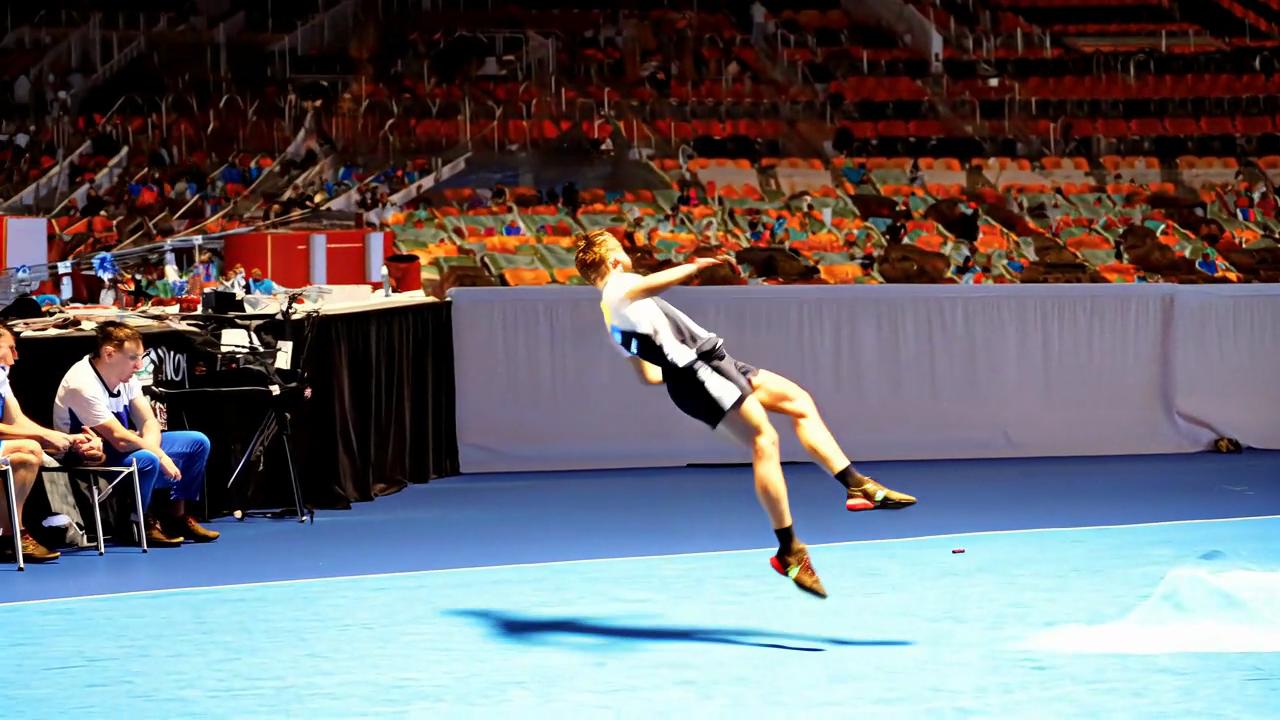}\hfill%
  \includegraphics[width=\imagewidth]{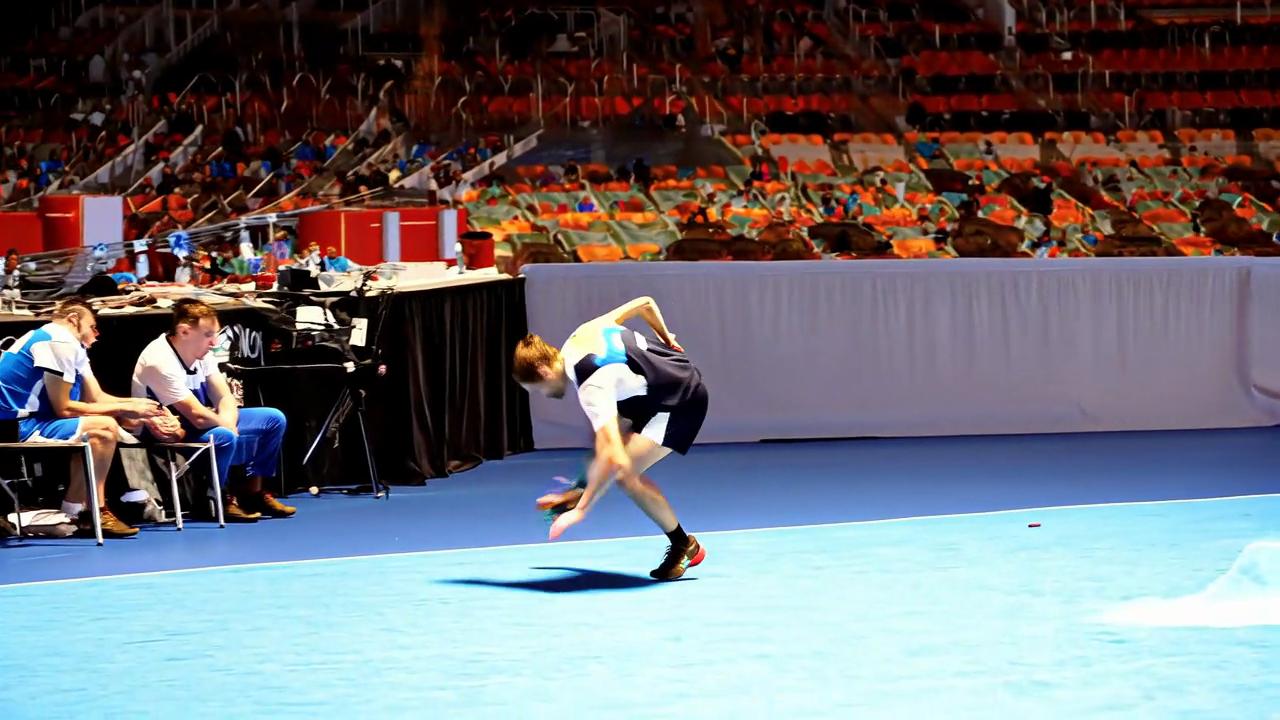}\hfill%
  \includegraphics[width=\imagewidth]{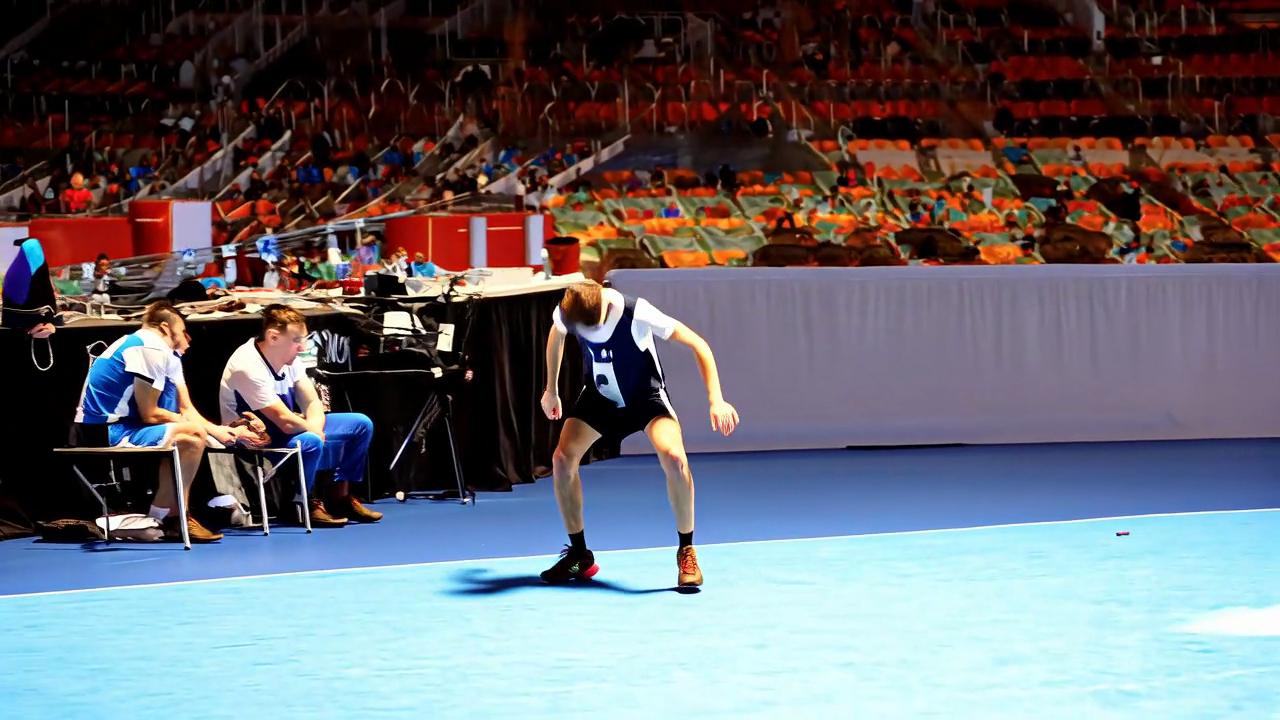}\\%
  \includegraphics[width=\imagewidth]{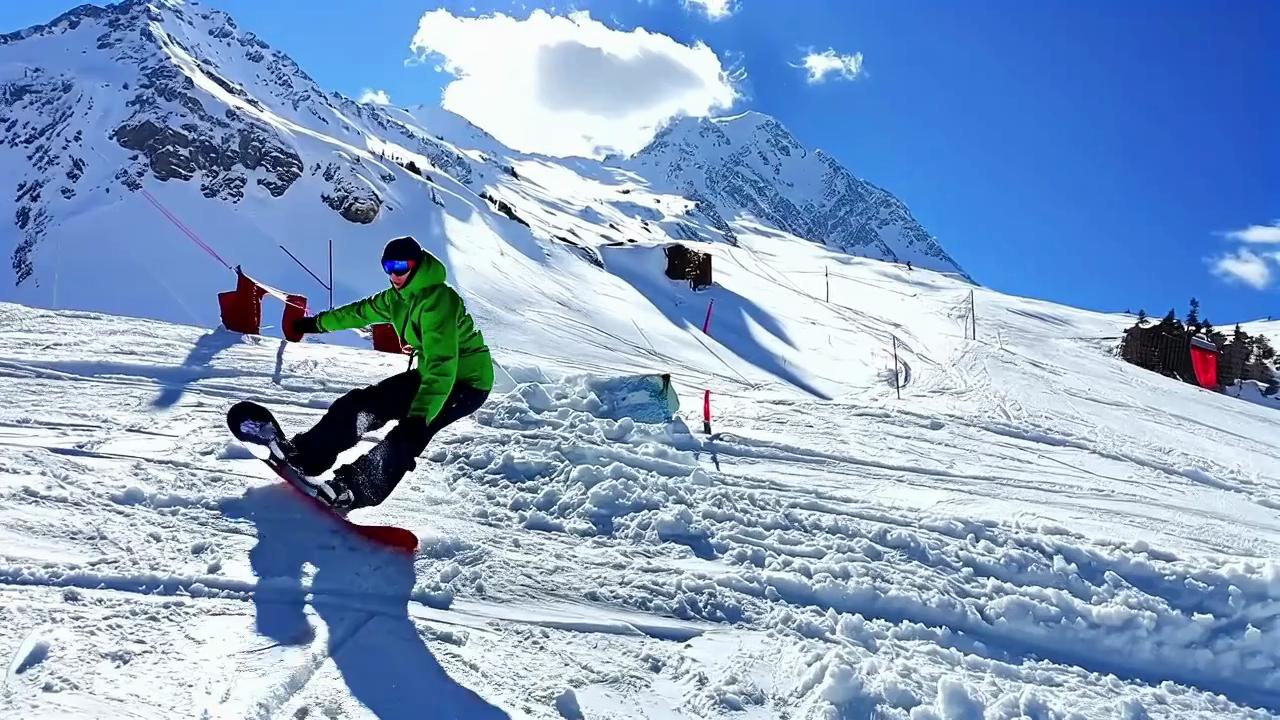}\hfill%
  \includegraphics[width=\imagewidth]{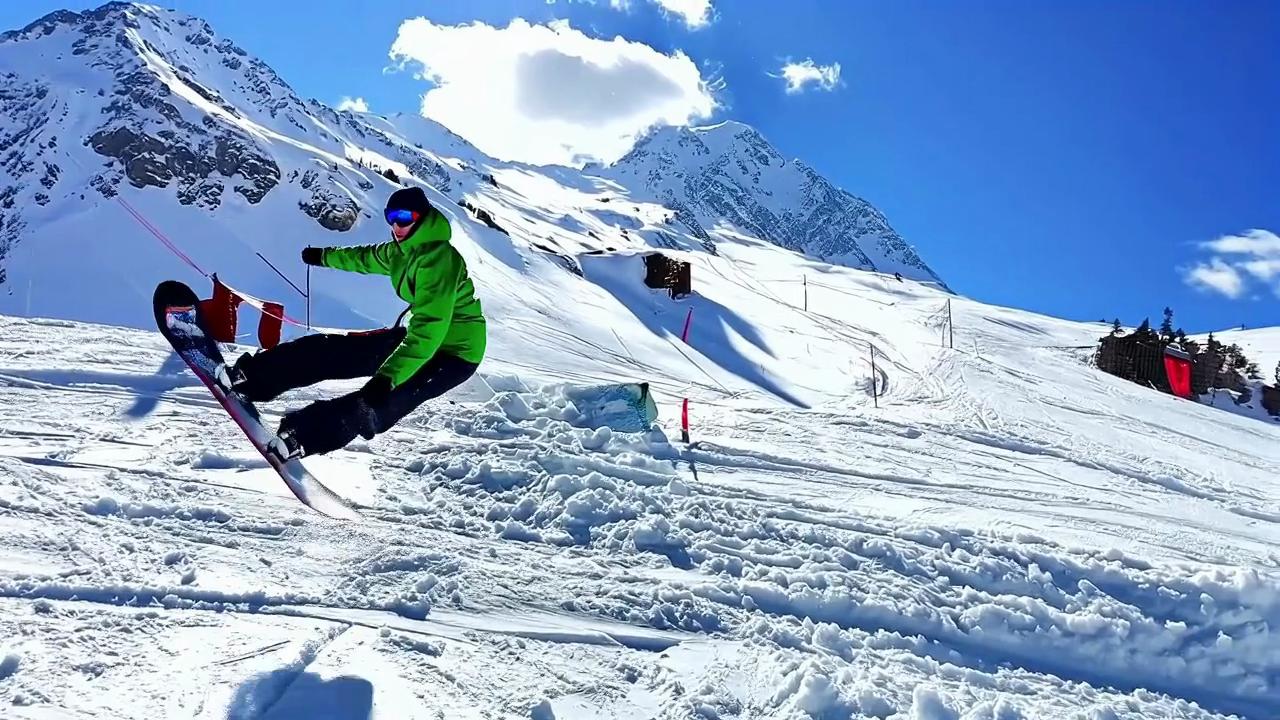}\hfill%
  \includegraphics[width=\imagewidth]{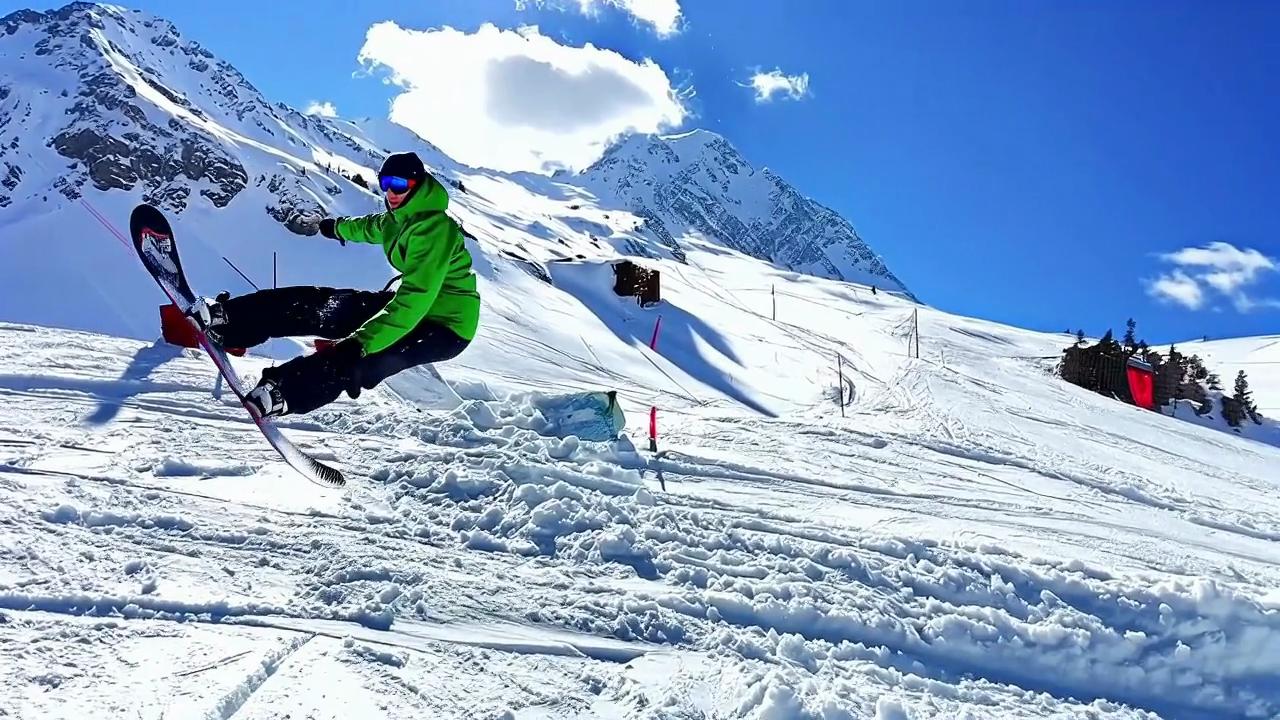}\hfill%
  \includegraphics[width=\imagewidth]{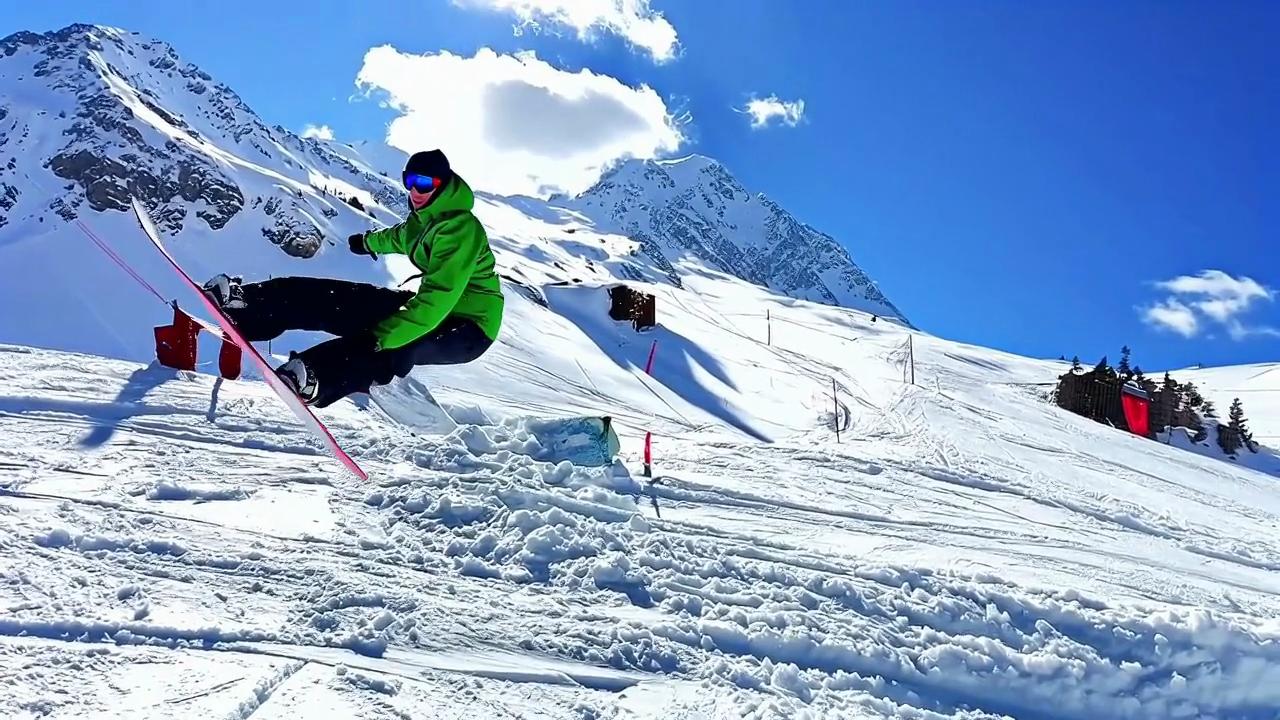}\hfill%
  \includegraphics[width=\imagewidth]{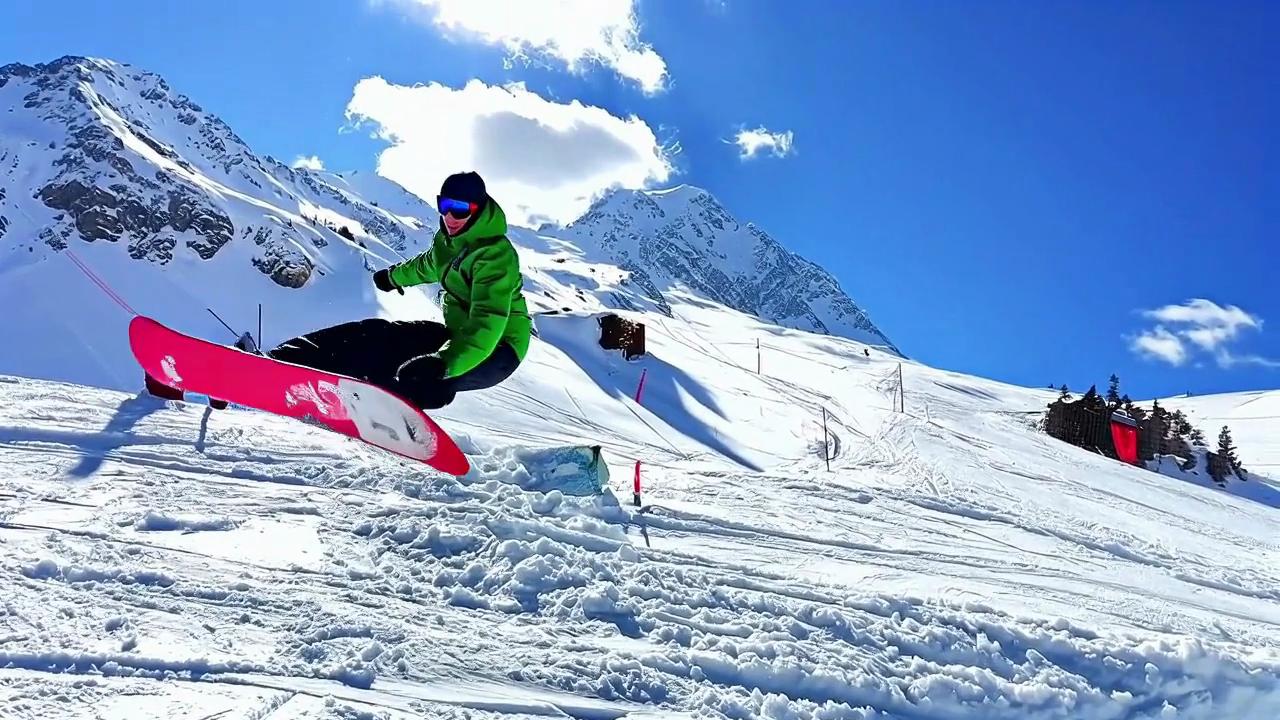}\hfill%
  \includegraphics[width=\imagewidth]{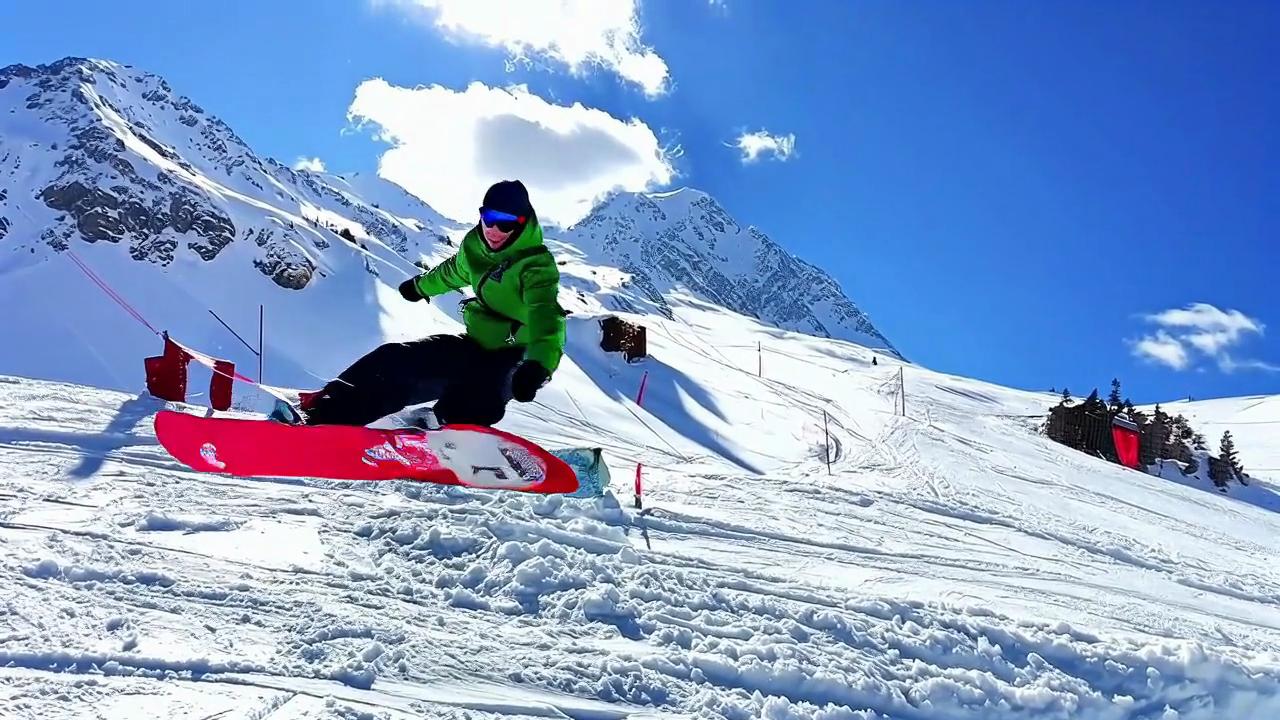}\hfill%
  \includegraphics[width=\imagewidth]{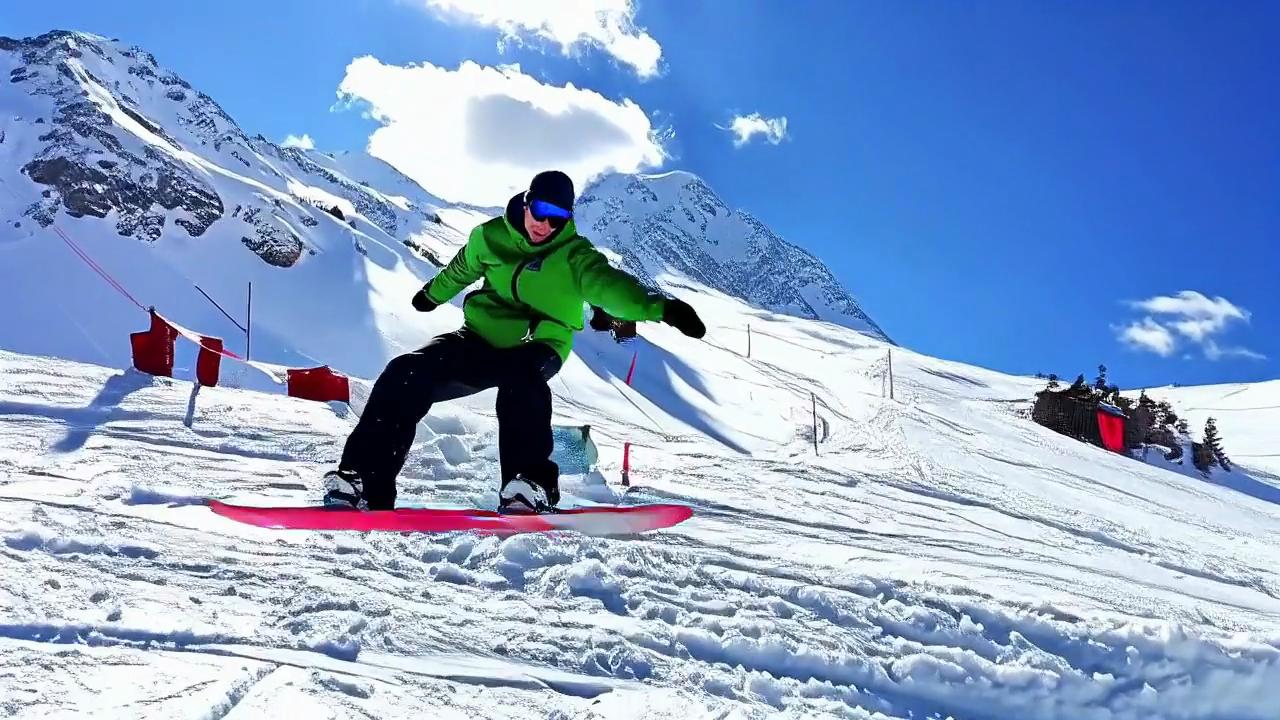}\hfill%
  \includegraphics[width=\imagewidth]{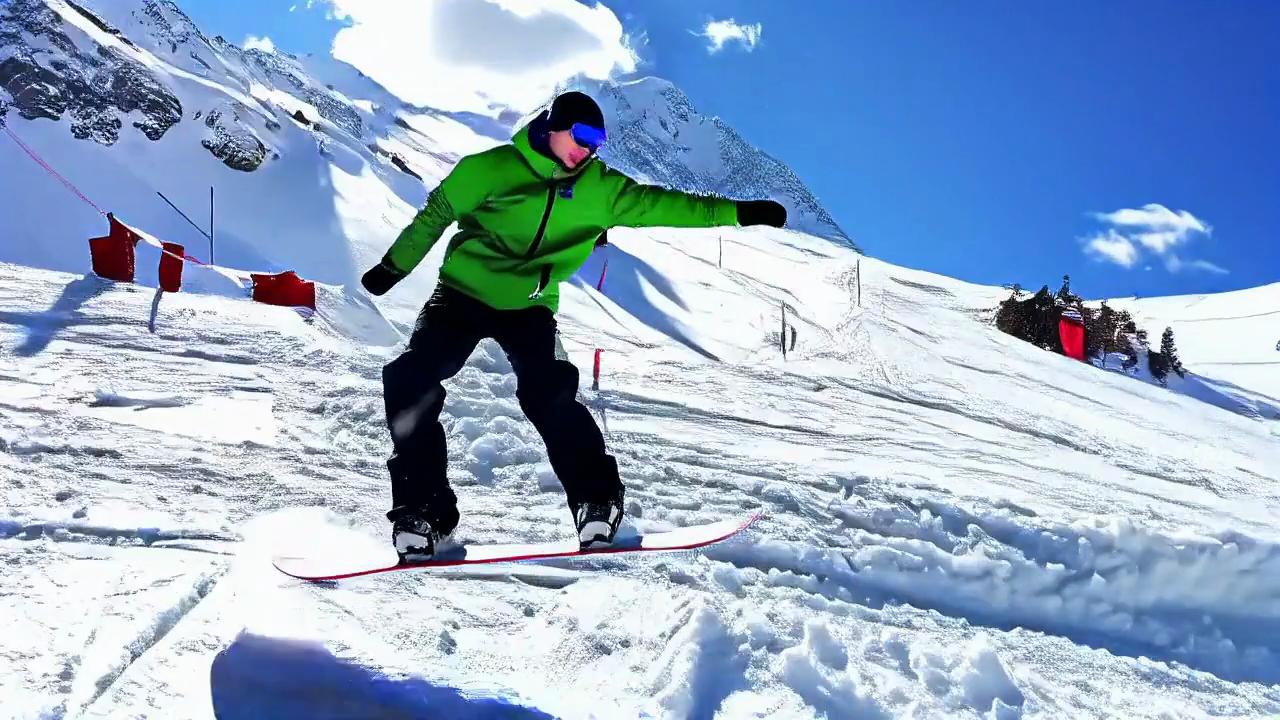}\\%
  \includegraphics[width=\imagewidth]{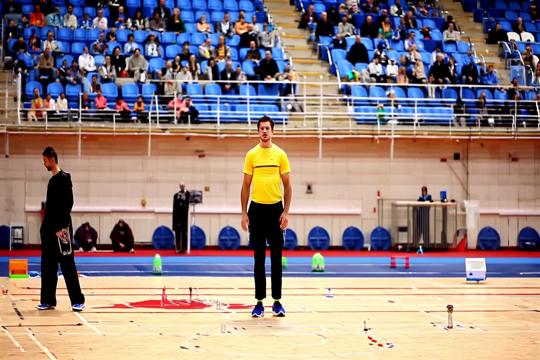}\hfill%
  \includegraphics[width=\imagewidth]{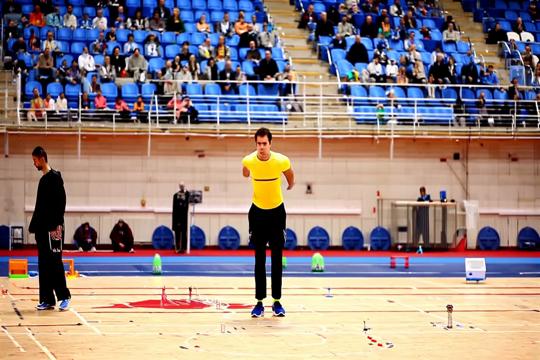}\hfill%
  \includegraphics[width=\imagewidth]{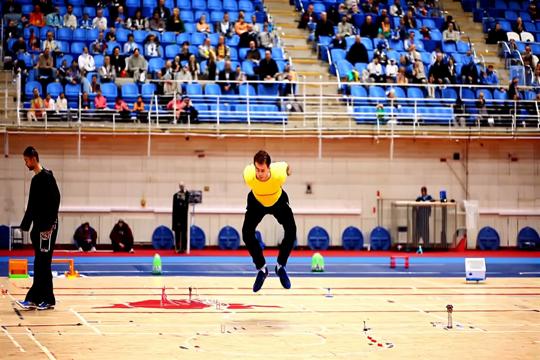}\hfill%
  \includegraphics[width=\imagewidth]{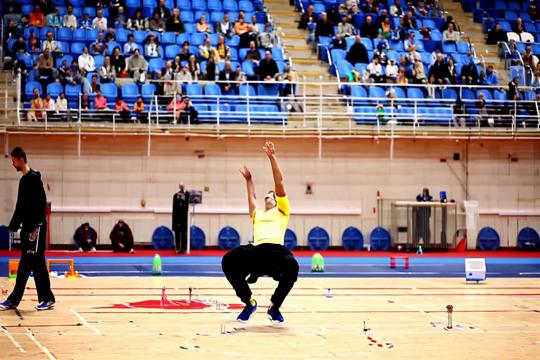}\hfill%
  \includegraphics[width=\imagewidth]{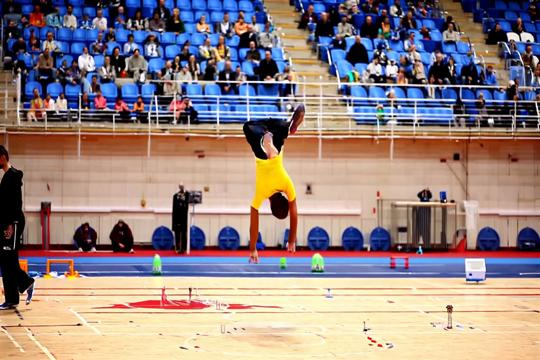}\hfill%
  \includegraphics[width=\imagewidth]{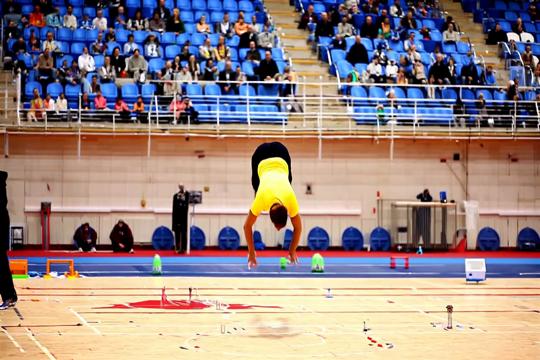}\hfill%
  \includegraphics[width=\imagewidth]{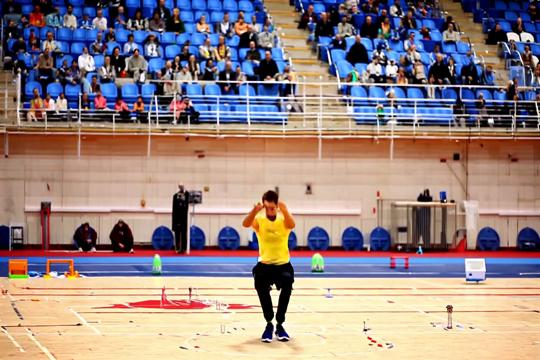}\hfill%
  \includegraphics[width=\imagewidth]{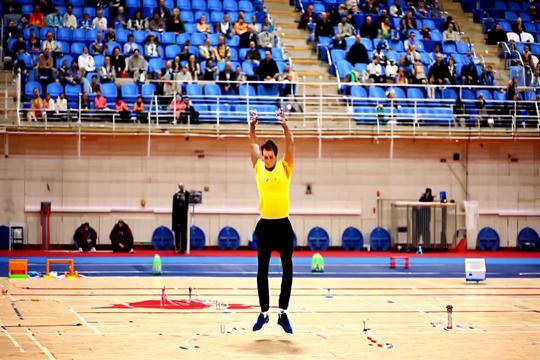}\\%
  \caption{Visualizations of the videos generated by our improved model, 
  trained using synthetic data.
Rows~1,2 highlight wide-angle camera motion;
rows~3 display layer decomposition;
and rows~4,5,6 demonstrate large human motion.}
  \label{fig:cases_vis}
\end{figure*}

% \TODO{rephrase this section emphasis more on physical fidelity, add sentences about we only focus on physics not aesthetics} 
To evaluate the effectiveness of synthetic videos for improving the physical fidelity of video generation models, 
we assess the trained model on three text-to-video tasks:
(a) large human motion (dancing and gymnastics), 
(b) camera spin shots, 
and (c) layer decomposition (\eg, a moving animal over a solid-colored background).
% These three tasks are representative of common challenges in reflecting physical fidelity for video generation models.
For each task, we use specific text prompts to test the model's ability to accurately generate the video content.
During evaluation, we focus on examining only the physical fidelity of the generated videos, and our criteria do not include aesthetics. 
% Nevertheless, we observe that models trained with synthetic data not only produce high visual quality but also achieve substantive quantitative improvements in physical fidelity.
\subsection{Implementation Details}
\noindent\textbf{Synthetic Video Dataset\ } 
Following the strategy we discussed in~\cref{sec:syn-design}, we first render 32,847 videos of static objects with diverse camera movements and scene setups using Blender and 18,364 videos of humans performing diverse motions captured in simple indoor scenes with different background colors using Unreal Engine for the experiments. Additionally, we plan to release over 1.5M synthetic videos on static objects and 300K synthetic videos on human motions that are outside of the scope of this research to facilitate future research.

\noindent\textbf{Experiment Setup\ }
To verify whether synthetic videos can benefit video generation models, 
we combine the resulting synthetic videos with real-world videos to train the video generation model with 8B parameters, 
pretrained with only real-world video data. 
In line with the tasks we aim to improve, we adopt the optimal strategy found in Sec.~\ref{sec:syn-design}. To evaluate our trained model, we create 10 prompts each for gymnastics, dancing, camera motion, and layer decomposition, resulting in 40 prompts in total. Please refer to~\cref{sec:prompt_details} for details.
We inference videos output with resolution of 1280$\times$720 and duration of 5s at 24 fps. We use the same negative prompts for all inference queries. We compare the outputs of our model trained with synthetic data against the outputs of both the original checkpoint and some of leading commercial models~\cite{kling, videoworldsimulators2024, gen3} at the same setting and follows the evaluation method in Sec.~\ref{sec:physics_metrics}. 

\subsection{Results}
\noindent\textbf{Large human motion\ }
Our model generates videos of humans performing dancing and gymnastics with significantly reduced limb collapse. 
As shown in~\Cref{tab:physics_human}, 
the user study indicates that our model produces fewer artifacts in generated videos compared with other models, including three leading generation models and our pretarined model named ``Base Model''. 
In particular, our video generation model greatly improves the success rate of gymnastics movements, 
while other video generation models generate significantly fewer successful cases.
The human pose estimation confidence scores, discussed in \cref{sec:physics_metrics}), further support the findings from the user study.
Although the synthetic videos used for training have less realistic shading,
the model still learns correct human body deformation from the synthetic data and preserves the base model's realism.
\Cref{fig:cases_vis} visualizes frames of our generated videos.
We can see that our model produces visually plausible shadows, 
a feature that other video generative models have struggled to achieve.

\begin{table}
    \centering
    \footnotesize
    \begin{tabular}{lcc|cc}
    \toprule
    \multirow{2}{*}{Model} & \multicolumn{2}{c}{$\epsilon_\mathrm{conf}$} & \multicolumn{2}{c}{User Study} \\
     & Gym$\uparrow$ & Dance$\uparrow$ & Gym$\uparrow$ & Dance$\uparrow$ \\
    \midrule
    Kling 1.6 \cite{kling}            &  0.715 &  0.812 &  10\% & 43\%  \\
    Runway Gen-3$\alpha$ \cite{gen3}  &  0.672 &  0.809 &  4\%  & 14\%  \\
    Sora \cite{videoworldsimulators2024}                &  0.722 &  0.813 &  15\% & 44\%  \\
    Base Model          &  0.779 &  0.818 &  9\%  & 30\%  \\
    Our Model           &  \textbf{0.791} &  \textbf{0.837} &  \textbf{61\%} & \textbf{86\%}  \\
    \bottomrule
    \end{tabular}
    \caption{The average confidence score of human pose estimation and user study results on the large human motion task.}
    % \TODO{Kevin, can you confirm the caption and mention that the confidence scores are averaged upon what and how many cases?}}
    % It has been mentioned in the paper?
    \label{tab:physics_human}
    \vspace{-0.1in}
\end{table}

\begin{figure}
    \centering
    \includegraphics[width=1\linewidth]{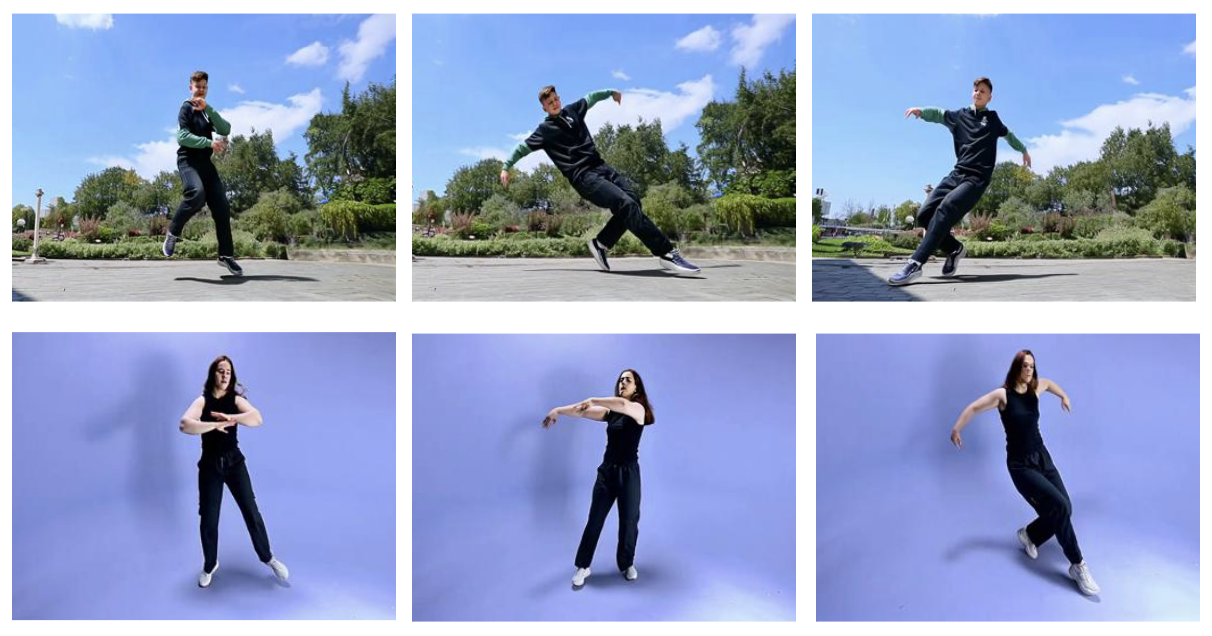}
    \caption{Visualization of video frames with large human motion generated by our model. 
    The shadow of human body follows the human motion.}
    \label{fig:shadow-visualization}
    \vspace{-0.1in}
\end{figure}

\noindent\textbf{Wide-angle camera rotation\ }
Our model can produce large camera spins around static objects and animals, 
as illustrated in \Cref{fig:cases_vis}. 
Our training set contains abundant synthetic videos featuring such camera rotations around daily objects. 
However, the objects used in our testing prompts -- including food, animals, and landscapes -- lie entirely outside the distribution of our training data. Nonetheless, the model successfully learns the general concept of extensive camera movements and adapts it to previously unseen objects while preserving a high level of realism.
As reported in~\Cref{tab:physics_objects}, 
the user study indicates that our model's success rate in producing the intended camera motion is significantly higher than that of other methods.
Furthermore, 3D reconstruction metrics, discussed in \cref{sec:physics_metrics}), confirm that objects generated by our model exhibit the strongest geometric consistency across different video frames. 
Our approach yields the largest number of feature points and the shortest track lengths,
indicate that our videos have the largest camera motion and meanwhile maintains best 3D consistency. 
When these feature points are projected back onto 2D images, our model's error $\hat{\epsilon}_\mathrm{proj}$ is more than twice as small compared with other approaches, demonstrating the enhanced physical fidelity of our generated videos.

% find that during camera movement, the objects in the generated video maintain better 3D consistency than videos of other models.
% Objects appeared in our model possess significantly more points to restore in the video and much shorter track length, while keeping relatively same reconstruction error. When only counting limit number of points, we can achieve much better reconstruction. \TODO{Xingyu, could you please rephrase and elaborate these two sentence, perhaps also add more on the implication of better 3D conssitentcy? Thanks.}
% \xy{The objects generated by our model exhibit strong geometric consistency across different frames in the video, with significantly more feature points and shorter track lengths (i.e., the average number of frames in which a single 3D feature point appears). When these feature points are projected back onto 2D images, the error $\hat{\epsilon}_\mathrm{proj}$ produced by our model is more than twice as small.}

\begin{table}
    \centering
    \footnotesize
    \setlength{\tabcolsep}{3pt}
    \begin{tabular}{lcccc|c}
    \toprule
    % \multirow{2}{*}{Models} & \multicolumn{4}{c}{physics metrics}                      & \multicolumn{1}{c}{user study} \\
      Model           & $\mathcal{N}\uparrow$ & $\mathcal{T}\downarrow$  & $\epsilon_\mathrm{proj}$$\downarrow$ & $\hat{\epsilon}_\mathrm{proj}$$\downarrow$ & User Study$\uparrow$ \\
    \midrule
    Kling 1.6~\cite{kling}          & 13,328    &  36.34 &  \textbf{0.972} & 0.298 & 20\% \\
    Runway Gen-3$\alpha$~\cite{gen3} & 13,199    &  36.21 &  1.181 & 0.361 & 26\% \\
    Sora~\cite{videoworldsimulators2024}               & 14,443    &  33.62 &  1.244 & 0.318 & 25\% \\
    Base Model         & 16,548    &  31.84  &  1.159 & 0.437 & 20\% \\
    Our Model          & \textbf{42,895}    &  \textbf{12.93} &  1.077 & \textbf{0.135} & \textbf{80\%} \\
    \bottomrule
    \end{tabular}
    \caption{3D reconstruction metrics and user study results on the large camera motion task.    
    Note that the re-projection error $\epsilon_\mathrm{proj}$ is computed over \emph{all} extracted feature points, 
    whereas $\hat{\epsilon}_\mathrm{proj}$ only considers the 1{,}000 points with the smallest error in each case.
    The latter metric offers a fairer comparison for methods that produce a significantly higher volume of feature points.
    }
    \label{tab:physics_objects}
    \vspace{-0.1in}
\end{table}

\eat{
\begin{wrapfigure}[7]{r}[0pt]{0.2\textwidth}
\begin{center}
\vspace{-0.7cm}
\hspace{-0.7cm}
\footnotesize
\begin{tabular}{lc}
    \toprule
    Model & User Study  \\
    \midrule
      Kling 1.6~\cite{kling}             & 4\%  \\
      Runway gen-3$\alpha$~\cite{gen3}   & 1\%  \\
      Sora~\cite{videoworldsimulators2024}                  & 4\%  \\
      Base Model     & 26\% \\
      Our Model        & \textbf{84\%} \\
      \bottomrule
    \end{tabular}
\end{center}
\end{wrapfigure}
}
\noindent\textbf{Layer decomposition\ }
% Our model appears to exhibit enhanced understanding of real physical world after trained with synthetic data. 
As shown in~\Cref{tab:physics-object},
While the baseline models largely fail, our model can produce outputs with a clear separation of the foreground object and the background when tasked to generate videos on pure color backgrounds. This decomposition is beneficial for compositing objects onto arbitrary backgrounds. Similarly to the large camera motion scenario, our model shows this capability to objects not present in the training dataset. 
\Cref{fig:cases_vis} shows an example in which the requested object appears cleanly over a green background,
suggesting that the model has learned to decompose the scene effectively.
Furthermore, the model can even generate dynamic objects and human motion in layers, 
decomposing the scene in both spatial and temporal dimensions. 
Neither the original pretrained model nor other commercial models achieve such a clear separation of layers.

\begin{table}
    \centering
    \begin{tabular}{lc}
    \toprule
    Model & Layer Decomposition$\uparrow$  \\
    \midrule
      Kling-1.6~\cite{kling}            & 4\%  \\
      Runway-gen3$\alpha$~\cite{gen3}   & 1\%  \\
      Sora~\cite{videoworldsimulators2024}                  & 4\%  \\
      Base Model     & 26\% \\
      Our Model        & \textbf{84\%} \\
      \bottomrule
    \end{tabular}
    \caption{User study results on the layer decomposition task. With synthetic data augmentation, our model greatly outperforms leading commercial models and the original pretrained model.
    }
    \vspace{-0.05in}
    \label{tab:physics-object}
\end{table}

\subsection{Ablation Studies}
\label{sec:4_3}
\noindent\textbf{Ablations on synthetic captions\ } We perform experiments of different synthetic caption setups to verify our design in~\cref{sec:3_2}. Experiments in~\Cref{tab:dance-caption} studies if the fine-grained captions help video generation model better learn human motion than their generic counterparts, typically from a zero-shot VLM inference. We observe that for various dance moves (``Uprock'', ``Spin'', ``Freeze''), having fine-grained caption (\Cref{fig:caption-showcase-c}) greatly reduce the video generation model to generate videos that include collapse and distortion of human body during large motions. Table~\ref{tab:special-tags} summarizes the experiments on embedding special tags in synthetic captions to distinguish synthetic videos from the real videos. We found that without special tags, the video generation model is much likely to output videos of animated visual or collapsed human body. We further added the special tags in negative prompts, but found only marginal improvements.

\begin{table}
\centering
\small
\begin{tabular}{l|c|c|c}
\toprule
Caption Type & Uprock$\uparrow$ & Spin$\uparrow$ & Freeze$\uparrow$ \\ 
\midrule
a) Generic  & 2\%  & 16\%  & 0\% \\
b) Fine-grained  & 98\%  & 84\%  & 66\% \\
\bottomrule
\end{tabular}
\caption{Fine-grained captions on human motion achieve better successful rate than generic captions on the large human motion task. ``Uprock'', ``Spin'', ``Freeze'' are particular dance moves.}
\label{tab:dance-caption}
\vspace{-0.1in}
\end{table}

\begin{table}
\centering
\small
\begin{tabular}{l|c}
\toprule
Caption Type & Dance Move \\ 
\midrule
a) No Special Tags  & 12.5\% \\
b) Special Tags  & 90\% \\
c) Special Tags+Special NP  & 92.5\% \\
\bottomrule
\end{tabular}
\caption{Experiment results on the effect of special tags in synthetic data captioning. Without special tags to differentiate the visual style of the synthetic videos, the video generated models will more likely to generate animated characters or collapsed human motions after training. Also, adding the special tags in negative prompts during generation will help although marginally.}
\label{tab:special-tags}
\vspace{-0.1in}
\end{table}

\noindent\textbf{Ablations on training with synthetic videos\ } To examine the mix rate of the synthetic and real videos. we perform the experiment summarized in~\Cref{tab:mix-rate}. We found that higher mix rate share the same effect as longer training steps and over-training the model on synthetic data will not lead to more performance increase. Instead, more patterns from synthetic data will appear in the final output. We also verify the design of \emph{SimDrop} in~\Cref{tab:sim-drop}. We found that using the captions in training the reference model to prompt them will achieve the best result in terms of the visually preferred cases rated by humans. It also reports the impact of the hypereparameter $\alpha$ value.
%and an value over 0.3 will introduce highly corrupt visual patterns that make visuals unrealistic to compare. We let human evaluate which video is more preferred over the others.

\vspace{-0.2cm}
\begin{table}
\centering
\small
\begin{tabular}{c|cccc}
\toprule 
\diagbox{Percentage}{\#Steps}  & 3000 & 5000 & 10000 & 15000 \\
\midrule
10\% synthetic videos  & 20\% & 25\% & 40\% & 60\%\\
50\% synthetic videos  & 55\% & 75\% & \textbf{85\%} & 80\% \\
\bottomrule
\end{tabular}
\caption{Ablation results on synthetic data mix rate and training steps. Here we measure the success rate which the trained foundation model generates videos that follows the prompts but does not include visual patterns in the synthetic videos. We found that large proportion and longer training steps help transferring the properties in synthetic videos to the video generation model. However, performance will saturate and failure cases will include visual patterns of synthetic data.}
\vspace{-0.3cm}
\label{tab:mix-rate}
\end{table}

\begin{table}[h!]
\centering
\begin{tabular}{c|c|c|c|c}
\toprule
 $\alpha$ & Good & Same & Bad & G-B$\uparrow$\\
\midrule
0.1 & 26.32\% & 71.05\% & 2.63\% & 23.69\%\\
0.2 & 39.47\% & 52.63\% & 7.89\% & 31.58\%\\
\bottomrule
\end{tabular}
\caption{Experiment results on \emph{SimDrop}. Here, we compare the output videos with SimDrop with the models without SimDrop. Evaluators will choose the best out of two videos side-by-side. We then compute the winning/same/losing rate against the baseline.}
\label{tab:sim-drop}
\vspace{-0.3cm}
\end{table}

\section{Related Work}

\textbf{Video generation\ } 
Conditional video generation is a challenging task aiming to synthesize temporally coherent and visually realistic video sequences from structured inputs such as images and text prompts. Current video generation models can be broadly categorized into Generative Adversarial Networks (GANs) ~\cite{goodfellow2014generative, li2019storygan, balaji2019conditional, brooks2022generating}, autoregressive models~\cite{kondratyuk2023videopoet, yu2023magvit, yu2023language, kumar2019videoflow}, and diffusion models~\cite{opensora, hong2022cogvideo, yang2024cogvideox, ehtesham2024movie, kong2024hunyuanvideo, gupta2024photorealistic}. These architectures in video generation usually inherit their success in image generation ~\cite{zhang2022styleswin, chang2022maskgit, esser2024scaling, peebles2023scalable}. In recent years, rapid advancements in video generation, represented by Sora~\cite{videoworldsimulators2024}, have been significantly driven by the availability of large-scale web-collected video datasets and the development of scalable model architectures such as DiT~\cite{esser2024scaling}. State-of-the-art commercial models ~\cite{kling, veo2, gen3, kong2024hunyuanvideo} have demonstrated the ability to generate highly realistic videos. These models leverage extensive training data to improve motion fluency, scene reality and overall aesthetic quality in generating videos. 

\noindent\textbf{Physics in video generation\ } Despite the effort in scaling data and model size, problems remain in the physics of generated videos after researchers' evaluation~\cite{motamed2025generative, bansal2024videophy, kang2024far}. Yet for video generation models, physics appears learnable directly from video data~\cite{garrido2025intuitive, chari2019visual, wu2017physics, meng2024towards} and is crucial for these foundation models to serve as world models~\cite{videoworldsimulators2024, agarwal2025cosmos, wang2024worlddreamer, du2023learning}. Therefore, there are growing number of works~\cite{liu2025generative, ba2019blending} in improving physics-grounding in video generation and beyond~\cite{menapace2022playable, bear2021physion}. They mainly propose model modifications by adding additional supervisory signals~\cite{liu2024physgen, chefer2025videojam, kadambi2023incorporating, xue2024phyt2v}, and mainly tailored for a certain aspect of physics such as motion~\cite{materzynska2024newmove, Lv_2024_GPT4Motion, chefer2025videojam} or sound in the videos~\cite{Su_2023_CVPR}. While such methods show more physically coherent results, they often require modifications to the diffusion architecture itself and rely on manually specified control signals. Our work focuses on physical fidelity and differentiates by proposing a data-centric approach without modifying the diffusion model architecture and harness the potential of 3D rendering engines~\cite{de2022next}. Our method build synthetic video data that can benefit video generation models regardless of their architectures and improves on diverse aspects of physics fidelity.

%Recently, a few works have attempted to enhance the physical realism of generated videos~\cite{liu2024physgen, chefer2025videojam, xue2024phyt2v}. 
%These methods can be broadly categorized into two approaches: 
%(1) introducing extra control signals (\eg, camera pose, human pose) into the diffusion process and 
%(2) adding additional channels (\eg, optical flow) to the input/output of the diffusion model. 
%While such methods show more physically coherent results, they often require modifications to the diffusion architecture itself and rely on manually specified control signals. 

\noindent\textbf{Synthetic data in AI\ }
Synthetic Data from simulation engines has been widely applied in advancing many fields of AI, such as autonomous driving~\cite{zhou2024simgen, xie2025vid2sim} and embodied agents~\cite{shridhar2020alfred, puig2018virtualhome, zhao2024epo}, or scene generation~\cite{shang2024urbanworld}. At the intersection of synthetic data and video, most work focus on understanding~\cite{yang2024depth, zhang2024video, Kim2022Video} and only a few early work~\cite{ba2019blending} explore how synthetic video data can help video generation in particular tasks such as camera control~\cite{bai2024syncammaster} or motion~\cite{Lv_2024_GPT4Motion, geng2024motion}. 
We are the first work to systematically study how synthetic videos from simulation engines can help improve the physics fidelity of video generation model.

\section{Conclusion}

In this study, we investigate how to use synthetic video data generated by CGI production pipelines (Blender~\cite{Blender} and Unreal Engine~\cite{UE5}) to enhance physical fidelity of video generation models. We verify our method on three tasks necessitating realistic physical behavior, where our model achieves superior results through synthetic data enhancement. Our results demonstrate that the physical fidelity of video generation can be enhanced using synthetic video. Note that while our method improves physical fidelity and aligns more closely with human perception, it still lacks an understanding of the underlying principles of physics, leaving significant room for further improvement.

Going beyond, future work may consider generating more intricate physical effects~\cite{liu2025generative}, including complex interactions among multiple objects and physically based fluid simulations. Moreover, while we only focus on the RGB color channel in this work, the synthetic rendering pipeline offer much more information(\eg, depth, normals, alpha masks) that could serve as supervisory signals, otherwise not easily obtainable in real datasets. 
%In addition, noticeable differences remain in the generated videos from the model enhanced with synthetic data. We hope these findings would facilitate research in this realm.
%This omission may limit the extent to which synthetic data can contribute to improved fidelity.
%A current limitation of our method is that it requires synthesizing videos for each task, which might be inefficient if we want to support a broad range of applications. 
%In addition, although the generated videos from our model are not obviously inferior to those produced by other approaches, noticeable differences remain. 
%This finding indicates that synthetic videos can still undesirably affect certain aspects of the final appearance.

\section{Acknowledgment}
We thank Ceyuan Yang, Liangke Gui, and Shanchuan Lin for their insightful discussions on this project. We also appreciate Zhibei Ma and Renfei Sun for their support in building the engineering foundation.

{
    \small
    \bibliographystyle{ieeenat_fullname}
    \bibliography{main}
} 

\clearpage
\appendix
\section{Appendix}

\subsection{Details of Synthetic Data Generation}
\label{sec:engine_details}

\begin{table*}
\centering
\footnotesize
\begin{tabular}{c|p{0.2\linewidth}|p{0.2\linewidth}|p{0.5\linewidth}}
\toprule
        & Property Name & Choice & Description \\
        \hline        
\multirow{8}{*}{\rotatebox[origin=c]{90}{Camera}} 
        & Camera Focus Type & Follow & The camera focus follows the object. \\
        \cline{3-4}
        &                   & Fixed  & The camera focus is static in the world space. \\
        \cline{2-4}
        & Camera Focus Position & Upper, Center, Lower  & The camera focus is at the upper/center/lower part of the object.  \\
        \cline{2-4}
        & Camera Movement Type  & Truck, Dolly, Pedestal, Tilt, Pan, Spin, Following, Zoom & The basic camera movement types. \\
        \cline{2-4}
        & Camera Movement Value & Scalar                                                   & How much the camera moves.       \\
        \cline{2-4}
        & Camera Initial Position & 3D Position                                            & The initial position of the camera. \\
        \cline{2-4}
        & Camera Focal Length    & Scalar                                                  & The scalar controls how much percentage of the object is visible on the screen. \\
\hline
\multirow{11}{*}{\rotatebox[origin=c]{90}{Light and Environment}} 
        & Scene Type & Env & The environment is given by a HDR environmental map. The map will also be used as the light source.  \\
        \cline{3-4}
        &            & Basic & The environment is an indoor room which color is controlled by ``Scene Color'' and has two light sources. \\
        \cline{3-4}
        &            & Empty & The environment is empty but has two light sources or one environmental map as the light source.\\
        \cline{2-4}
        & Scene Color & RGB color & The color for the indoor room when presented. \\
        \cline{2-4}
        & Light Position & 3D position & The position of the light when presented. \\
        \cline{2-4}
        & Light Color & Scalar & The color temperature of the light when presented. \\
        \cline{2-4}
        & Light Intensity & Scalar & The intensity of the light when presented.\\
        \cline{2-4}
        & Ambient Light Intensity & Scalar & Ambient light intensity. The ambient light exists when the lights are used. \\
\hline
\multirow{3}{*}{\rotatebox[origin=c]{90}{Render}}
        & Background Color & RGBA color  & The background color of the location where the scene is empty. \\
        & Render Engine    & Blender/Unreal & \\
        & Render Quality   & High/Low    & The quality of the rendering. We have two presets of rendering setting.     \\
\bottomrule
\end{tabular}
    \caption{The parameters used for controlling our rendering pipeline.}
    \label{tab:synthetic-parameters}
\end{table*}

\begin{figure*}[t]
    \centering
    \includegraphics[width=1\linewidth]{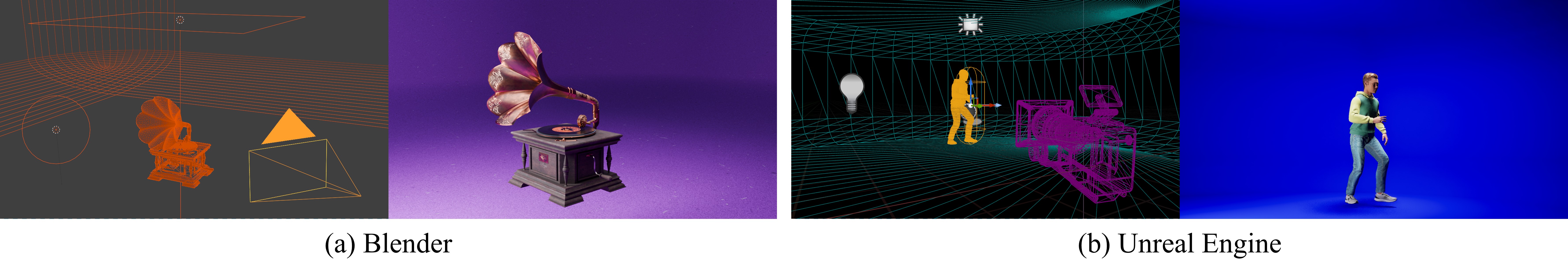}
    \caption{3D scene setup in Blender and Unreal Engine. The wireframes and corresponding rendering outputs.}
    \label{fig:3D-scene}
\end{figure*}

Following a standard CGI production pipeline for creating videos, 
our synthetic video generation framework comprises two main modules: 
(1) \emph{3D scene setup} and (2) \emph{rendering}. 
Below, we provide a detailed overview of these modules and the specific parameters that govern them.

\subsubsection{3D Scene Setup}
As discussed in \cref{sec:engine}, we focus on generating videos featuring a single object per scene.
To achieve this, we build a procedural 3D scene generator driven by a carefully chosen set of parameters, 
enabling the production of a wide variety of synthetic videos.
A typical 3D scene is composed of four main components: 
(1) the 3D object, (2) the camera, (3) the lighting conditions, and (4) the environment.
We adopt this composition in our generator.
Each component in our generator is controlled by a set of parameters, which we detail below.

\noindent\textbf{3D Object.} 
As we target single-object videos, 
we seek to include 3D assets that are both high-quality and highly varied. 
To this end, we collect assets from 
Objaverse 1.0~\cite{deitke2023objaverse}, 
Digital Twin Catalog~\cite{DTC}, 
Blender Market~\cite{BlenderMarket}, 
and Metahuman~\cite{Metahuman}. 
These sources collectively provide diverse asset categories and styles. 
We further filter assets from Objaverse based on categories, polygon count, view count, 
user ratings, and VLM to ensure overall quality. 
For other sources, we retain all assets since they are already curated with high fidelity.

\noindent\textbf{Camera.}
We represent the camera using a set of parameters that capture real-world usage scenarios (see Table~\ref{tab:synthetic-parameters}).
These parameters include:
\begin{itemize}
    \item \emph{Camera movement type:} Determines the camera’s trajectory around the object. 
    In our experiments, we select one movement type at a time and quantify its extend using a parameter ``Camera Movement Value''.
    \item \emph{Initial position and focus:} Specifies where the camera starts and how it focuses on the primary object.
    \item \emph{Focal length:} Adjusts the camera’s field of view relative to how much of the screen the object occupies.
\end{itemize}
Such parameterization allows us to mimic various camera behaviors from the real world.

\noindent\textbf{Lighting and Environment.}
For simplicity, 
we jointly model the environment and its lighting conditions (see Table~\ref{tab:synthetic-parameters}).
Our parameterization supports three main configurations:
\begin{itemize}
\item \emph{HDR environment map:} Provides both the background and primary light source. 
We use environment maps from Poly Haven~\cite{PolyHaven}.
\item \emph{Solid-color indoor room:} Uses two light sources (Figure~\ref{fig:3D-scene}) for illumination: 
one positioned above the object and another placed elsewhere in the scene.
\item \emph{Empty scene:} Lit by either an environment map or two lights for more controlled illumination with empty surroundings.    
\end{itemize}
Although these settings may appear simple, they cover a wide range of lighting conditions and backdrop variations, thereby maintaining diversity while keeping the primary object prominent.

\subsubsection{Rendering Setup}
We employ two open-source rendering engines to generate high-quality video outputs:
\begin{itemize}
\item \emph{Unreal Engine (Lumen):} We use Unreal Engine 5.4.4 with Lumen as our renderer with maximal render-quality settings to achieve realistic rendering effects~\cite{UE5}.
\item \emph{Blender (Cycles):} We use Blender 4.2 and Cycles renderer configured with carefully chosen parameters to balance rendering speed and visual fidelity~\cite{Blender}
\end{itemize}
These engines offer robust rendering pipelines and physically based shading models, 
ensuring that our synthetic data closely reflects real-world lighting conditions.

\begin{figure}
    \centering
    \includegraphics[width=1\linewidth]{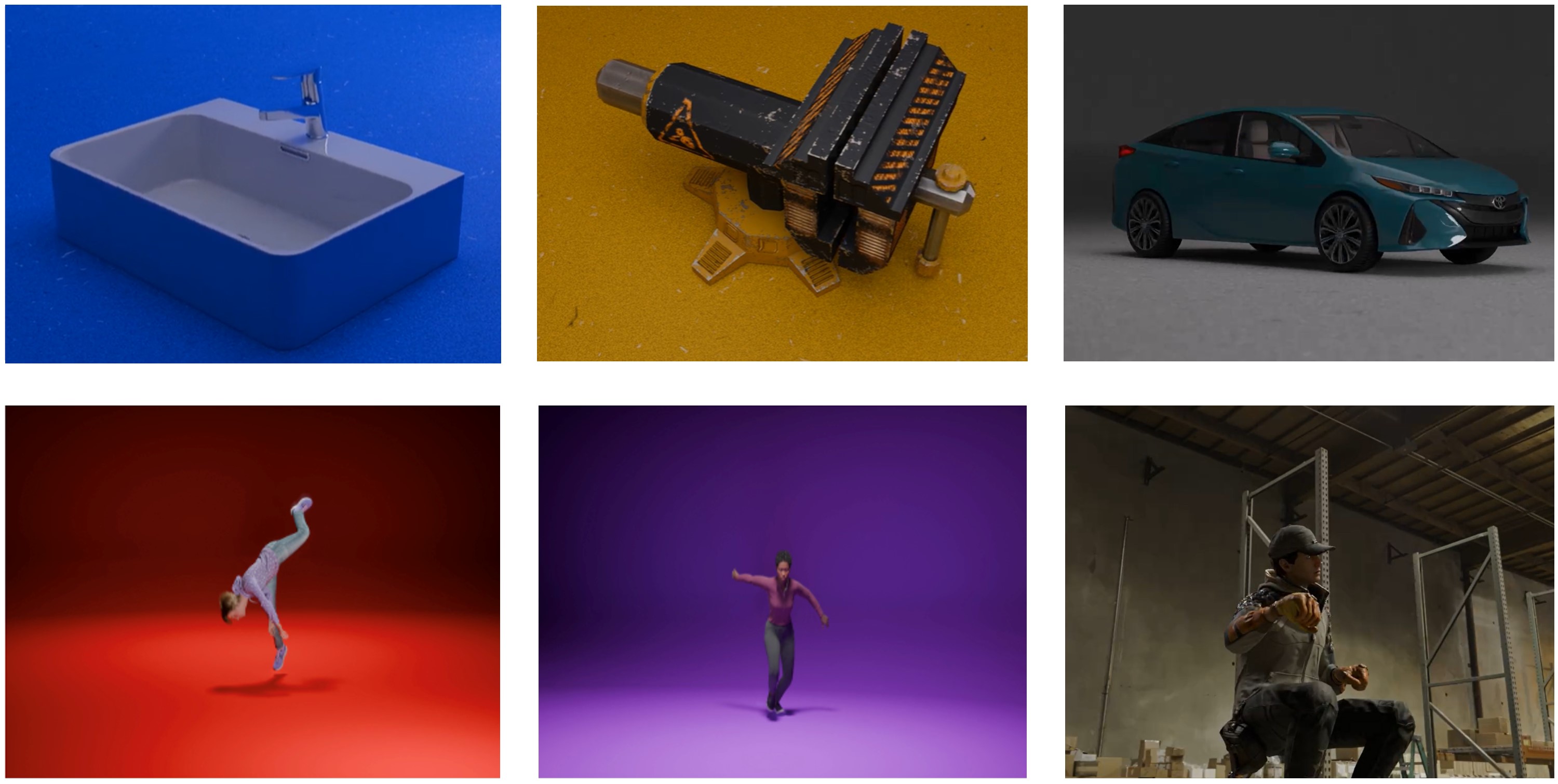}
    \caption{Examples of our synthetic video data. We render the synthetic videos with diverse background to alleviate the potential biases in synthetic videos.}
    \label{fig:data-showcase}
    % \vspace{-0.15in}
\end{figure}

\subsubsection{Random Sampling of Parameter Space}
\label{sec:engine_details_random_sample}
To produce a large and diverse set of synthetic videos, 
we define a configuration (``config'') file containing all relevant parameters described above.
Figure~\ref{fig:data-showcase} show some examples of synthetic videos with diverse setups.
Our 3D scene generator parses this config file and sets up the scene.
Then, the rendering engines render the scene into a video.
For large-scale generation, 
we employ random sampling over each parameter’s prescribed probability distribution, 
guided by the key insights from Sec.~\ref{sec:syn-design}. 
Each sampling step produces a unique config file, 
which is then rendered into a separate synthetic video. 
This process enables us to generate a vast set of diverse synthetic videos with minimal manual intervention.

% \TODO{Bohan/Ziyu: A brief introduction to 3D rendering engine. what are object assets, human assets, and motion assets and how they can be composed to generate a video? put the detailed dataset in appendix}
% \kevin{can include the scope of assets here? We use objaverse, DTC, xxx (number) unreal human character and human motion assets to create synthetic videos.}

\subsection{More Ablation Experiments and Visualizations}
\label{sec:ablation_details}
\begin{figure}
    \centering
    \includegraphics[width=1\linewidth]{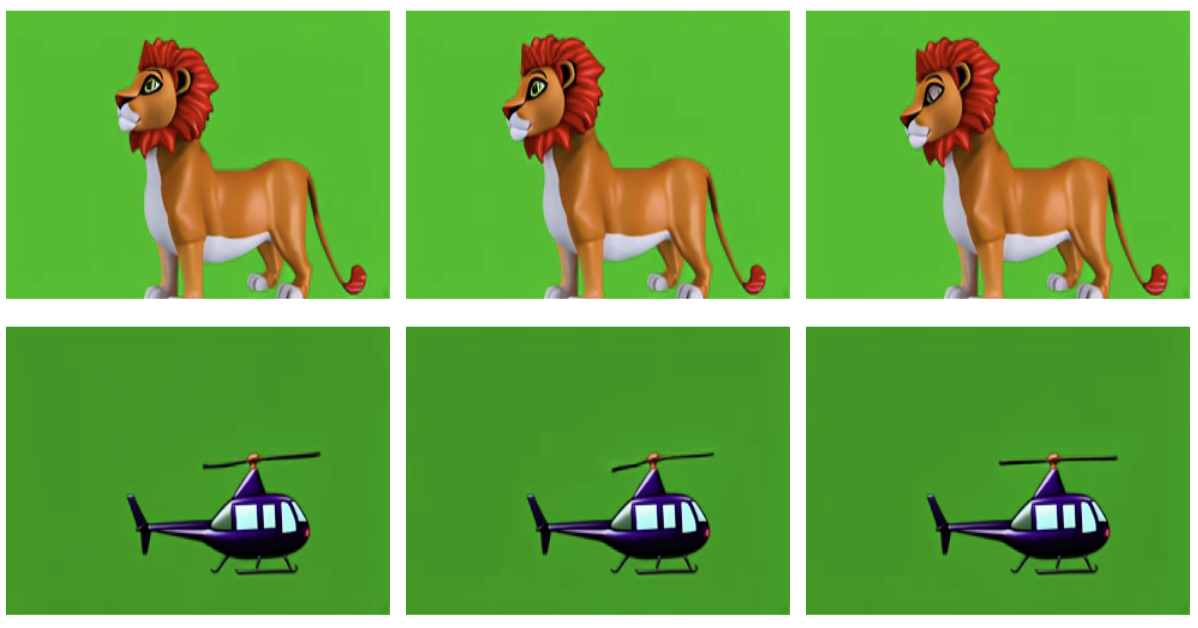}
    \caption{Example outputs from video generation models trained on synthetic datasets with low-quality assets. The resulting objects frequently exhibit cartoonish or animated characteristics, diverging from the intended original visual style.}
    \label{fig:bad-asset-layer}
\end{figure}

\begin{figure}
    \centering
    \includegraphics[width=1\linewidth]{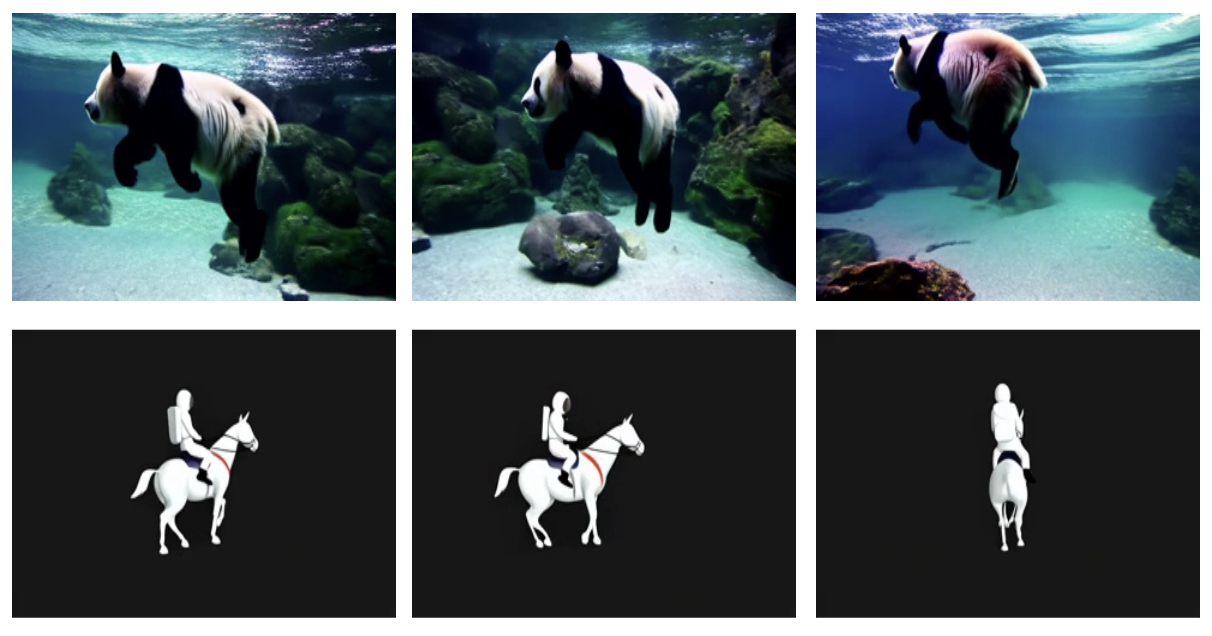}
    \caption{Visualization of generated outputs from video generation models trained with synthetic videos of low quality assets in large camera motion task. The objects in these generated videos more likely to appear static or animated.}
    \label{fig:bad-asset-spin}
\end{figure}

\begin{figure}
    \centering
    \includegraphics[width=1\linewidth]{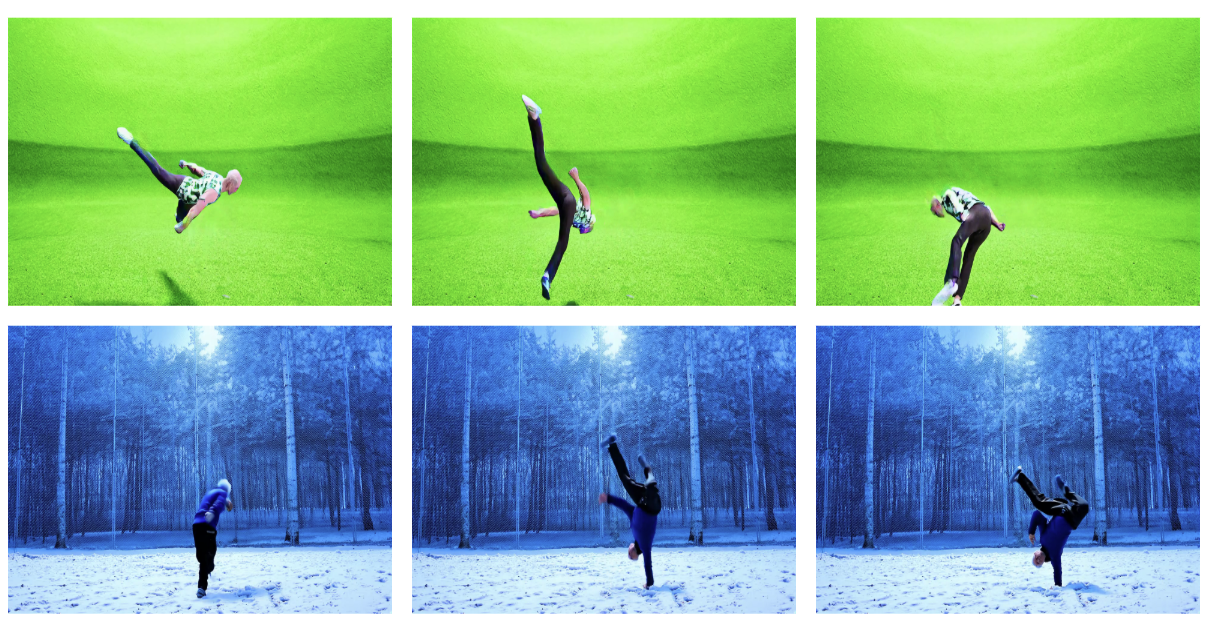}
    \caption{Visualization of over training video generation models trained with synthetic videos. Visual patterns such as color tone are more likely to appear in generated videos.}
    \label{fig:overtrain}
\end{figure}

\begin{figure}
\centering
\includegraphics[width=1\linewidth]{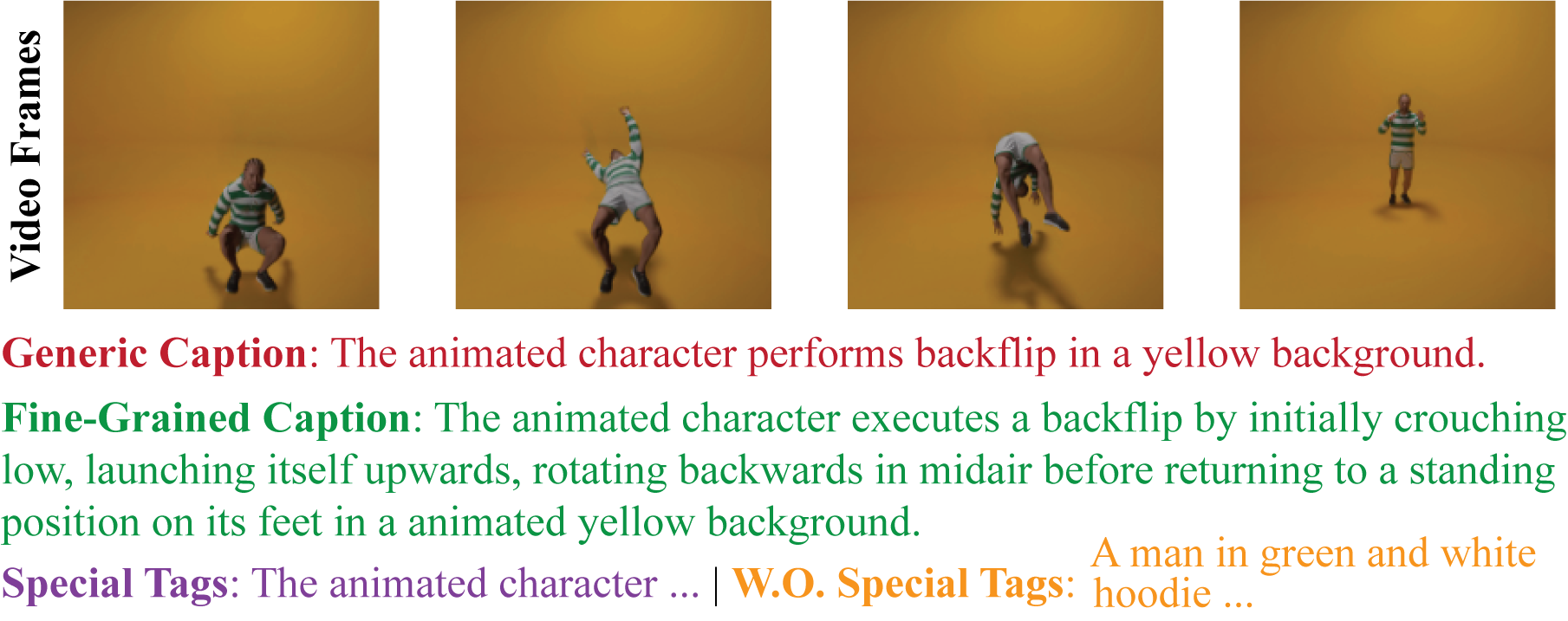}
\vspace{-0.7cm}
\caption{A comparison of generating captions for synthetic videos
using existing methods (Generic Caption) and our method (Fine-Grained Caption).
We also show a comparison of captions with special tags and without special tags.}
\vspace{-0.5cm}
\label{fig:caption-showcase-c}
\end{figure}

\begin{figure}
    \centering
    \includegraphics[width=1\linewidth]{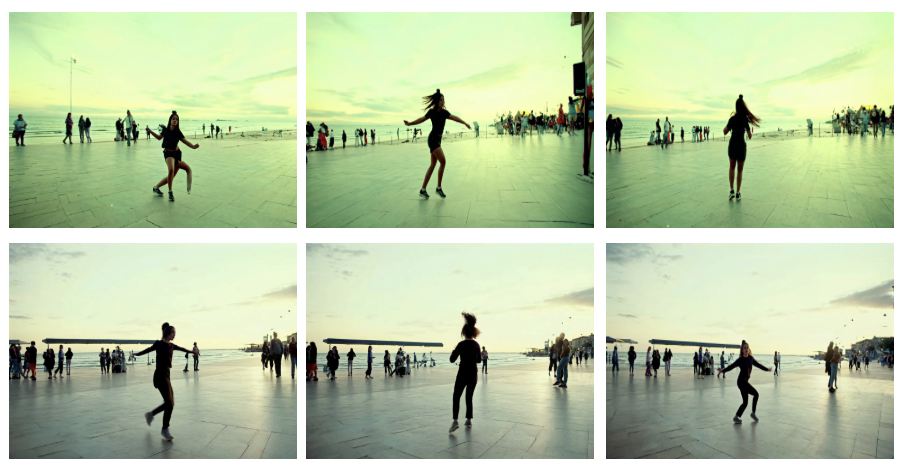}
    \caption{A comparison showcasing the effect of \emph{SimDrop}. Row 1 is the result without \emph{SimDrop} and Row 2 is the video with the method. The color tone in row two is significantly more better and without color pattern from the synthetic data.}
    \label{fig:sim-drop1}
\end{figure}

\begin{figure}
    \centering
    \includegraphics[width=1\linewidth]{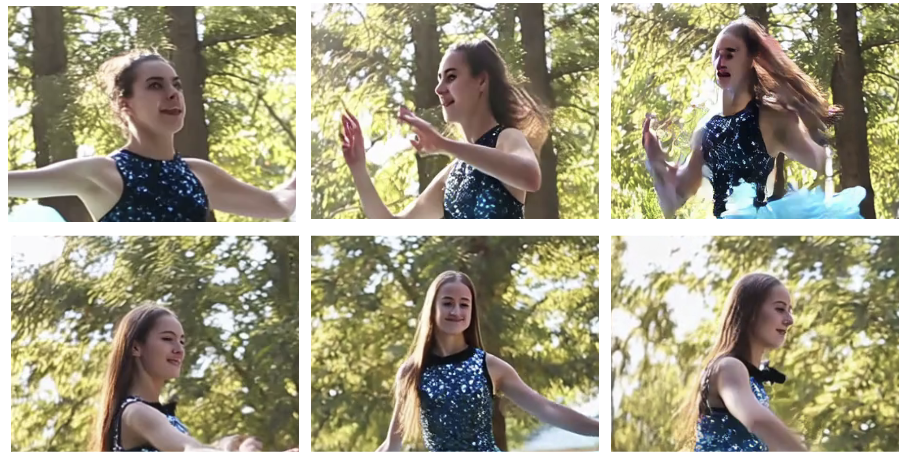}
    \caption{A comparison showcasing the effect of \emph{SimDrop}. Row 1 is the result without \emph{SimDrop} and Row 2 is the video with the method. The human faces in row two is significantly more realistic and appealing.}
    \label{fig:sim-drop2}
\end{figure}

\begin{figure}
    \centering
    \includegraphics[width=1.0\linewidth]{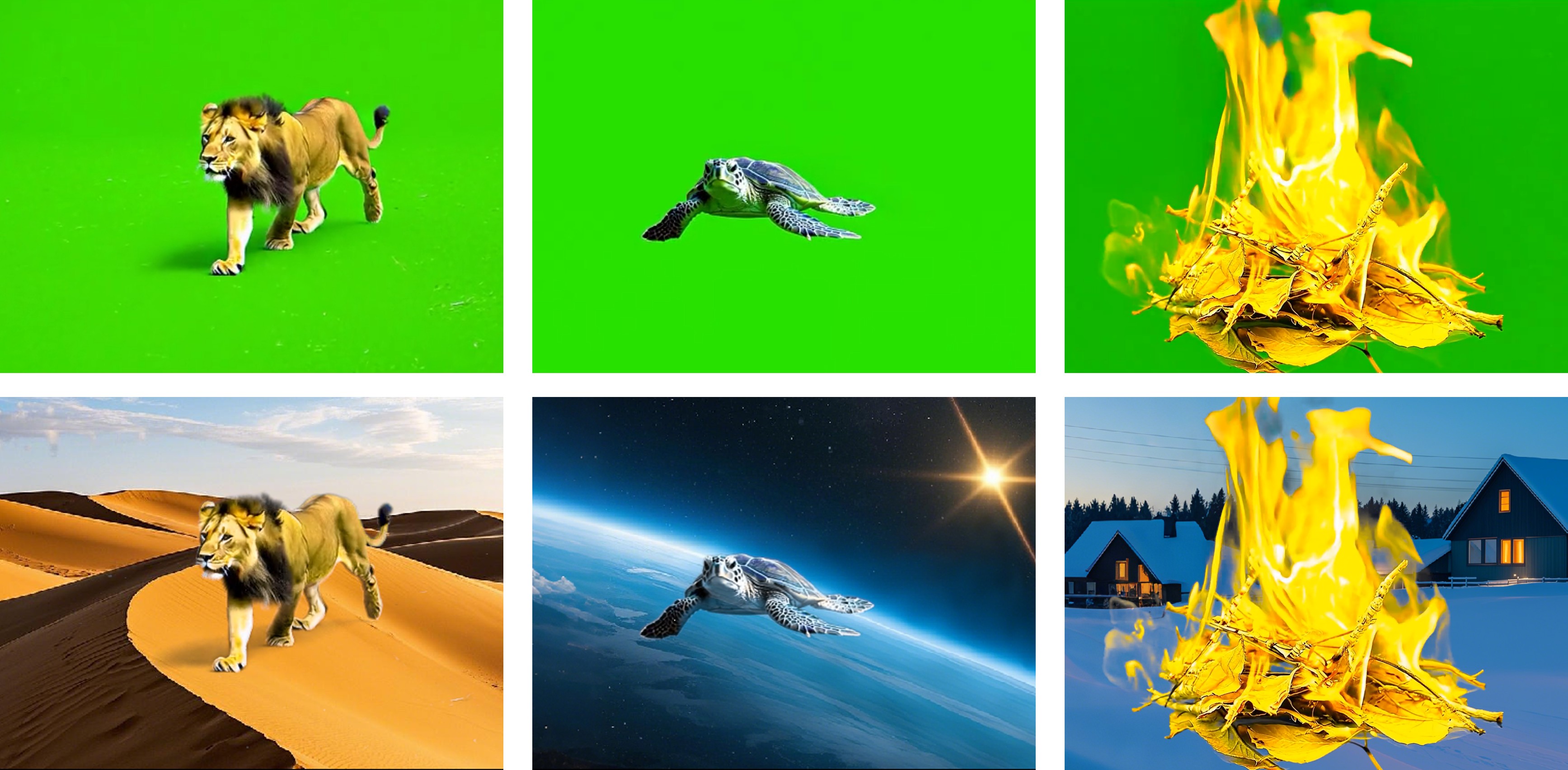}
    \caption{Example of background editing. Our layer generation enables easy background replacement via green-screen matting.}
    \label{fig:change_bg}
\end{figure}

\begin{figure*}
  \centering
  \setlength{\imagewidth}{.123\linewidth}
  \includegraphics[width=\imagewidth]{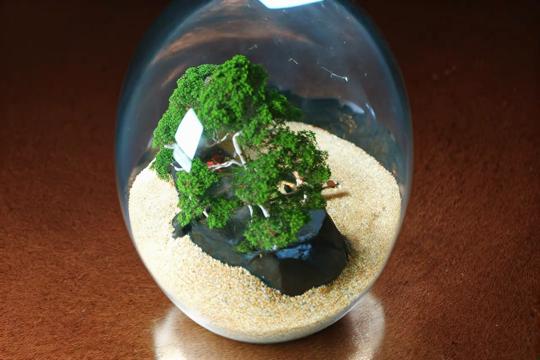}\hfill%
  \includegraphics[width=\imagewidth]{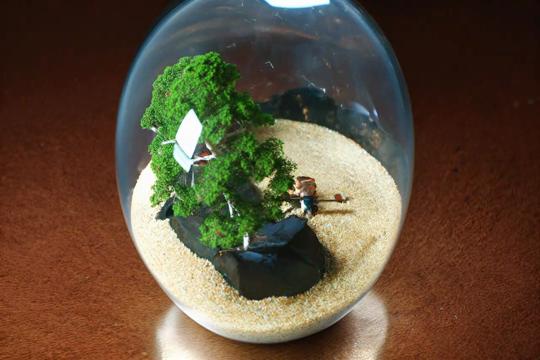}\hfill%
  \includegraphics[width=\imagewidth]{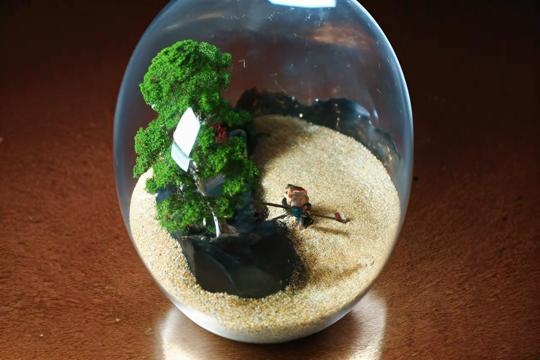}\hfill%
  \includegraphics[width=\imagewidth]{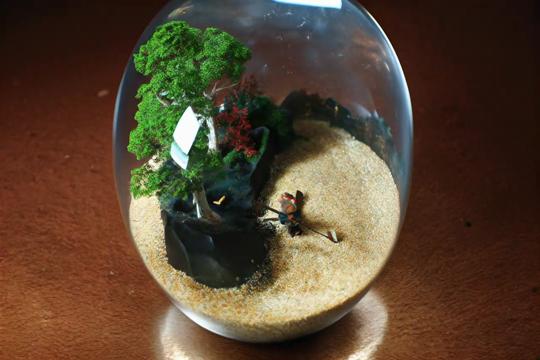}\hfill%
  \includegraphics[width=\imagewidth]{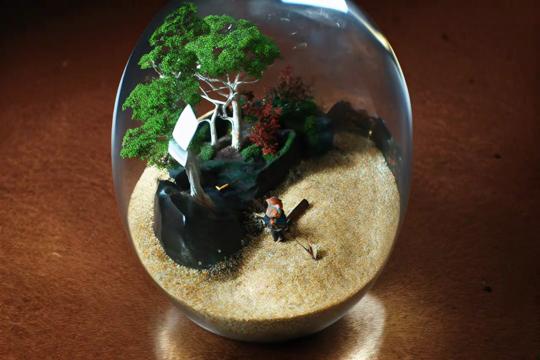}\hfill%
  \includegraphics[width=\imagewidth]{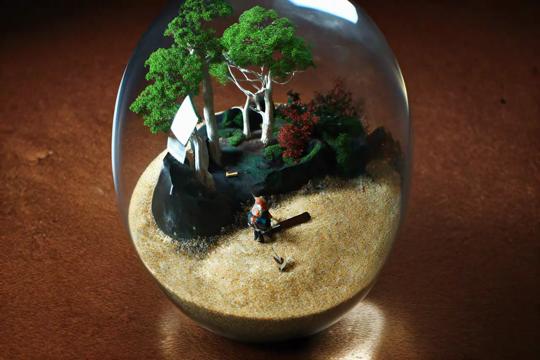}\hfill%
  \includegraphics[width=\imagewidth]{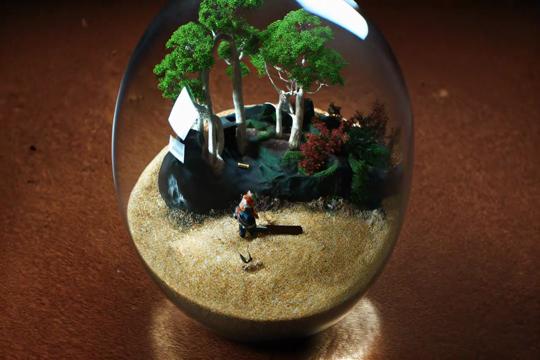}\hfill%
  \includegraphics[width=\imagewidth]{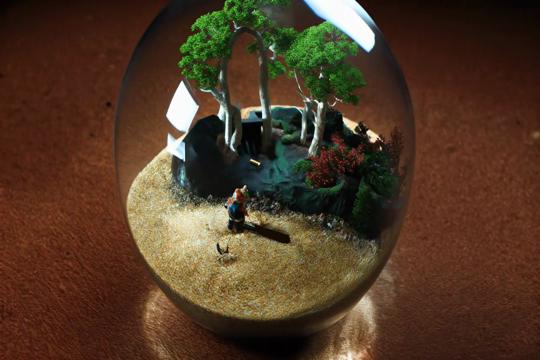}\\%
  \includegraphics[width=\imagewidth]{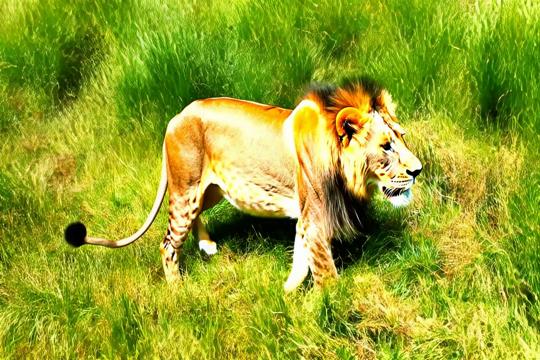}\hfill%
  \includegraphics[width=\imagewidth]{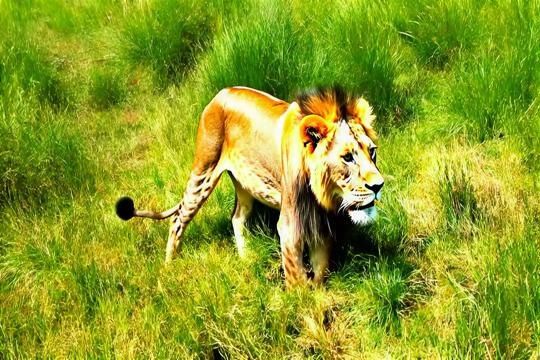}\hfill%
  \includegraphics[width=\imagewidth]{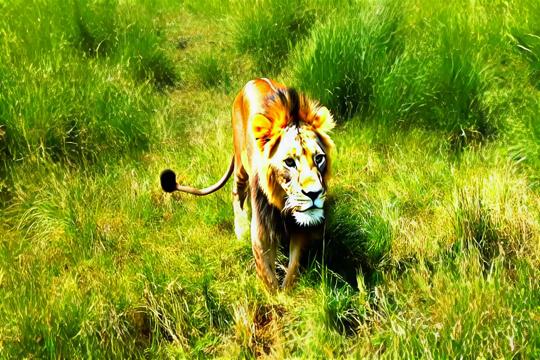}\hfill%
  \includegraphics[width=\imagewidth]{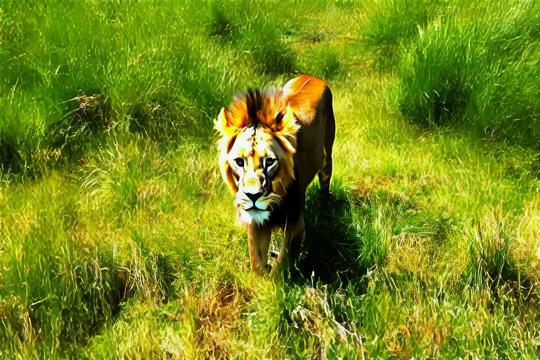}\hfill%
  \includegraphics[width=\imagewidth]{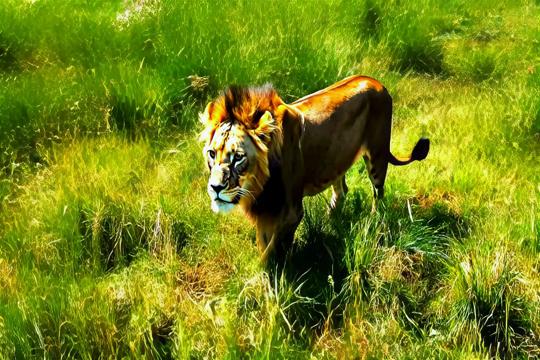}\hfill%
  \includegraphics[width=\imagewidth]{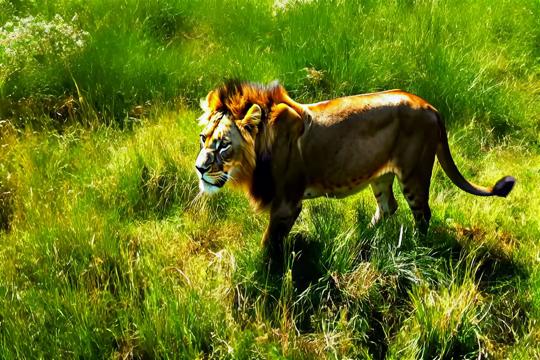}\hfill%
  \includegraphics[width=\imagewidth]{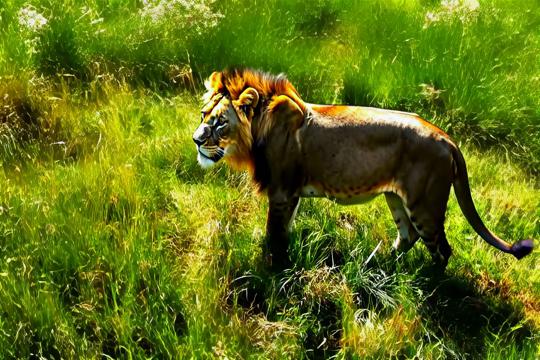}\hfill%
  \includegraphics[width=\imagewidth]{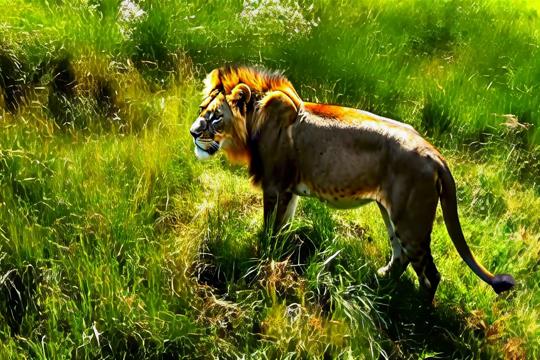}\\%
  \includegraphics[width=\imagewidth]{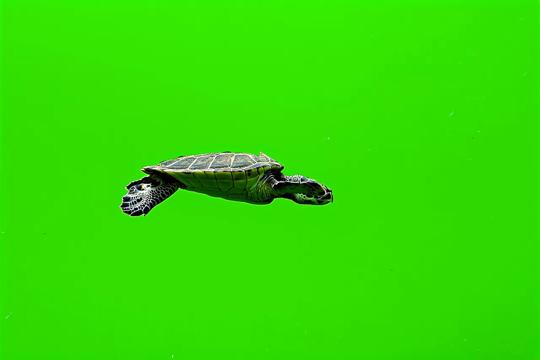}\hfill%
  \includegraphics[width=\imagewidth]{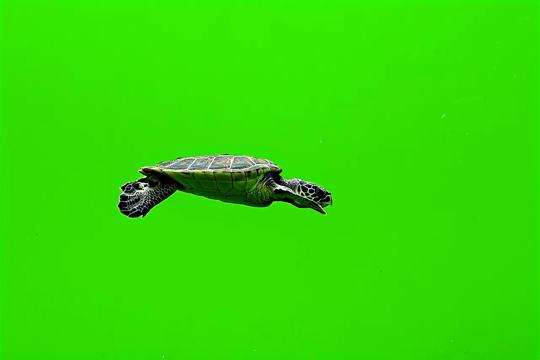}\hfill%
  \includegraphics[width=\imagewidth]{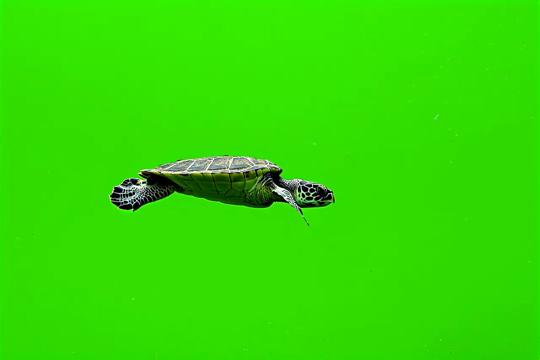}\hfill%
  \includegraphics[width=\imagewidth]{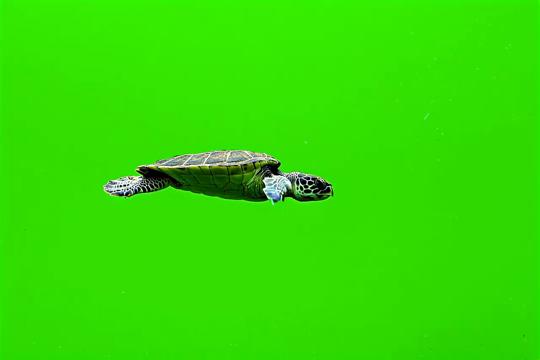}\hfill%
  \includegraphics[width=\imagewidth]{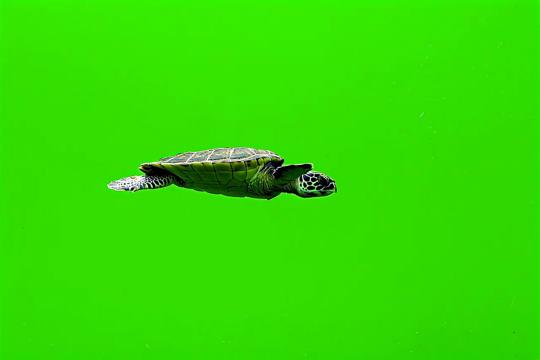}\hfill%
  \includegraphics[width=\imagewidth]{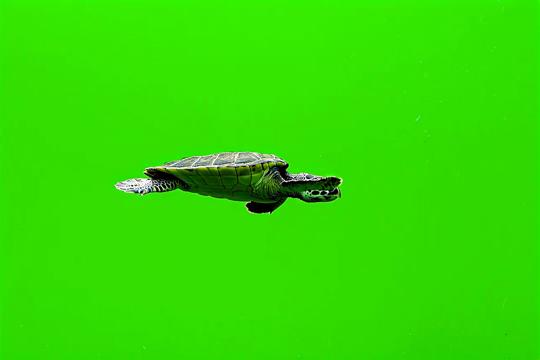}\hfill%
  \includegraphics[width=\imagewidth]{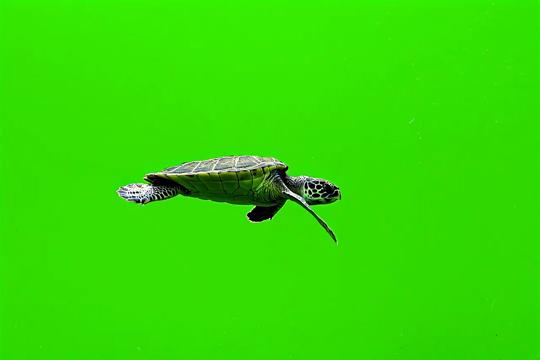}\hfill%
  \includegraphics[width=\imagewidth]{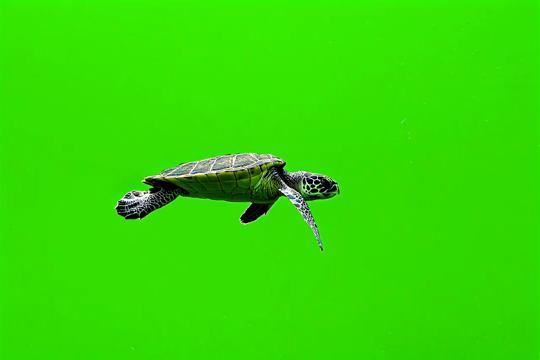}\\%
  \includegraphics[width=\imagewidth]{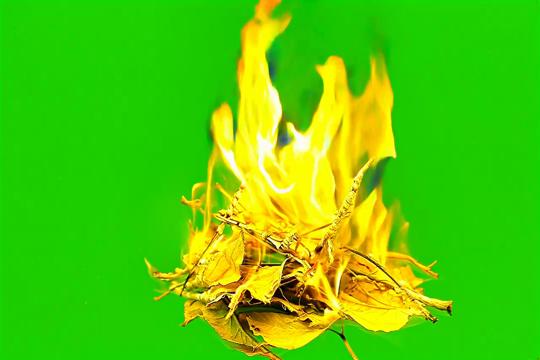}\hfill%
  \includegraphics[width=\imagewidth]{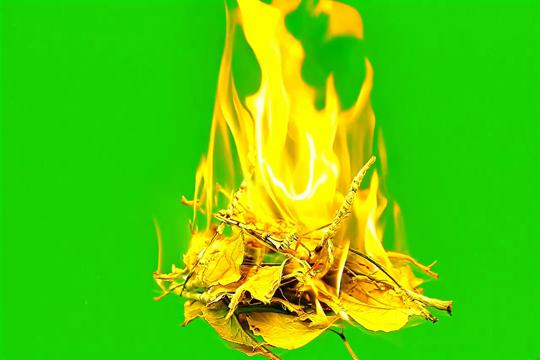}\hfill%
  \includegraphics[width=\imagewidth]{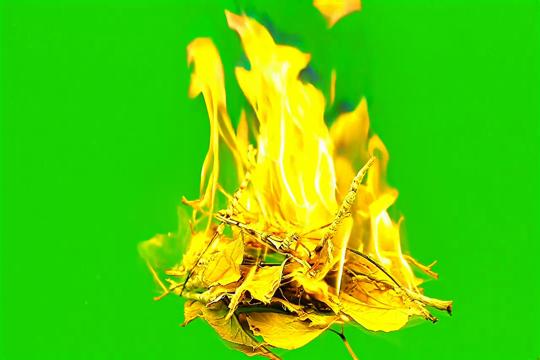}\hfill%
  \includegraphics[width=\imagewidth]{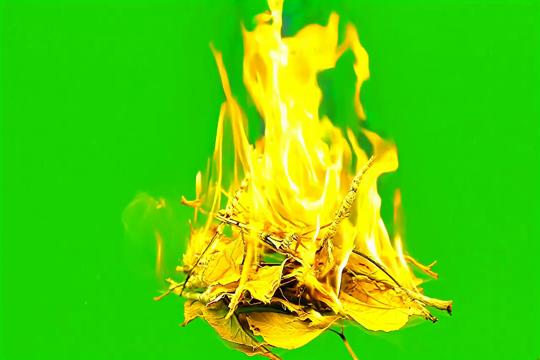}\hfill%
  \includegraphics[width=\imagewidth]{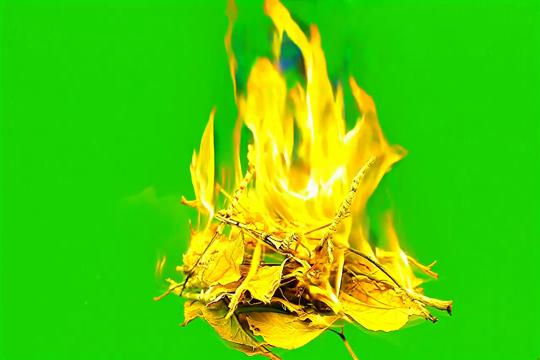}\hfill%
  \includegraphics[width=\imagewidth]{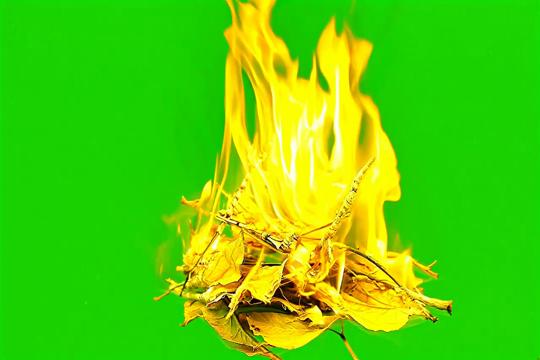}\hfill%
  \includegraphics[width=\imagewidth]{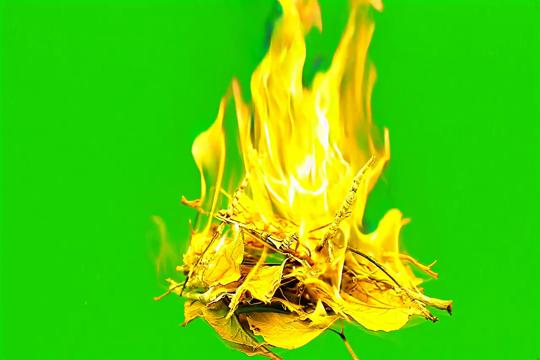}\hfill%
  \includegraphics[width=\imagewidth]{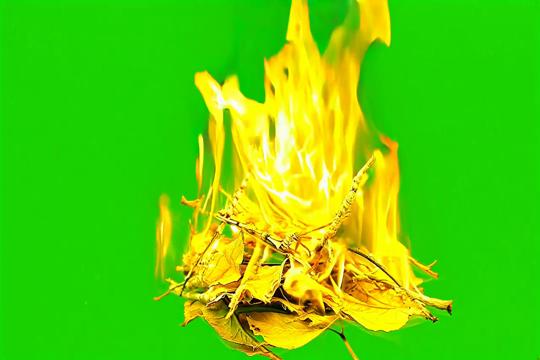}\\%
  \includegraphics[width=\imagewidth]{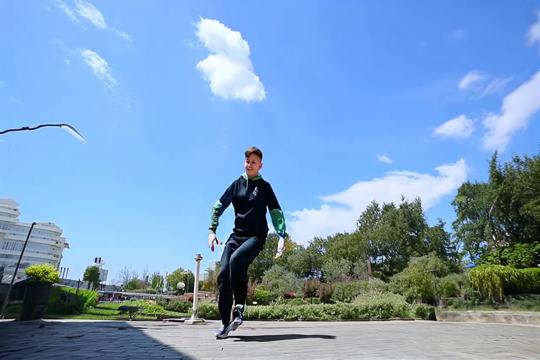}\hfill%
  \includegraphics[width=\imagewidth]{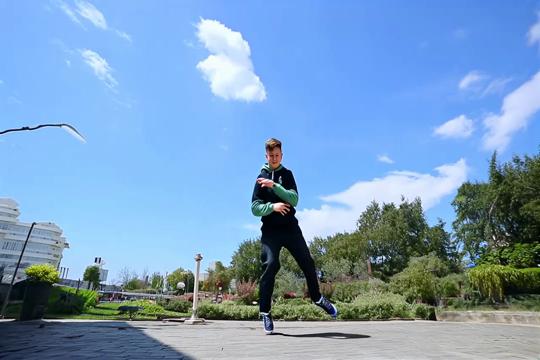}\hfill%
  \includegraphics[width=\imagewidth]{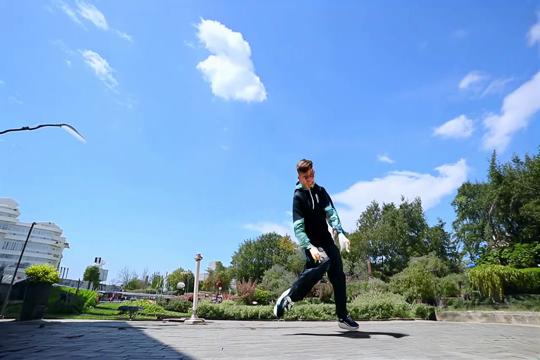}\hfill%
  \includegraphics[width=\imagewidth]{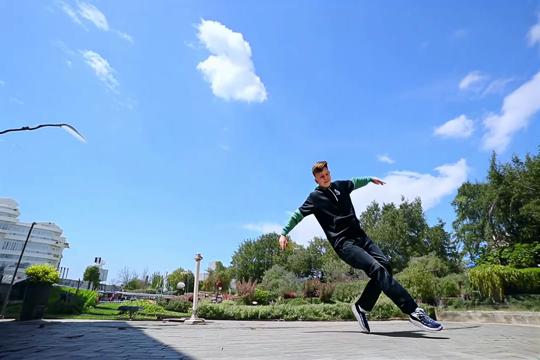}\hfill%
  \includegraphics[width=\imagewidth]{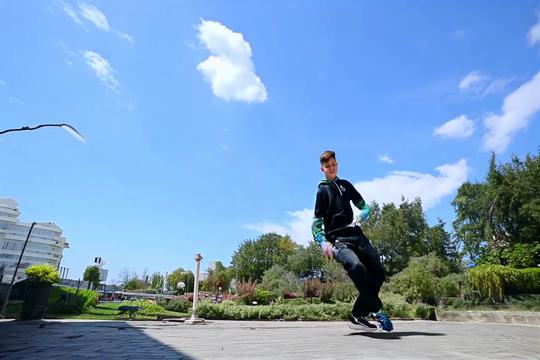}\hfill%
  \includegraphics[width=\imagewidth]{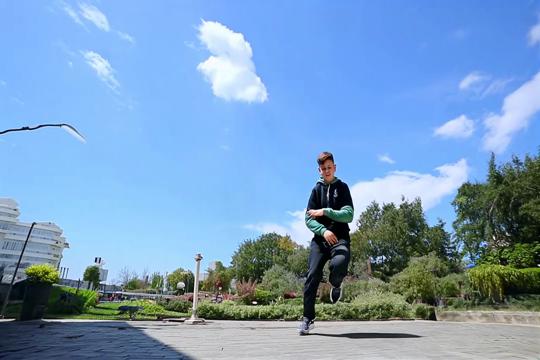}\hfill%
  \includegraphics[width=\imagewidth]{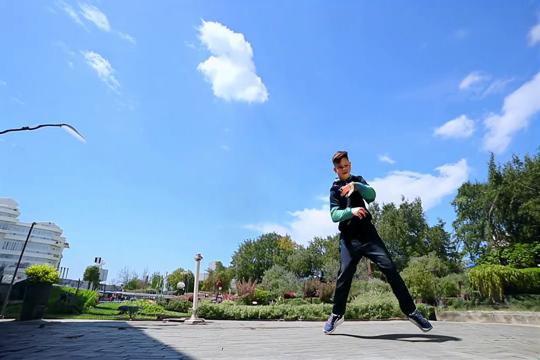}\hfill%
  \includegraphics[width=\imagewidth]{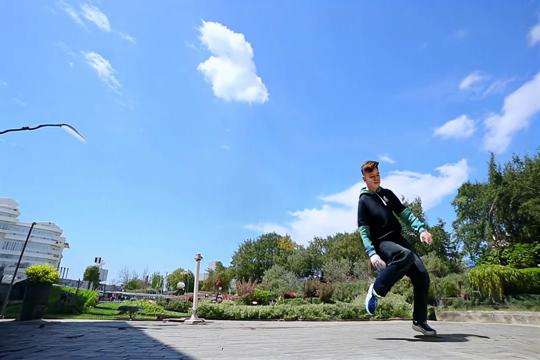}\\%
  \includegraphics[width=\imagewidth]{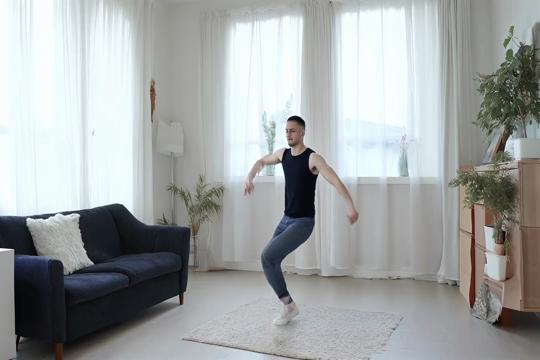}\hfill%
  \includegraphics[width=\imagewidth]{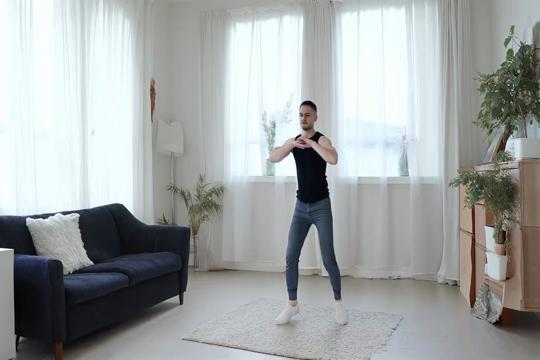}\hfill%
  \includegraphics[width=\imagewidth]{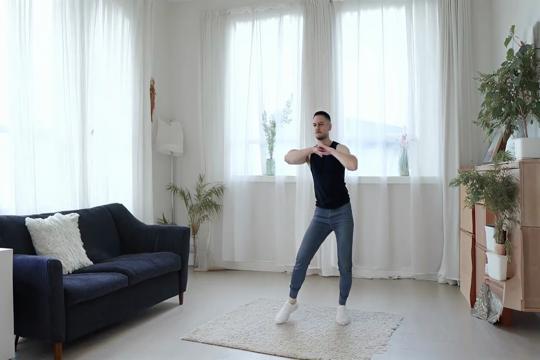}\hfill%
  \includegraphics[width=\imagewidth]{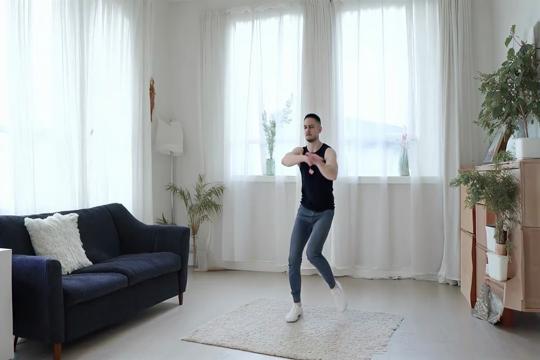}\hfill%
  \includegraphics[width=\imagewidth]{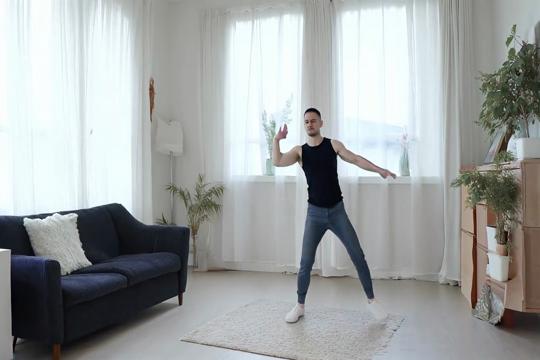}\hfill%
  \includegraphics[width=\imagewidth]{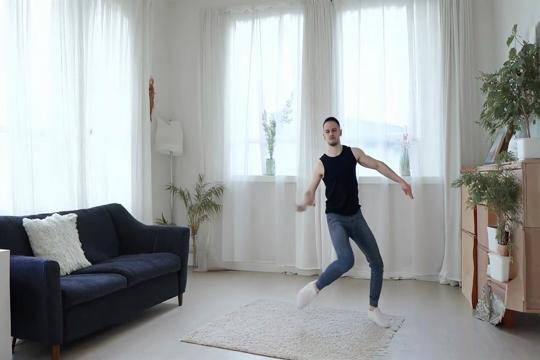}\hfill%
  \includegraphics[width=\imagewidth]{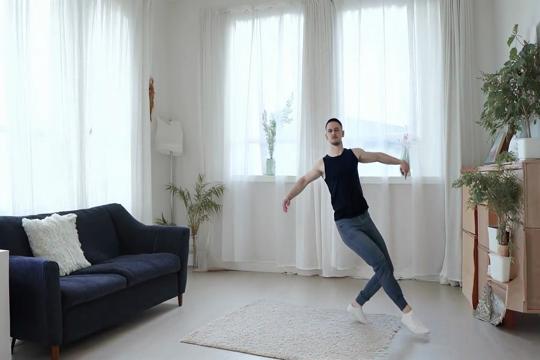}\hfill%
  \includegraphics[width=\imagewidth]{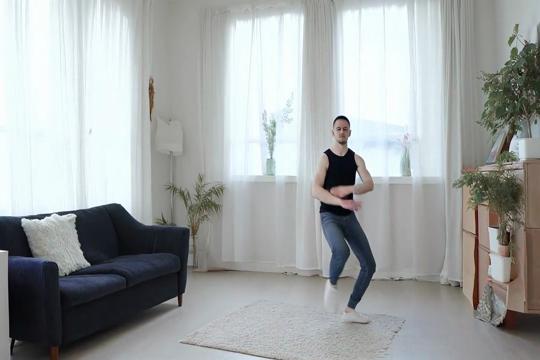}\\%
  \caption{More visualization of generated videos for large camera motion (row~1,2), layer decomposition (row~3,4), and large human motion(row~5,6).}
  \label{fig:frames_2}
\end{figure*}

In this section, we provide additional visualizations of the data curation experiments and the ablation studies. Figure~\ref{fig:bad-asset-layer} and Figure~\ref{fig:bad-asset-spin} show the effect of using poor quality asset and rendering respectively. Figure~\ref{fig:overtrain} shows the effect of excessive training on synthetic data. Color patterns are introduced into the generation model. Figure~\ref{fig:caption-showcase-c} gives an example of fine-grained and generic captions and an example of using special tags. Figure~\ref{fig:sim-drop1} and Figure~\ref{fig:sim-drop2} show the comparison between videos from generation with and without \emph{SimDrop}. Lastly, Figure~\ref{fig:change_bg} showcases the layer decompostion videos can use to separate out dynamic objects (\eg animals, fluids) to enable video matting.
Finally,~\Cref{fig:frames_2} shows more generated videos across all three tasks.

% \begin{figure}
%     \centering
%     \includegraphics[width=1\linewidth]{visualizations/bad_render.png}
%     \caption{Bad rendering comparison.}
%     \label{fig:bad-render}
% \end{figure}

% \begin{figure}
%     \centering
%     \includegraphics[width=1\linewidth]{visualizations/bad_quality.png}
%     \caption{Bad rendering comparison.\TODO{change top row}} 
%     \label{fig:bad-quality}
% \end{figure}

% \begin{figure}
%     \centering
%     \includegraphics[width=1\linewidth]{visualizations/caption.png}
%     \caption{A illustration of generating captions for synthetic videos. We make comparison between generic captions and detailed captions, and find that detailed caption is critical in improving human motion generation. We also illustrate a comparison between adding special tags and without special tags. We find that special tags are critical in bridging the appearance gap between real and synthetic videos while letting it learn the knowledge.}
%     \label{fig:caption-showcase}
% \end{figure}

\eat{
\begin{figure}
    \centering
    \includegraphics[width=1\linewidth]{figures/fig5x3.png}
    \caption{\TODO{Loss curve; several screenshots of rendering quality get negatively affected with different mix ratio}.}
    \label{fig:loss-curve}
\end{figure}
}

\subsection{Evaluation Prompts}
\label{sec:prompt_details}

\textbf{Large Human Motion} 

Dancing:
\begin{itemize}
    \item A dancer practicing at home
    \item In a street setting, a teenager is performing breakdance moves, including leaning back, balancing on one leg, and rhythmically moving arms.
    \item An attractive man energetically dances, featuring lively movements. He crosses his arms and vigorously moves his legs, imitating horse riding and other whimsical actions.
    \item A young woman gracefully pirouettes on one foot, her other leg bent elegantly and arms outstretched for balance and flair. She transitions through various spins, showcasing a dynamic dance routine that blends elements of northern soul dancing. She dances in a bustling urban plaza, or a serene beach at sunset, or a lively street festival, or, a beautifully lit dance studio. Each setting captures the fluidity and energy of her movements, adding depth and variety to her performance.
    \item A young woman is performing breakdance moves, including leaning back and balancing on one leg while engaging arms rhythmically.
    \item A woman dancing on grassland during sunset
    \item On a beach, an Ultraman from Japanese TV show is spinning around on one foot while keeping other leg bent and arms extended for balance and style. It performs multiple spins, emphasizing a dance move commonly associated with northern soul dancing.
    \item In a bright dance room, a young woman is performing a dance with enthusiastic movements The person crosses arms and moves legs energetically, mimicking riding a horse and performing other playful gestures.
    \item A young woman is performing a breakdance move, starting with a dynamic step and then transitioning into a series of fluid body movements and rhythmic steps.
    \item A handsome man initiates with a dynamic step followed by a series of fluid body motions and rhythmic steps.
\end{itemize}
Gymnastics:
\begin{itemize}
    \item In a bright dance room, A man executes a backflip by initially crouching low, launching himself upwards, rotating backwards in midair before returning to a standing position on his feet.
    \item In a well-lit dance studio, A woman performs a gymnastics moves to flip her body. Her backflip is to first crouch low, then rotating upwards and backward in midair, eventually landing back in a standing position.
    \item A man performs a backflip by first squatting down, then launching itself into the air, flipping backward, and finally landing back onfeet on grassland under sunshine.
    \item In a sunny grassland, a woman executes a backflip by initially crouching, then springing into the air, rotating backward, and ultimately landing on her feet.
    \item A female athlete performs a backflip by first squatting down, then launching itself into the air, flipping backward, and finally landing back onfeet during the floor execrise event at the Olympic Games.
    \item During the floor exercise event at the Olympic Games, a male athlete performs a stunning backflip. He begins by squatting down low, gathering his strength and focus. With a powerful burst of energy, he launches himself into the air, his body gracefully arching as he flips backward. The sunlight glints off his muscular form as he completes the rotation, and he lands solidly on his feet, his expression a mix of concentration and triumph.
    \item A man Moves with dynamic energy, shifting from a standing position to a deep crouch, then rotating her body mid-air before landing upright on the sunlit grassland.
    \item A woman is moving dynamically, transitioning from a standing position to a deep crouch and then rotating body mid-air before returning to an upright stance on grassland under sunshine.
    \item During the floor exercise event at the Olympic Games, a female athlete moves with dynamic precision. She transitions from a standing position to a deep crouch, then launches herself into the air, rotating her body mid-flight before landing gracefully back on her feet.
    \item At the Olympic Games' floor exercise event, a male athlete showcases his agility by swiftly dropping into a deep crouch from a standing position. He then propels himself into the air, executing a mid-air rotation, and lands back on his feet with precision and grace.
\end{itemize}

\textbf{Large Camera Motion}
\begin{itemize}
    \item A lion standing on the grass. spin shot.
    \item An astronaut riding a horse, high definition, 4k. spin shot.
    \item A panda swimming underwater. spin shot.
    \item Video of sailboat on a lake during sunset. spin shot.
    \item Variety of succulent plants on a garden. spin shot.
    \item A birthday cake in the plate. spin shot.
    \item Big cargo ship passing on the shore. spin shot.
    \item Time lapse video, sunrise of the Great Wall. spin shot.
    \item A tree with Halloween decoration. spin shot.
    \item A Labrador dog wearing glasses and casual clothes is lying on the bed reading. spin shot.
\end{itemize}

\textbf{Layer Decomposition}
\begin{itemize}
    \item A lion standing in a green background.
    \item A lion running in a green background.
    \item Turtle swimming in a green background.
    \item An african penguin walking in a green background.
    \item Variety of succulent plants in a green background.
    \item Leaves swaying in the wind in a green background.
    \item A stack of dried leaves burning in a green background.
    \item Big cargo ship like in the movies passing in a green background.
    \item Helicopter landing in a green background.
    \item A young woman is performing breakdance moves, including leaning back and balancing on one leg while engaging arms rhythmically in a light blue background.
\end{itemize}

\subsection{Human Evaluation Details}
Our user study videos are available on the project website. We invite the community to also rate the videos.
\label{sec:human_details}

\textbf{Large Human Motion} For large human motions, we asks our human raters to examine how many out of the generated videos in each video show no collapse in human body structure. Specifically, we ask them to focus on the limbs and torso areas. 
%For commercial models, we are not sure if our prompt is directly feed into the video generation model or passed through a rephraser. Therefore, we made little modification to explicitly ask for the full body video to ensure fair comparison. If the full body is shown no more than 2 seconds out of 5 seconds in the video. Then the video is automatically marked as failed. In addition, we asked the human raters to evaluate if the person in the video looks animated, which is a potential pitfall for our approach.If any of the following question is yes, please mark the video as 0
The detailed rules are as following:
1. Does the video include the full body of the person (all four limbs) for more than 2 seconds?
2. Is the video bascially showing what is specified by the prompt, including background and motion?
3. Does the person in the video looks animated?
4. Is there limbs or torso addition/missing from the video?
5. Is there transition of body parts that are obviously unnatural (e.g. switching body parts at the same location)? 

Please Note:
1. DO NOT focus your judgement on these part of the human body: hands, feet, or face
2. DO NOT judge the asethetics or naturalness of the human motion, please just focus on human body integrity

\textbf{Large Camera Motion} For Large camera motion, we instruct the human raters to focus on the object and the degree which the picture rotates. The detailed rules are as following:
If any of the following question is yes, please mark the video as 0
1. If the object appear in the video is corrupt, unnatural, or animated
2. If the background is not of pure color as instructed by the prompt

\textbf{Layer Decomposition} For layer decompostion, we instruct the human raters to focus on the object and the background quality. The detailed rules are as following:
If any of the following question is yes, please mark the video as 0
1. If the object appear in the video does not spin at all.
2. If the object appear in the video spins but the background does not move with the object
3. If the object appear in the video corrupts, becomes unnatural or looks animated.

%\subsection{Additional Visualizations}

\end{document}